\documentclass[usenatbib,longnamesfirst,usedcolumn]{mn2e} % Normal MNRAS format
\usepackage{longtable} % Allows table to span multiple pages
\usepackage{graphicx}

\title[Open clusters Czernik 14, Haffner 14, Haffner 17 and King 10]
      {A Comprehensive study of open clusters Czernik 14, Haffner 14, Haffner 17 and King 10 using multicolour photometry and Gaia~DR2 astrometry}

\author[Bisht et al.]
   {D. Bisht$^{1}$,\thanks{E-mail: dbisht@ustc.edu.cn; zhuqf@ustc.edu.cn; rkant@aries.res.in; alokdurgapal@gmail.com; geetarangwal91@gmail.com}
Qingfeng Zhu$^{1}$,  
R.\ K.\ S.\ Yadav$^{2}$,   
Alok Durgapal$^{3}$, and    
Geeta Rangwal$^{3}$
\\\\
    $^{1}$ Key Laboratory for Researches in Galaxies and Cosmology, University of Science and
           Technology of China, Chinese Academy of Sciences,\\
           Hefei, Anhui, 230026, China\\ 
    $^{2}$Aryabhatta Research Institute of Observational Sciences, 
           Manora Peak, Nainital 263129, India\\
    $^{3}$Department of physics, DSB campus, KU Nainital, India\\
}

\begin{document}

\date{\today}

%\pagerange{\pageref{firstpage}--\pageref{lastpage}} \pubyear{2007}

\maketitle

\label{firstpage}

\begin{abstract}

This paper presents an investigation on the four open clusters Czernik 14, Haffner 14, Haffner 17 and King 10 located near the Perseus arm of
Milky Way Galaxy using {\bf Gaia DR2}, {\bf 2MASS}, {\bf WISE}, {\bf APASS} and {\bf Pan-STARRS1} data sets. We find normal
interstellar extinction in twelve photometric bands for these clusters. Likely cluster members are identified as 225, 353, 350
and 395 for Czernik 14, Haffner 14, Haffner 17 and King 10, respectively by using Gaia~DR2 proper
motion data. Radii are determined as 3.5, 3.7, 6.2 and 5.7 arcmin for Czernik 14, Haffner 14, Haffner 17 and King 10 respectively.
Mean proper motions in RA and DEC are estimated as ($-0.42\pm0.02$, $-0.38\pm0.01$),
($-1.82\pm0.009$, $1.73\pm0.008$), ($-1.17\pm0.007$, $1.88\pm0.006$) and ($-2.75\pm0.008$, $-2.04\pm0.006)$ ~mas~yr$^{-1}$ for Czernik 14,
Haffner 14, Haffner 17 and King 10, respectively. The comparison of observed CMDs with solar metallicity isochrones leads to an age 
of $570\pm60$, $320\pm35$, $90\pm10$ and $45\pm5$ Myr for these clusters. The distances
$2.9\pm0.1$, $4.8\pm0.4$, $3.6\pm0.1$ and $3.8\pm0.1$ kpc determined using parallax are comparable with the values derived by the 
isochrone fitting method. Mass 
function slopes are found to be in good agreement with the Salpeter value. The total masses are derived as 
$348$, $595$, $763$ and $1088~ M_{\odot}$ for the clusters Czernik 14, Haffner 14, Haffner 17 and King 10, respectively.
Evidence for the existence of mass-segregation effect is observed in each cluster. Using the Galactic potential model,
Galactic orbits are derived for the clusters. The present study indicates that all clusters under study fallow a circular path 
around the Galactic center.

\end{abstract}


% sec:intro
%________________________________________________________________

\section{Introduction} \label{sec:intro}

\begin{figure}
\begin{center}
\hbox{
\includegraphics[width=4.2cm, height=4.2cm]{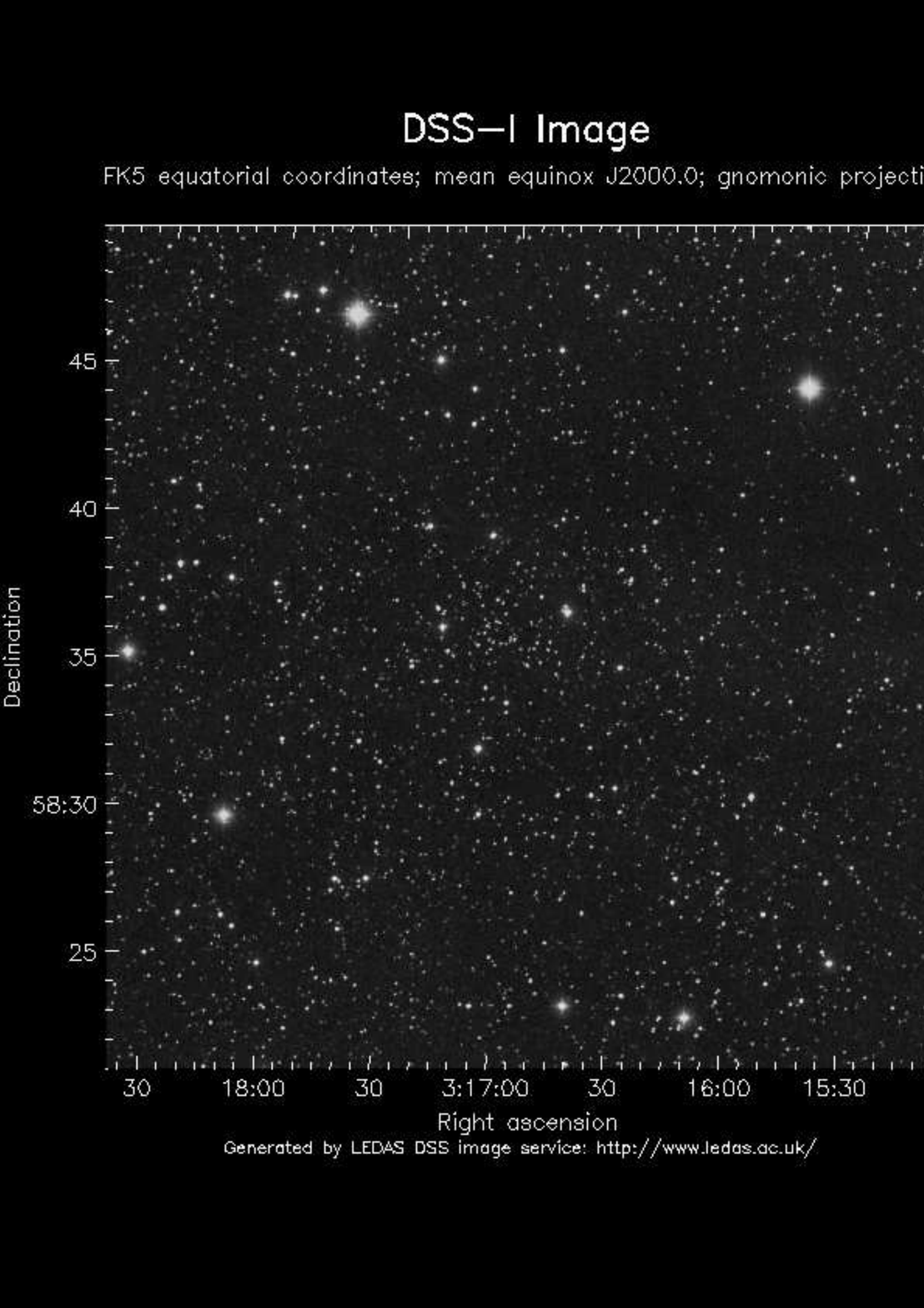}
\includegraphics[width=4.2cm, height=4.2cm]{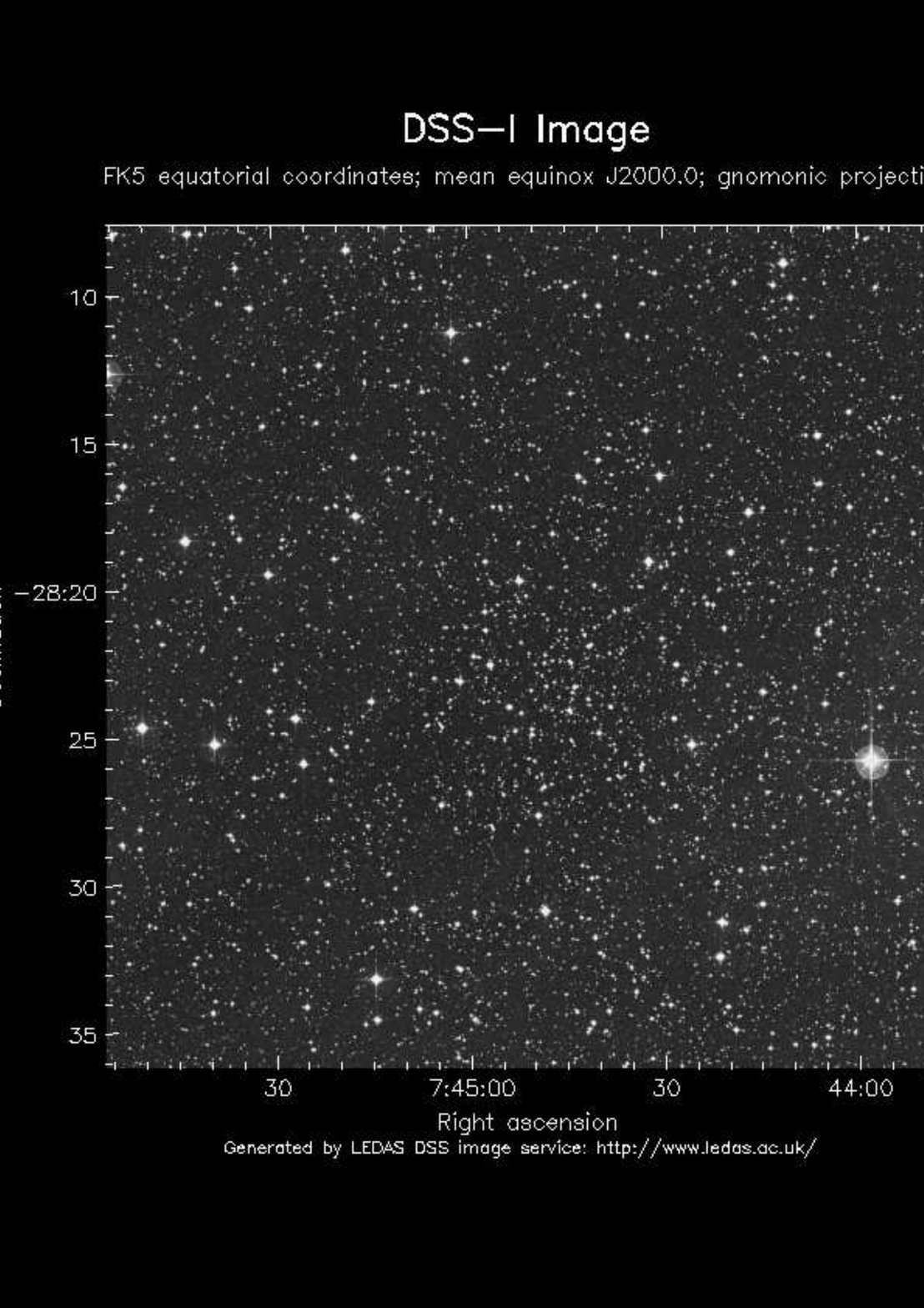}
}
\hbox{
\includegraphics[width=4.2cm, height=4.2cm]{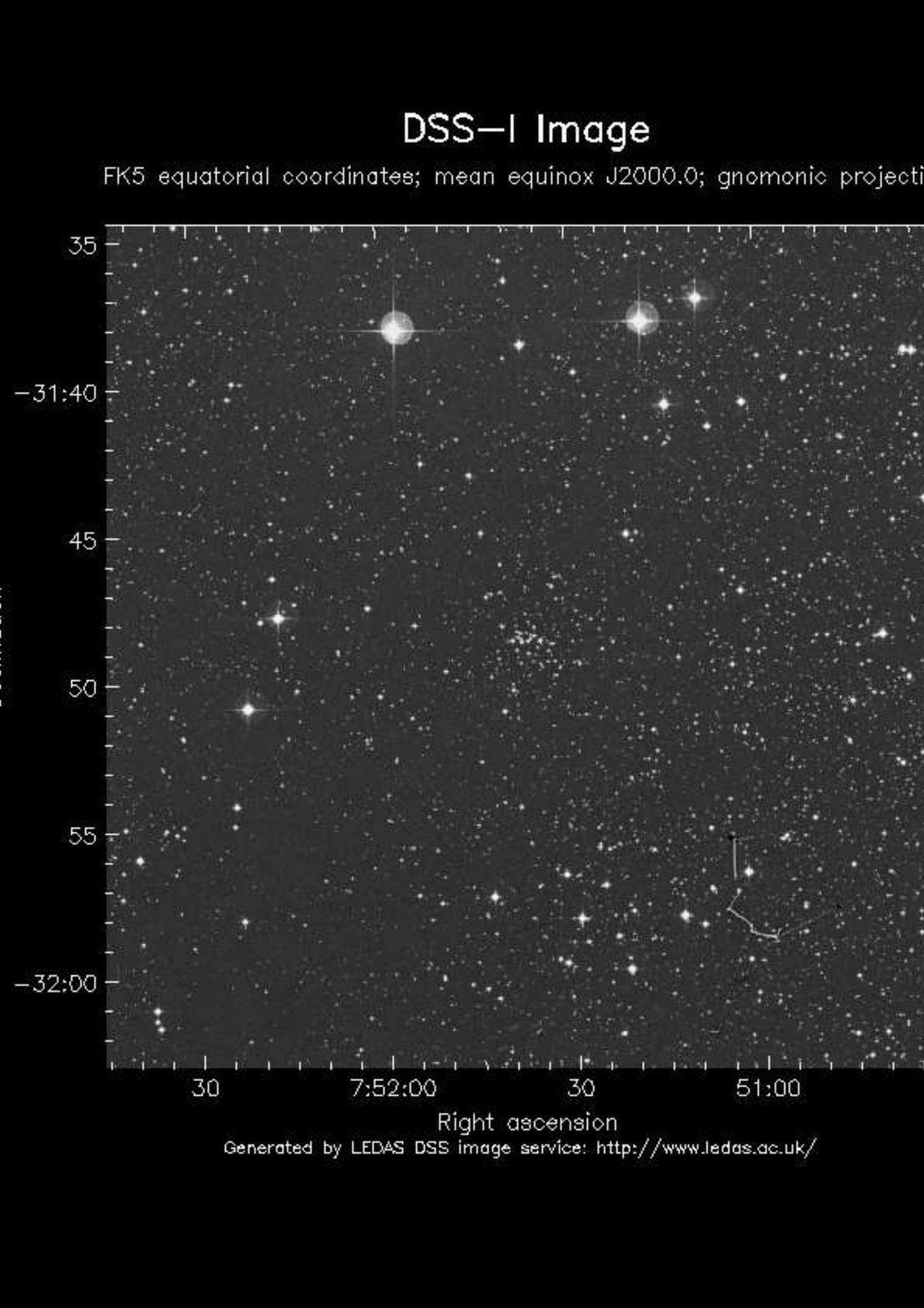}
\includegraphics[width=4.2cm, height=4.2cm]{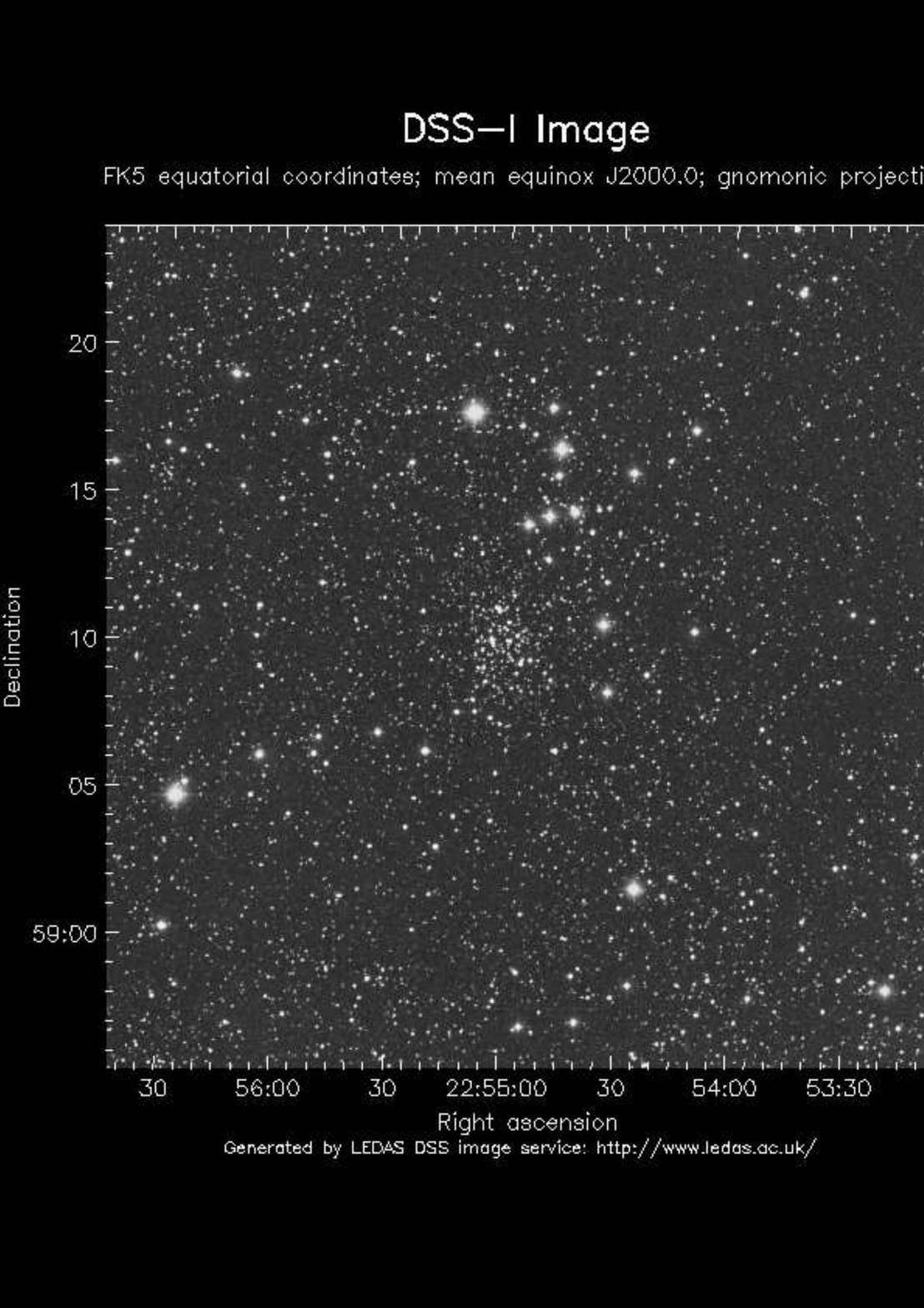}
}
\caption{Identification maps of clusters Czernik 14 (Top left), Haffner 14 (Top right), Haffner 17 (Bottom Left) and
King 10 (Bottom right) taken from LEDAS.}
\label{id}
\end{center}
\end{figure}

Open clusters (OCs) are convenient probes in studying the structure and evolution of the Milky Way Galaxy. Since OCs
are formed by the collapse and fragmentation of huge molecular cloud (e.g. Harris \& Pudritz 1994; Bate et al. 2003), they become
ready samples to study the stellar evolution history. The second Gaia data release (Gaia~DR2) provides accurate five parameter
astrometric data (Positions, proper motions and parallaxes) for more than 1.3 billion sources. Gaia~DR2 is very effective and 
it will be a showcase of how effectively these data can be used to identify cluster members, especially in the crowded regions.
Open clusters have been the subject of many studies over the past decades. They are very important and widely used to distinguish 
the Galactic disk properties, such as understanding the spiral arms of the
Milky Way (e.g. Bonatto et al. 2006; van den Bergh 2006), the stellar metallicity gradient (Janes 1979; Geisler et al. 1997;
Frinchaboy et al. 2013) and the age-metallicity relationship in the Galactic disk
(e.g. Carraro \& Chiosi 1994; Carraro et al. 1998; Salaris et al. 2004; Margrini et al. 2009). Young open clusters are essential
laboratories to understand about star formation scenario. The intermediate age clusters are very useful in testing
stellar isochrones and dynamical evolution of cluster stars. Over the last decade, the clusters inspection rate has been lifted exceptionally well.
This can be attributed to the appearance of multiple wide-ranging Near-Infrared (NIR) and Mid-Infrared (MIR) surveys, such as GLIMPSE, 
2MASS (Skrutskie et al. 2006), UKIDSS-GPS (Lucas et al. 2008), VISTA-VVV (Minniti et al. 2010) and $WISE$ (Wright et al. 2010).

The most impressive cause to employ 2MASS, WISE and Gaia~DR2 database is that they deliver us a robust tool to recognize star
clusters behind the broad hydrogen clouds. The 2MASS survey has been manifest to be a strong tool in the investigation of the
framework and stellar content of open clusters (Bonatto \& Bica 2003). Lately, many OCs have been discovered by exploring $2MASS$
data (Kronberger et al. 2006);  Froebrich et al. (2007); Koposov et al. (2008);  Glushkova et al. (2010)). It is expected
that the Gaia mission will fully transform our knowledge of the structure and dynamics of the Galaxy (Gaia 
Collaboration et al. 2016a). The latest version of the Gaia data (Gaia DR2) covers more than 1.3 billion sources
and was made public on 2018 April 24 (Gaia Collaboration et al. 2016a,b). Gaia DR2 catalogue contains three 
photometric bands  $(G, G_{BP}, G_{RP})$, the precise astrometry at the sub-milliarcsecond level and parallax
(Gaia collaboration et al., 2018b). 

The OCs mass function gives direction about the star formation process. Phenomenal works have been done on the mass function of OCs during
the last two decades (e.g. Baume et al. 2004, Durgapal and Pandey 2001, Pandey et al. 2001, Phelps and Janes 1993, Piatti et al. 2002,
Piskunov et al. 2004, Sagar and Griffiths 1998, Prisinzano et al. 2001, Scalo 1986, Scalo et al. 1998, Sung and Bessell 2004, Yadav and
Sagar 2002, 2004a, Bisht et al. 2017 \& 2019 and Geeta et al. 2019) . Despite all these remarkable works, the universality of initial 
mass functions is still a question of debate (Elmegreen 2000, Larson and Nakamoto 1999). Additionally, the mass segregation evaluation
in OCs gives proof about the spatial arrangement of massive and faint stars within the area of the cluster.

In the present study, we also focus on the study of Galactic orbits of these open star clusters. Orbits carry information about the history of path 
of the cluster. Orbits are given by the environment in which the cluster lives under the Galactic gravitational potential of the
Milky Way. Hence the orbits bring information about the clusters themselves, as well as about the Milky Way.

In this paper, we attempt to investigate the members, distances and mean proper motion of open
clusters Czernik 14, Haffner 14, Haffner 17 and King 10 using the high precision astrometry and photometry taken from the Gaia~DR2 catalogue.
The main goal of this article is to study the
fundamental parameters (age, distance, reddening, etc.) of the clusters, luminosity functions (LFs), mass functions (MFs)
and Galactic orbits of the target objects. We did not find a detailed study of these clusters in the literature. Available information
for these clusters is given below.

\begin{figure}
\centering
\includegraphics[width=8.5cm,height=8.5cm]{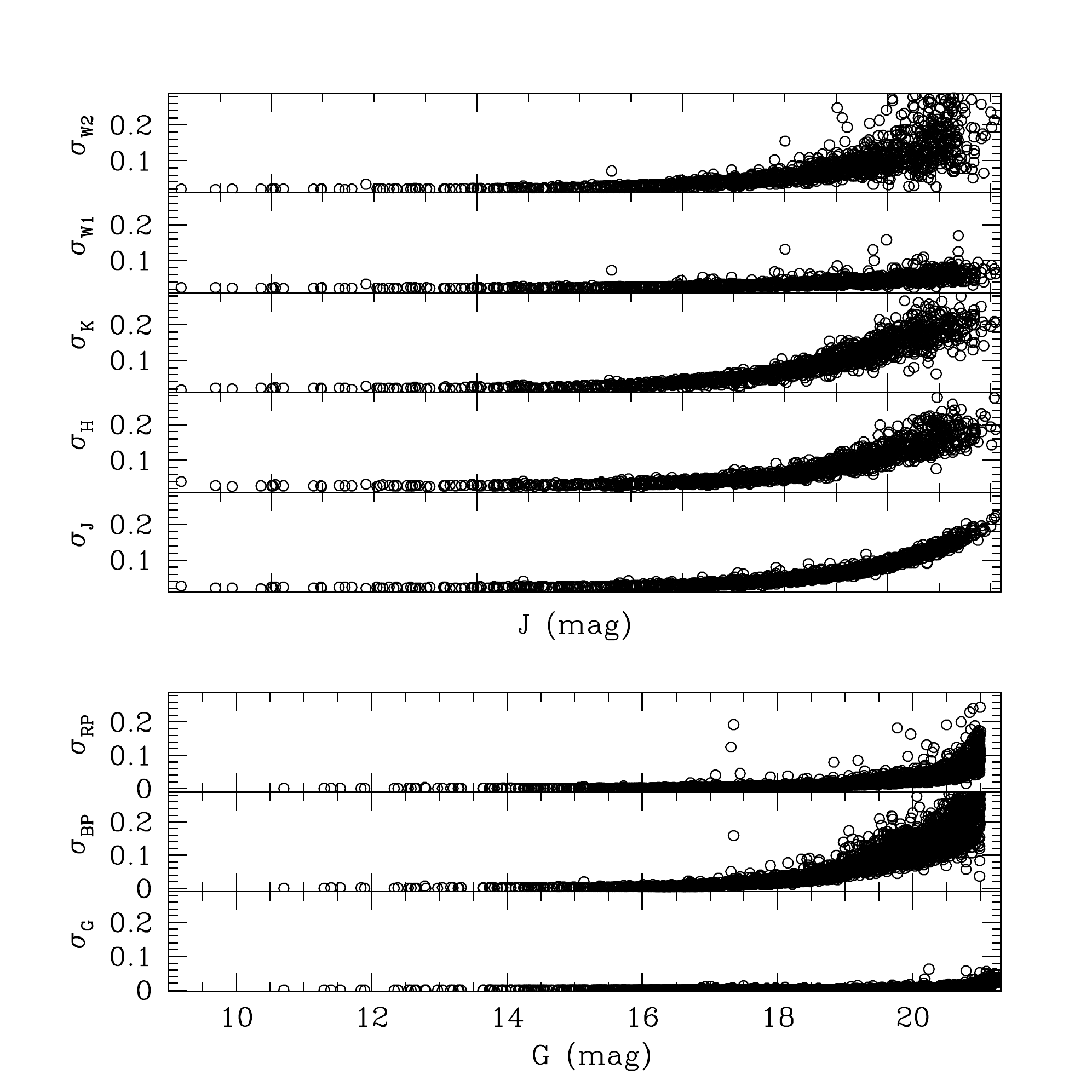}
\caption{Photometric errors in Gaia passbands $G$, $G_{BP}$ and $G_{RP}$ against $G$ magnitude in three lower panels while photometric errors
in $J$, $H$, $K$, $W_{1}$ and $W_{2}$ magnitudes against $J$ magnitude in upper five panels.}
\label{error}
\end{figure}

{\bf Czernik 14:} This cluster is positioned at $\alpha=3^{h} 16^{m} 54^{s}$ and 
$\delta=58^{\circ} 36^{\prime} 0^{\prime\prime}$ (J2000.0), corresponding to Galactic coordinates 
$l= 140^{\circ}.92$ and $b=0^{\circ}.9$. It is part of the Perseus arm in the second Galactic quadrant.

{\bf Haffner 14:} The location of this cluster is at $\alpha=7^{h} 44^{m} 51^{s}$ and $\delta=-28^{\circ} 22^{\prime} 0^{\prime\prime}$ (J2000.0), 
corresponding to Galactic coordinates $l= 243^{\circ}.98$ and $b=-2^{\circ}.1$.
This is close to Perseus arm in the third Galactic quadrant of the Milky Way.

{\bf Haffner 17:} This is positioned at $\alpha=7^{h} 51^{m} 37^{s}$ and
$\delta=-31^{\circ} 49^{\prime} 0^{\prime\prime}$ (J2000.0), corresponding to the galactic coordinates $l= 247^{\circ}.71$ and $b=-2^{\circ}.5$. 
As Haffner 14, it is also near to Perseus arm in the third Galactic quadrant. Pedreros (2000) has estimated basic parameters of this cluster 
using $UBV$ data. The interstellar reddening, distance modulus and cluster age was estimated as $1.26\pm0.04$, $12.3\pm0.2$ mag and 50 Myr.

{\bf King 10:} This is located at $\alpha=22^{h} 54^{m} 54^{s}$ and $\delta=59^{\circ} 10^{\prime} 0^{\prime\prime}$ (J2000.0), 
corresponding to Galactic coordinates $l= 108^{\circ}.48$ and $b=-0^{\circ}.4$. CCD $UBVRI$ photometric analysis is performed 
by Mohan et al. (1992). They found a variable reddening in the cluster with distance and age as 3.2 Kpc and $\le 50$ Myr. 
It is a part of the Perseus arm in the second Galactic quadrant.

This paper is organized as follows. Section 2 presents the different data sets used in this study. Section 3 describes the method for members
selection. In section 4, we derive different fundamental parameters of the clusters. Section 5 is devoted to the
luminosity and mass function of the clusters. The dynamical state of the clusters is described in section 6. In section 7, orbits of the
clusters are calculated. Finally, the conclusion of the present study is given in the last section.

\section{Data Used}

\begin{figure}
\centering
\includegraphics[width=6.5cm,height=6.5cm]{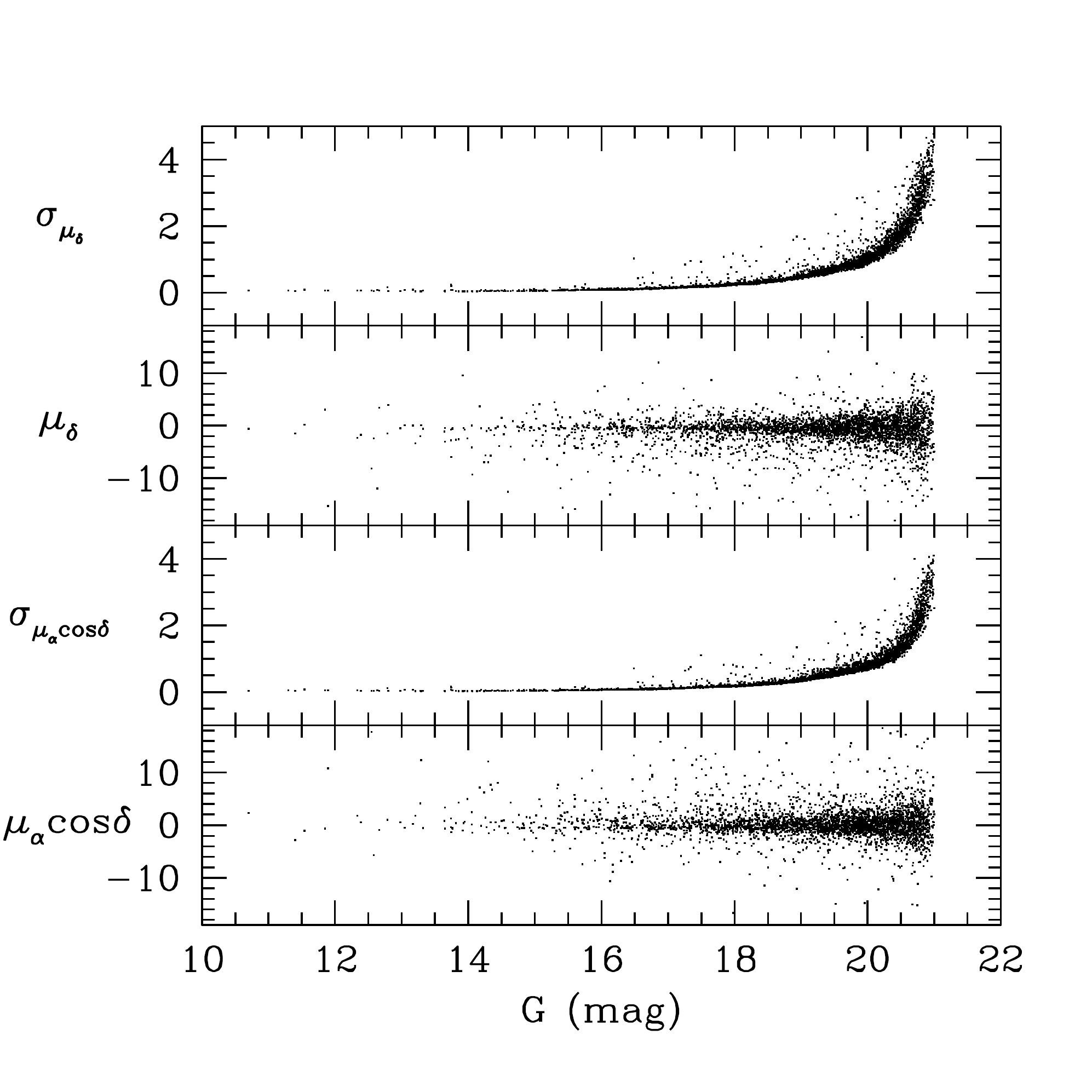}
\caption{Plot of Proper motions and their errors versus $G$ magnitude for the cluster Czernik 14 is shown as an example.}
\label{error_proper}
\end{figure}

We collected astrometric and photometric data from Gaia DR2 along with 
broad-band photometric data from APASS, Pan-STARRS1, 2MASS and WISE 
for clusters Czernik 14, Haffner 14, Haffner 17 and King 10. We cross-matched each catalog for the present analysis.
The descriptions of used data set are the following:

\begin{figure*}
\centering
\hbox{
\hspace{-0.5cm}\includegraphics[width=14.5cm, height=14.5cm]{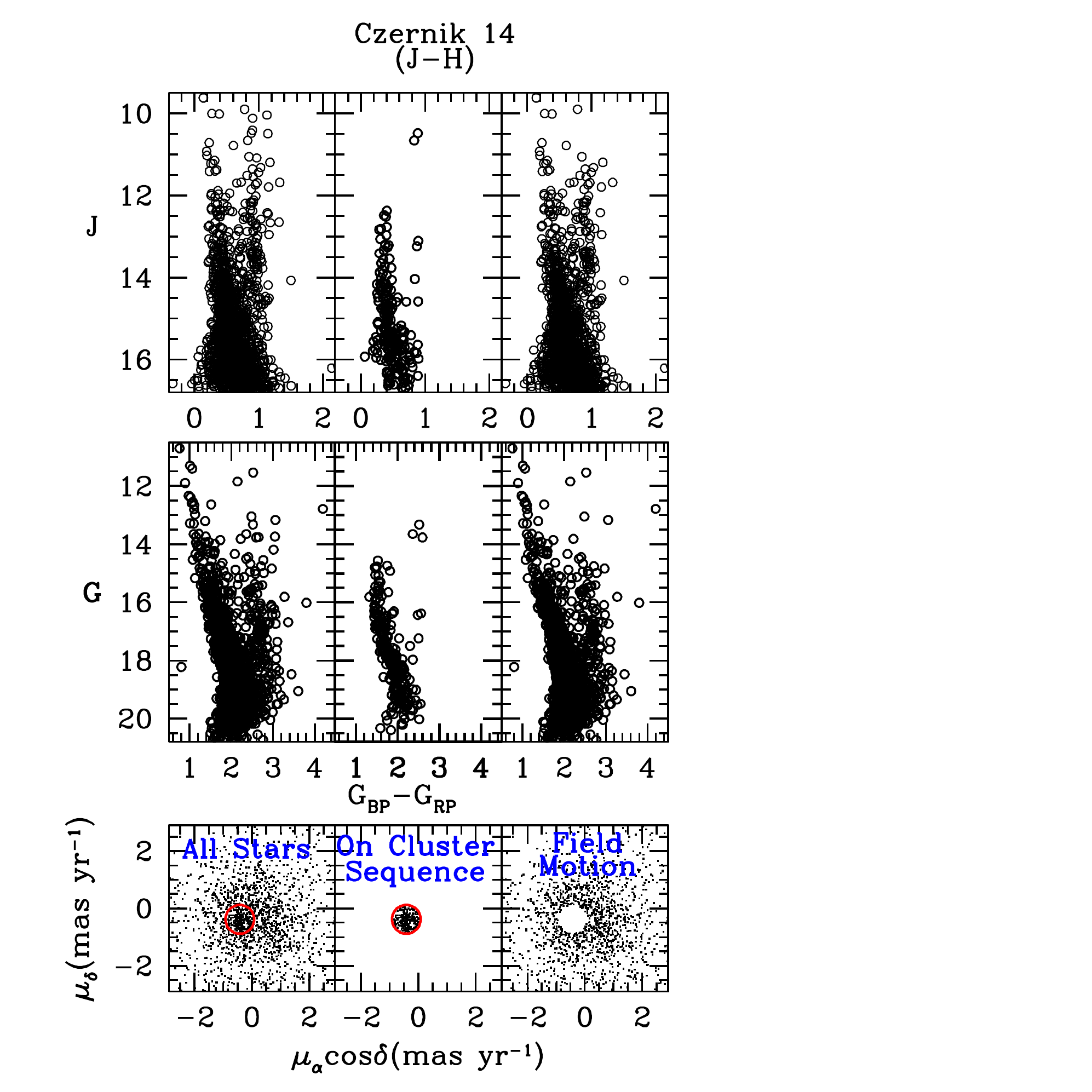}
\hspace{-6cm}\includegraphics[width=14.5cm, height=14.5cm]{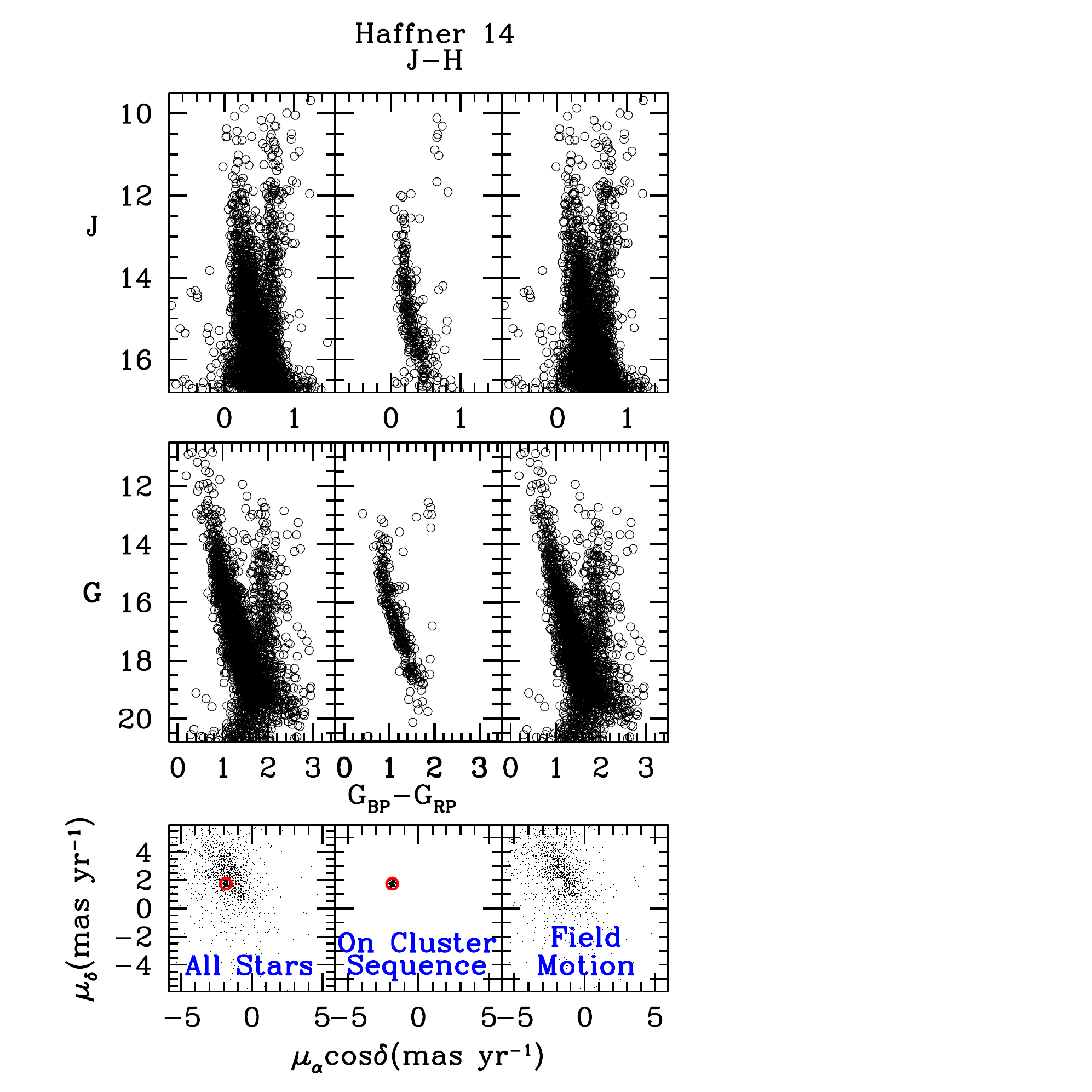}
}
\caption{(Bottom panels) Proper-motion vector point diagrams (VPDs) for clusters Czernik 14 and Haffner 14.
(Top panels) $J$ versus $J-H$ colour magnitude diagrams. (Middle panels) $G$ versus $(G_{BP}-G_{RP})$ colour magnitude diagrams.
For each cluster CMDs, (Left panel) The entire sample. (Center) Stars within the circle of $0.5$ and
$0.4~ mas~ yr^{-1}$ radius centered around the mean proper motion of Czernik 14 and Haffner 14 respectively. (Right) Probable
background/foreground field stars in the direction of these clusters. All plots show only stars with PM error smaller 
than $1~ mas~ yr^{-1}$ in each coordinate.}
\label{pm_dist}
\end{figure*}

\begin{figure*}
\centering
\hbox{
\hspace{-0.5cm}\includegraphics[width=14.5cm, height=14.5cm]{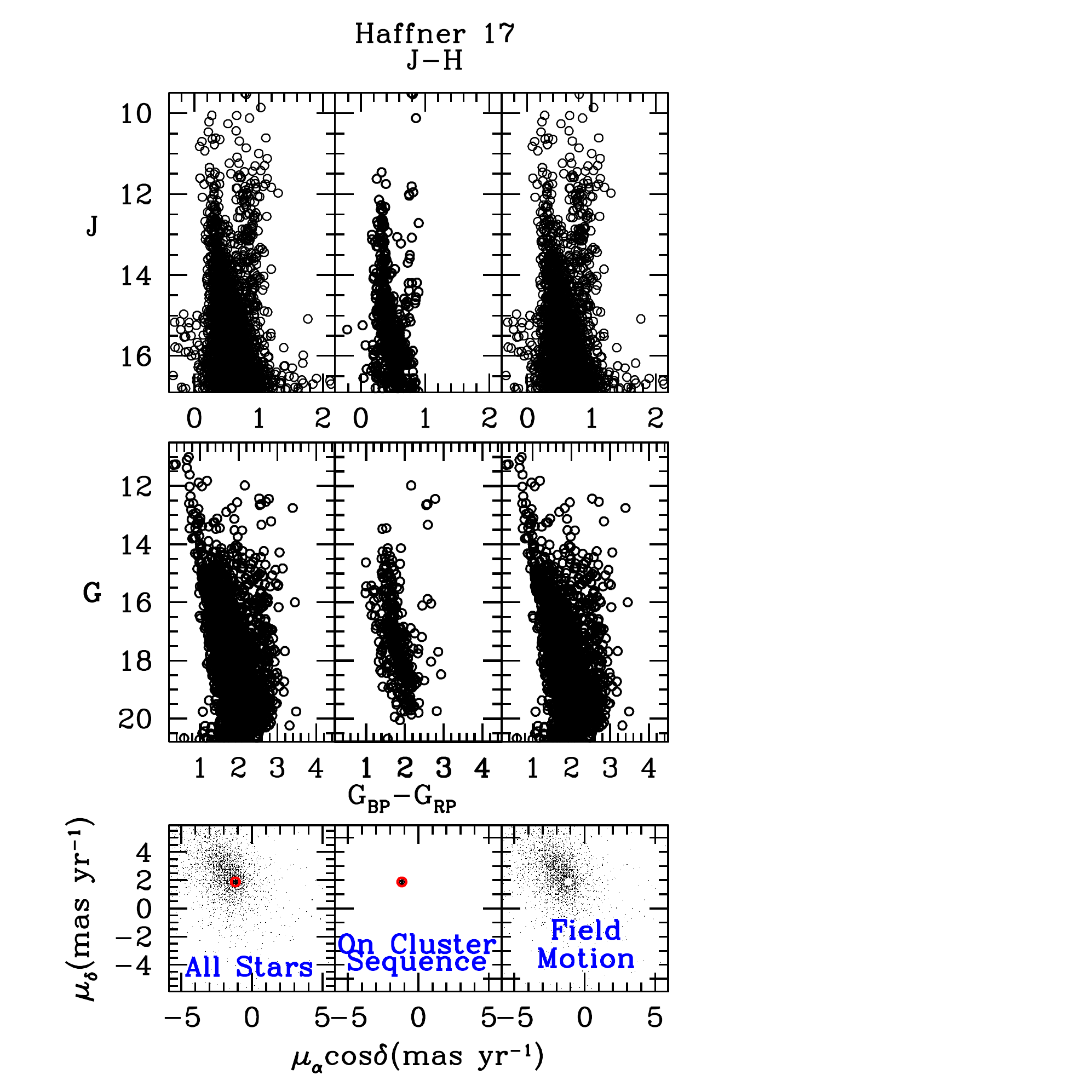}
\hspace{-6cm}\includegraphics[width=14.5cm, height=14.5cm]{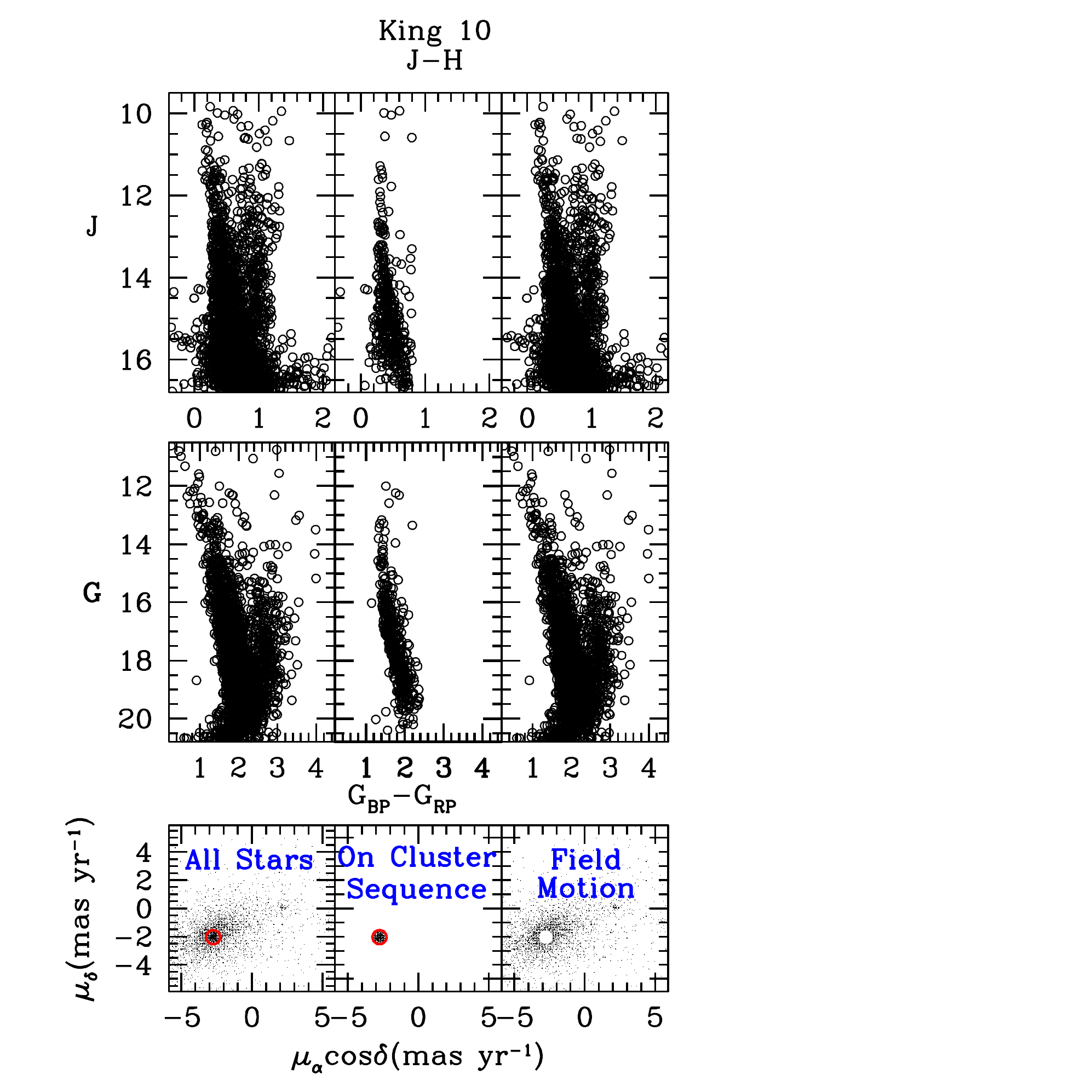}
}
\caption{Same as Fig \ref{pm_dist} for clusters Haffner 17 and King 10. Radii of the circles centered around the mean PM are 0.4 
and $0.5~mas~yr^{-1}$ as shown in the VPDs for Haffner 17 and King 10.}
\label{pm_dista}
\end{figure*}

\subsection{\bf The multi-dimensional Gaia DR2 data set}

We used Gaia~DR2 (Gaia Collaboration et al. 2018a) data for the astrometric analysis of the clusters.
Gaia data consist of five parametric astrometric solution-positions on the sky $(\alpha,
\delta)$, parallaxes and ($\mu_{\alpha} cos\delta , \mu_{\delta}$) with a limiting magnitude of $G=21$ mag. The completeness of the
Gaia survey has much improved now in comparison to the first data release. The Gaia~DR2 is essentially complete between G=12 to 17 mag. The $G$
band covers the whole optical wavelength ranging from 330 to 1050 nm, while $G_{BP}$ and $G_{RP}$ bands cover the wavelength range 330-680 nm
and 630-1050 nm, respectively (Evans et al. 2018). The central wavelengths are 673, 532, 797 nm for G, $G_{BP}$ and $G_{RP}$ bands respectively
(Jordi et al. 2010). Parallax uncertainties are in the range of $\sim$ 0.04 milliarcsecond (mas) for sources at $G\le14$ mag, $\sim$ 0.1 mas
for sources with $G\sim17$ mag, and 0.7 mas at G=20 mag (Lindegren et al. 2018). The uncertainties in the respective proper motion components
are up to 0.06 $mas~ yr^{-1}$ (for $G\le15$ mag), 0.2 $mas~ yr^{-1}$ (for $G\sim17$ mag) and 1.2 $mas~ yr^{-1}$ (for $G\sim20$ mag). The proper
motion and their corresponding errors are plotted against $G$ magnitude in Fig \ref{error_proper}. In this figure, errors in proper
motion components are $\sim 1.2$ at $G\sim20$ mag.

\subsection{\bf 2MASS data }

The 2MASS (Skrutskie et al. 2006) used two highly automated 1.3m telescopes (one at Mt. Hopkins, Arizona (AZ), USA and other at the 
Cerro Tololo Inter-American Observatory, Chile) with a 3-channel camera, each having a $(256\times256)$  array of HgCdTe detectors. 
This 2MASS photometric catalogue provides $J~ (1.25~ \mu m)$, $H~ (1.65~ \mu m)$  and $Ks~ (2.17~ \mu m)$ band photometry for millions
of galaxies and nearly a half-billion stars (Carpenter, 2001). The sensitivity of the 2MASS catalogue is 15.8 mag for $J$, 15.1 mag 
for $H$ and 14.3 mag for $Ks$ band at a signal-to-noise of 10. $VizieR$\footnote{vizier.u-starsbg.fr/viz-bin/VizieR?-source=II/246} was
used to extract $J$, $H$ and $K_{S}$ photometric data in circular areas centered on the clusters under study.
Identification maps for the clusters are taken from Leicester Database and Archive Service (LEDAS) and shown in Fig \ref{id}.

The stars with observational uncertainties $\ge$ 0.20 mag are excluded, and photometric completeness limit is applied on the 2MASS data 
to avoid the over-sampling at the lower parts in the cluster's colour-magnitude diagrams (Bonatto et al. 2004). The errors given in 2MASS
catalogue for $J$, $H$ and $Ks$ band are plotted against $J$ magnitudes in Fig \ref{error}. This figure shows the mean error in $J$,
$H$ and $Ks$ band is $\le$~0.05 at $J\sim$ 13.0 mag. The errors become $\sim$ 0.09 at $J\sim$ 15 mag.

\subsection{\bf WISE data}

The WISE database is a NASA Medium Class Explorer mission that conducted a digital imaging survey of the entire
sky in the mid-IR bands. The effective wavelength of mid-IR bands are $3.35 \mu m$ (W1), $4.60 \mu m $(W2),
$11.56 \mu m$ (W3) and $22.09 \mu m$ (W4) (Wright et al. 2010). This data has been taken from ALLWISE source
catalogue for the clusters under study. This catalogue has achieved $5\sigma$ point source sensitivities
better than 0.08, 0.11, 1 and 6 mJy at 3.35, 4.60, 11.56 and 22.09 $\mu m$, which is expected to be more than
$99\%$ of the sky. These sensitivities are 16.5, 15.5, 11.2 and 7.9 for W1, W2, W3 and W4 bands correspond to
vega magnitudes.

\begin{figure}
\begin{center}
\centering
\hbox{
\includegraphics[width=4.2cm, height=4.2cm]{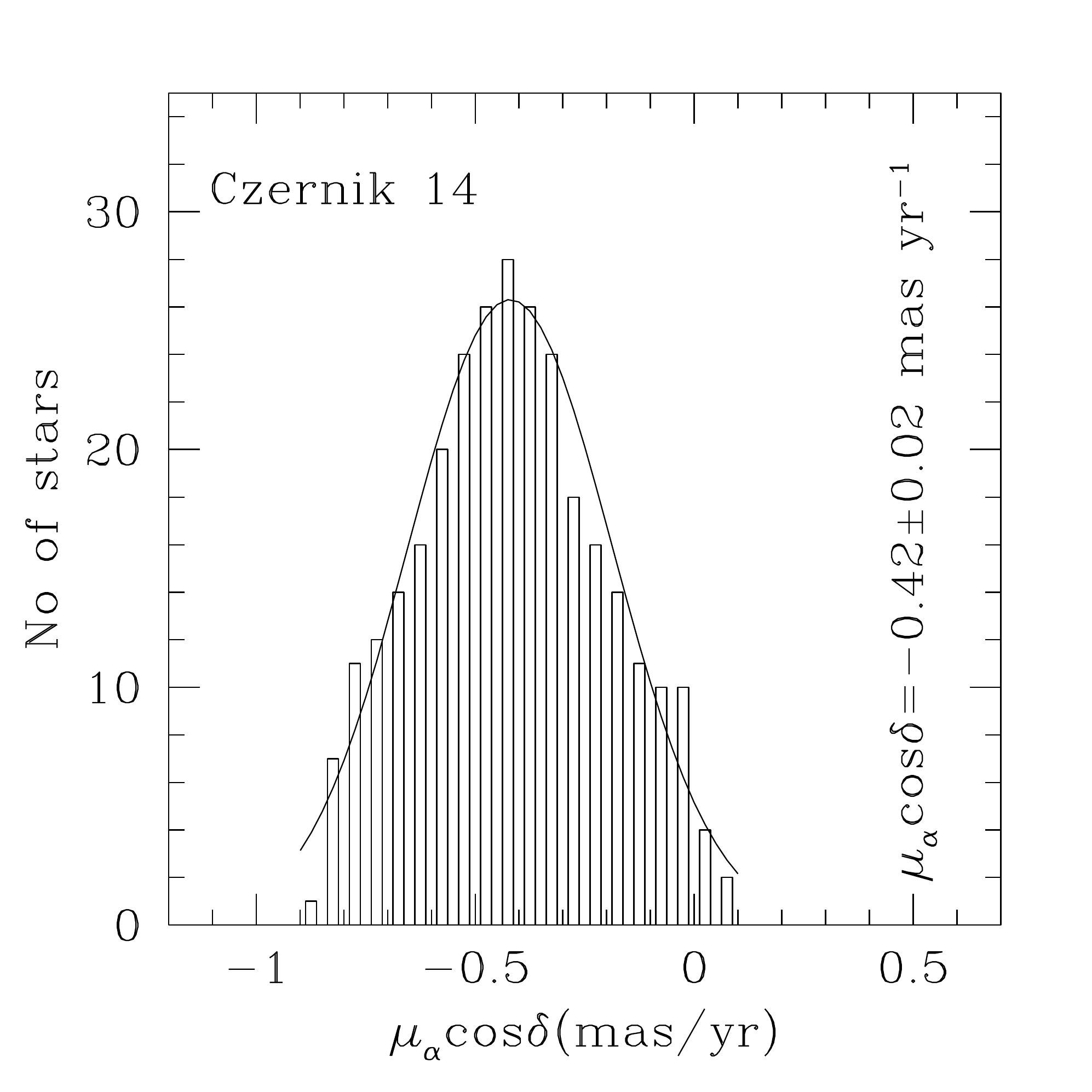}
\includegraphics[width=4.2cm, height=4.2cm]{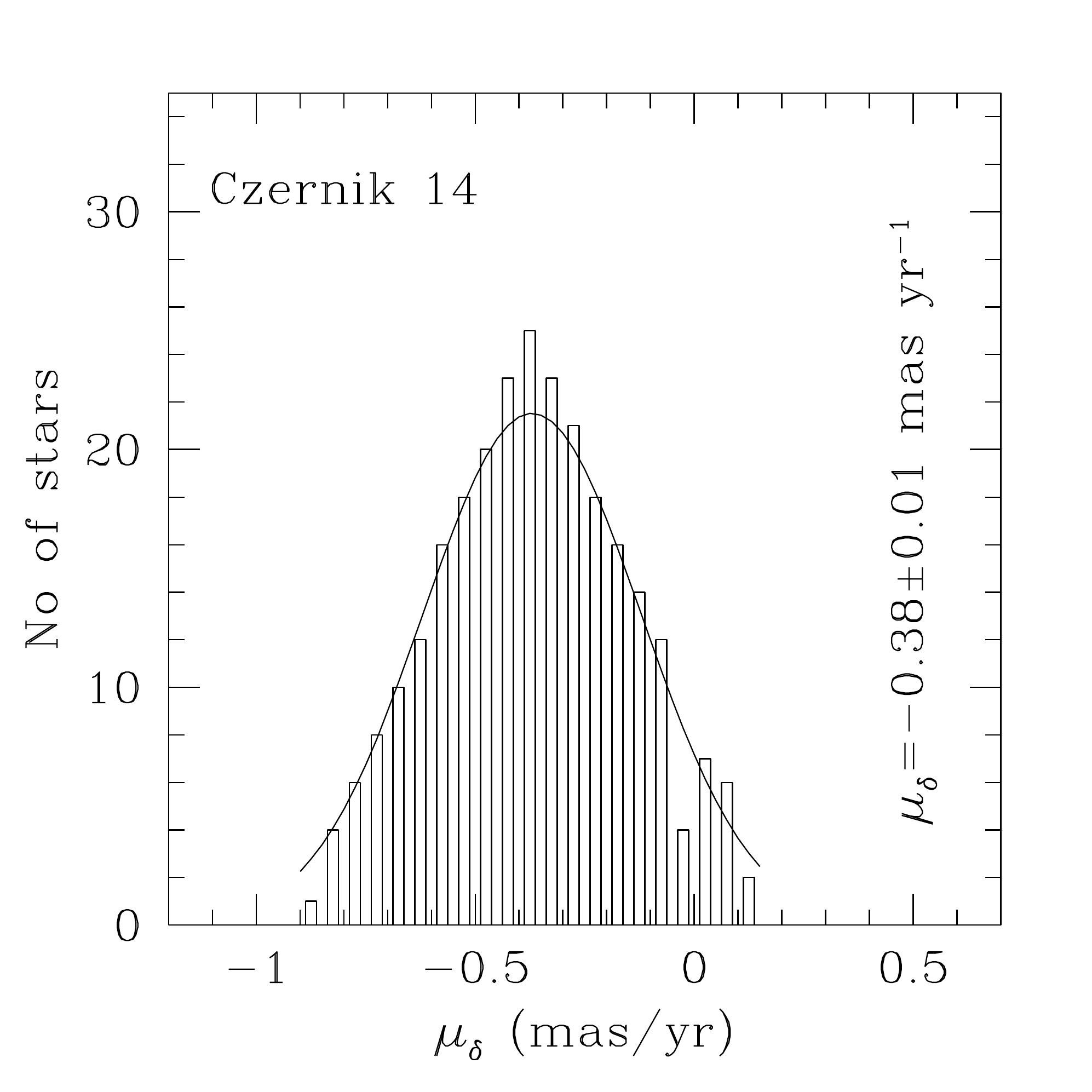}
}
\hbox{
\includegraphics[width=4.2cm, height=4.2cm]{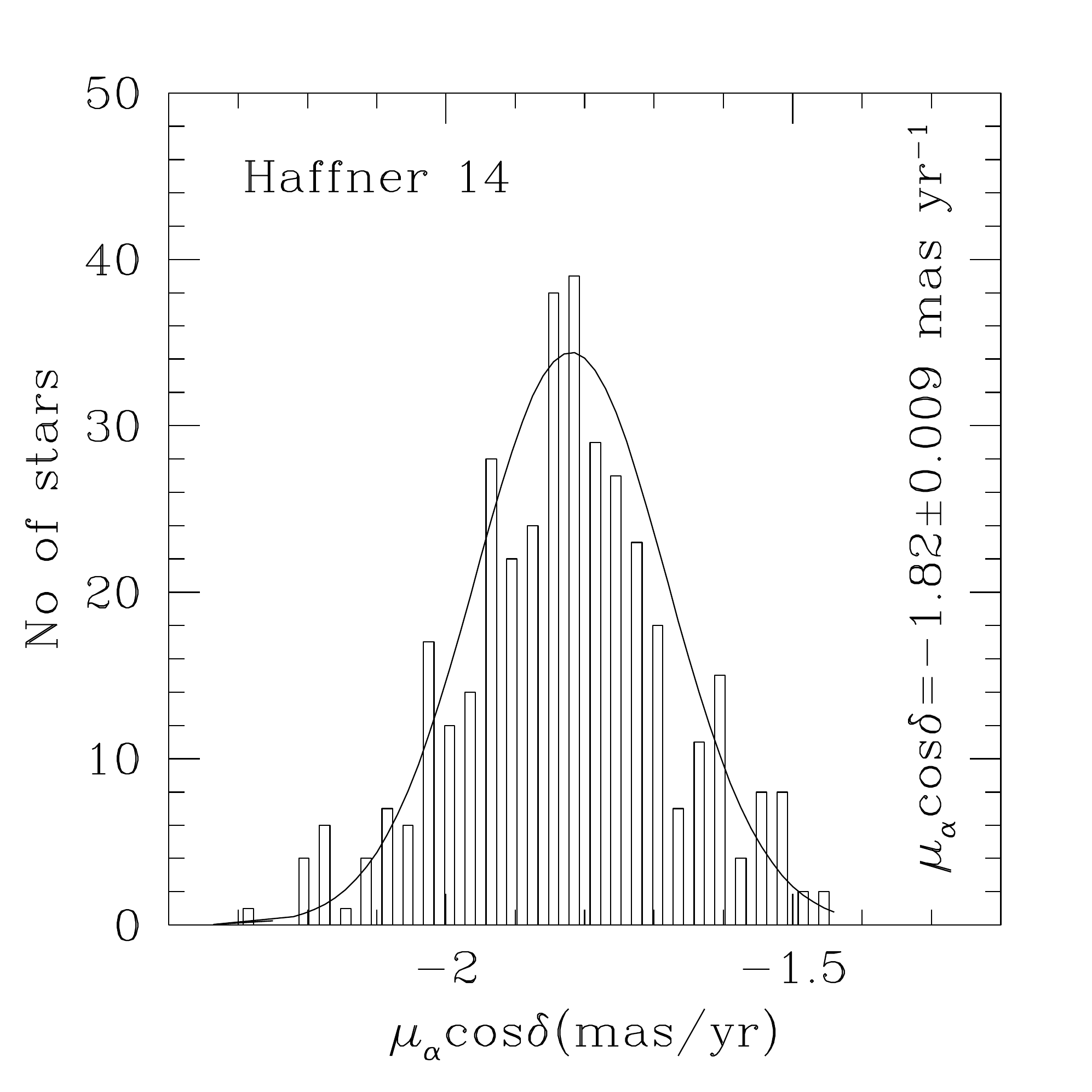}
\includegraphics[width=4.2cm, height=4.2cm]{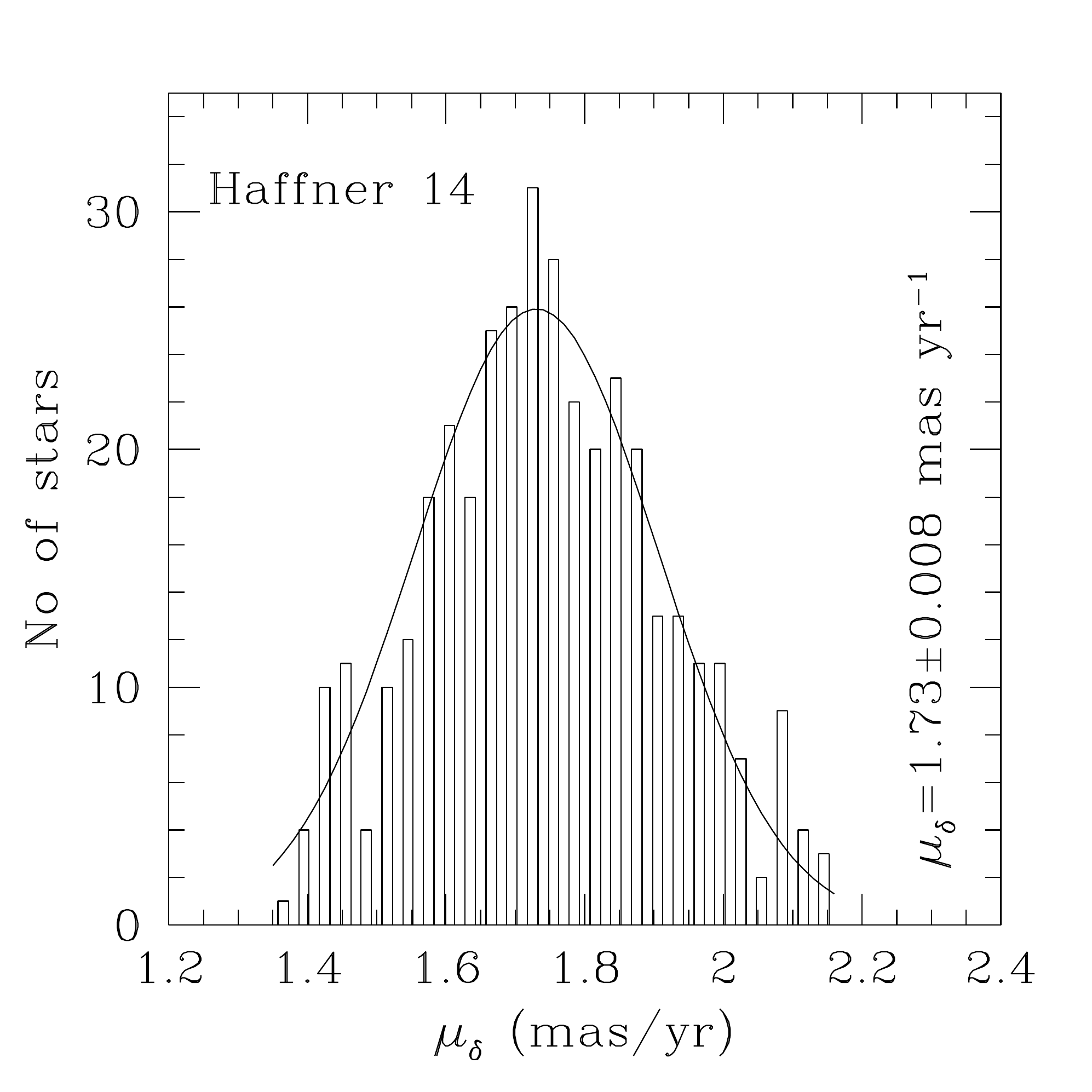}
}
\hbox{
\includegraphics[width=4.2cm, height=4.2cm]{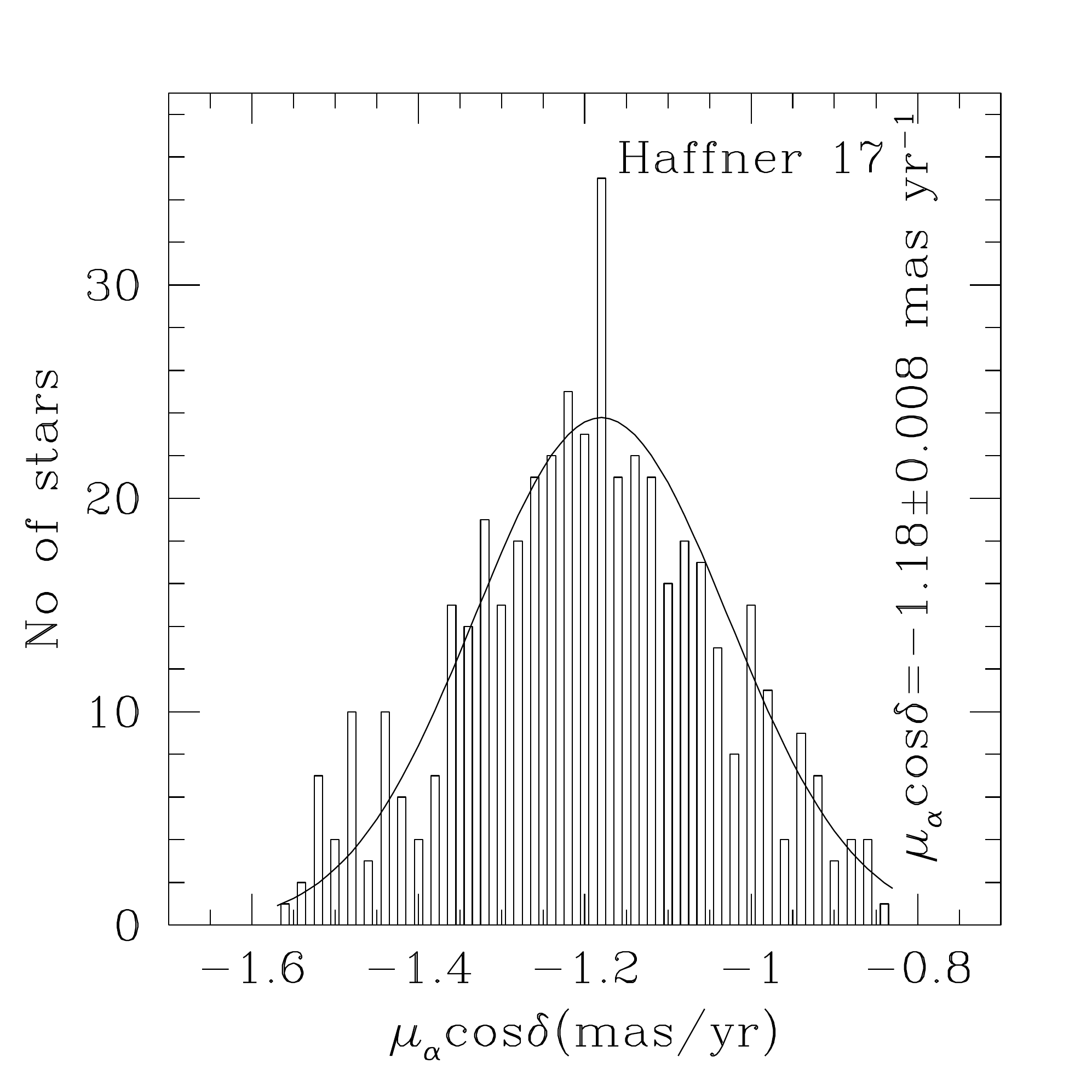}
\includegraphics[width=4.2cm, height=4.2cm]{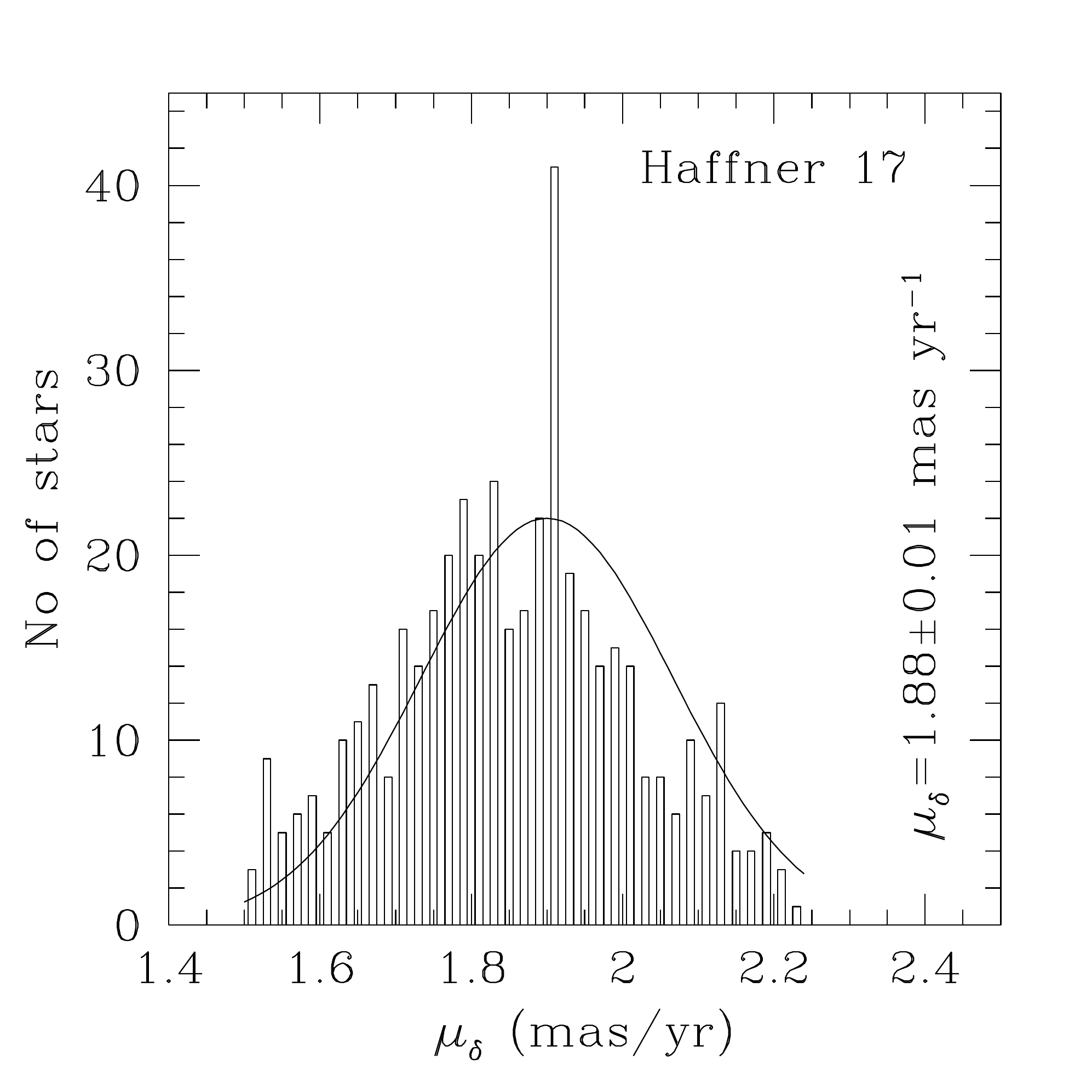}
}
\hbox{
\includegraphics[width=4.2cm, height=4.2cm]{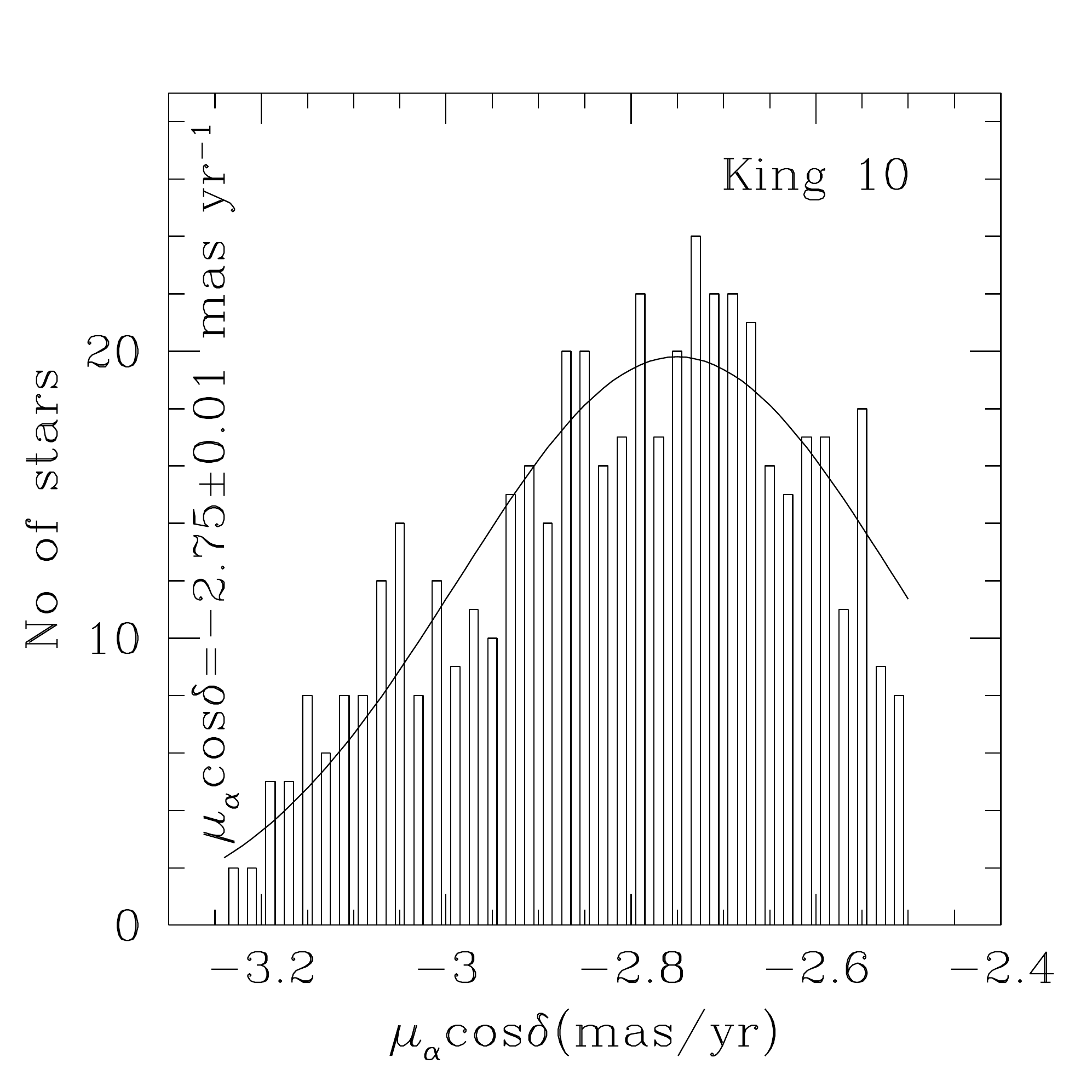}
\includegraphics[width=4.2cm, height=4.2cm]{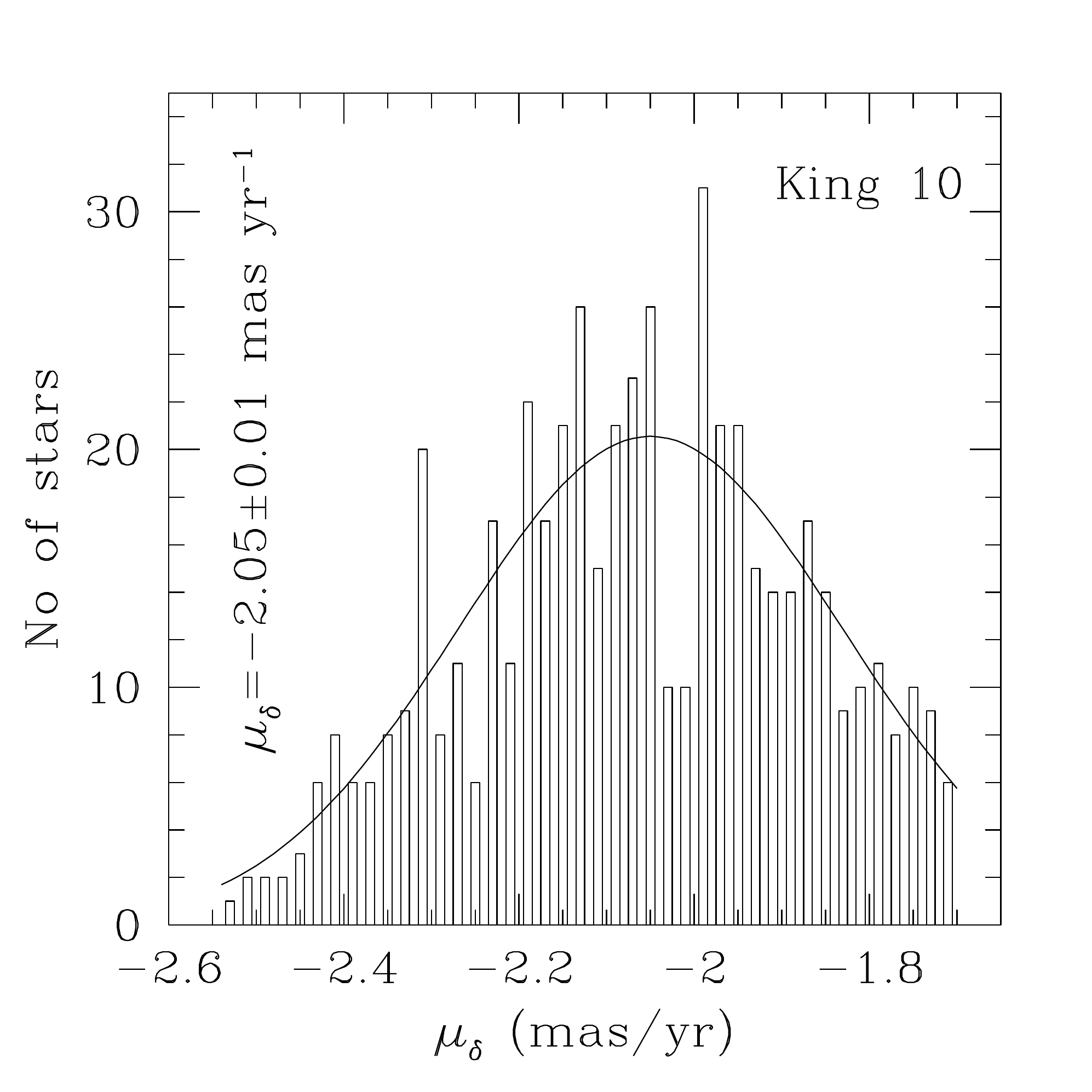}
}
\caption{Proper motion histograms in 0.1 $mas~ yr^{-1}$ bins in $\mu_{\alpha} cos{\delta}$ and $\mu_{\delta}$ of the clusters.
The Gaussian function fit to the central bins provides the mean values in both directions as shown in each panel.
}
\label{pm_dist1}
\end{center}
\end{figure}

\subsection{\bf APASS data}

The American Association of Variable Star Observers (AAVSO) Photometric All-Sky Survey (APASS) is organized in five filters:
$B$, $V$ (Landolt) and $g^{\prime}$, $r^{\prime}$, $i^{\prime}$ proving stars with $V$ band magnitude range from 7 to 17 mag
(Heden \& Munari 2014). DR9 is the latest catalogue and covers about $99\%$ sky (Heden et al. 2016). We have extracted this
data from $http://vizier.u-strasbg.fr/viz-bin/VizieR?-source=II/336$.

\subsection{\bf Pan-STARRS1 data}

\begin{figure}
\begin{center}
\hbox{
\includegraphics[width=4.2cm, height=4.2cm]{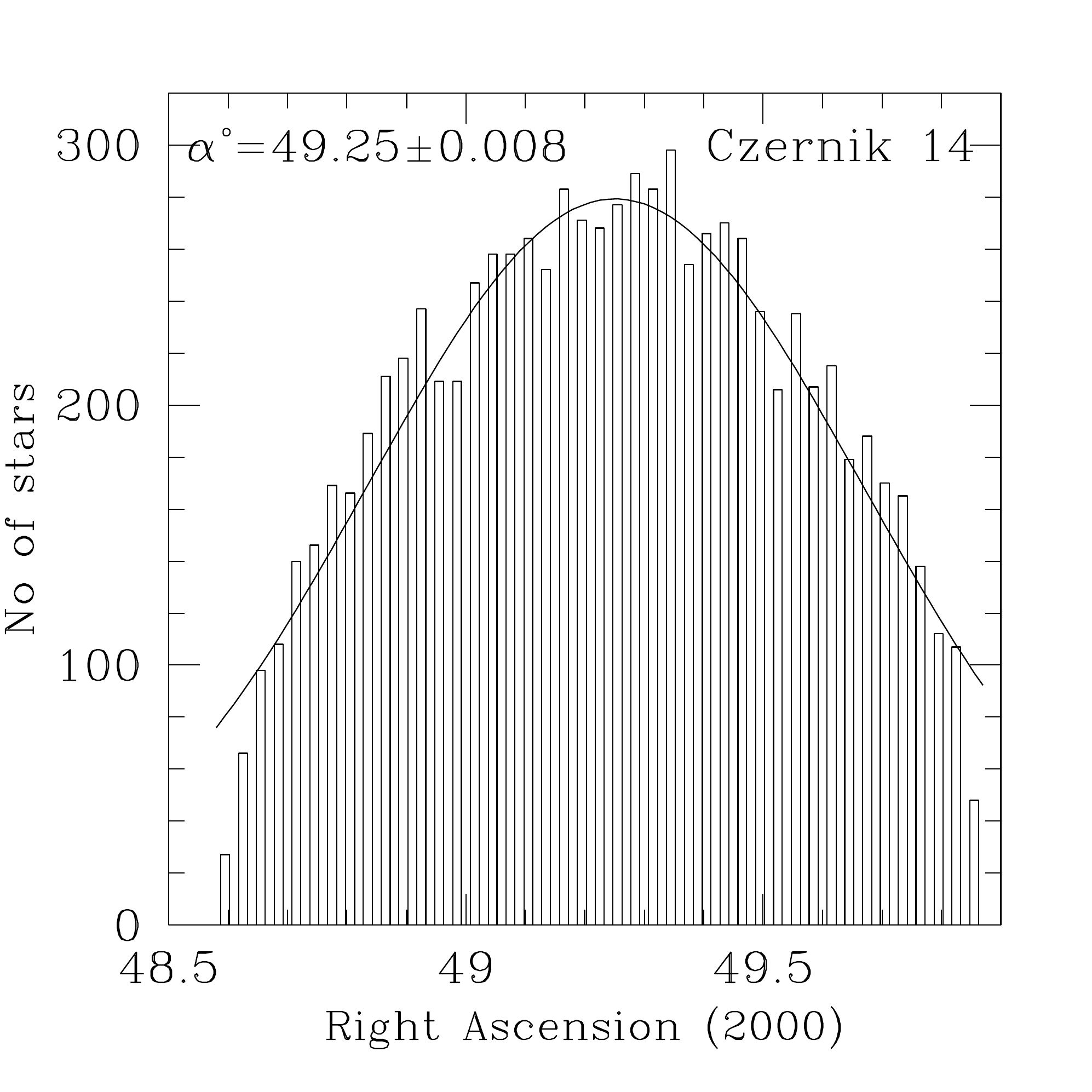}
\includegraphics[width=4.2cm, height=4.2cm]{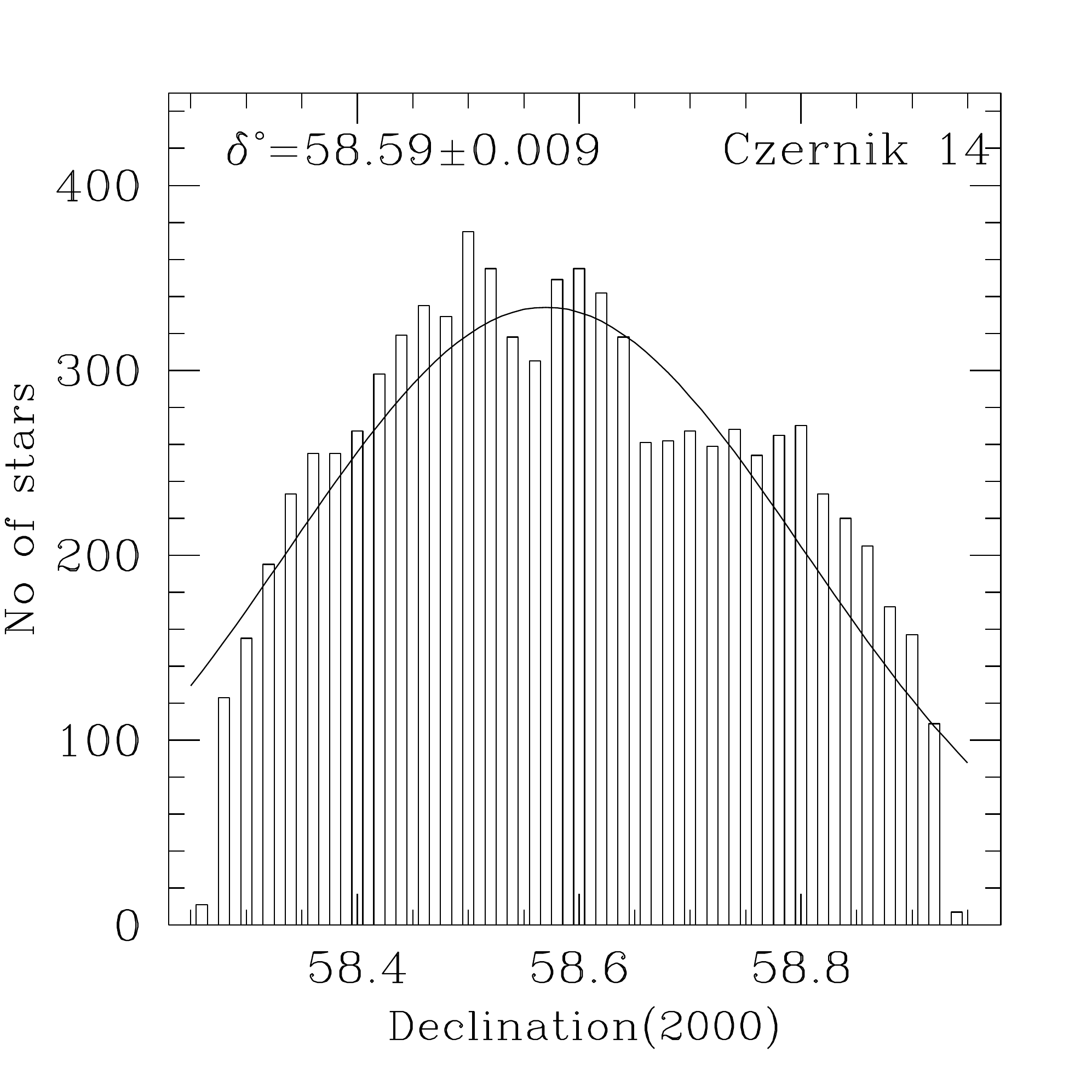}
}
\hbox{
\includegraphics[width=4.2cm, height=4.2cm]{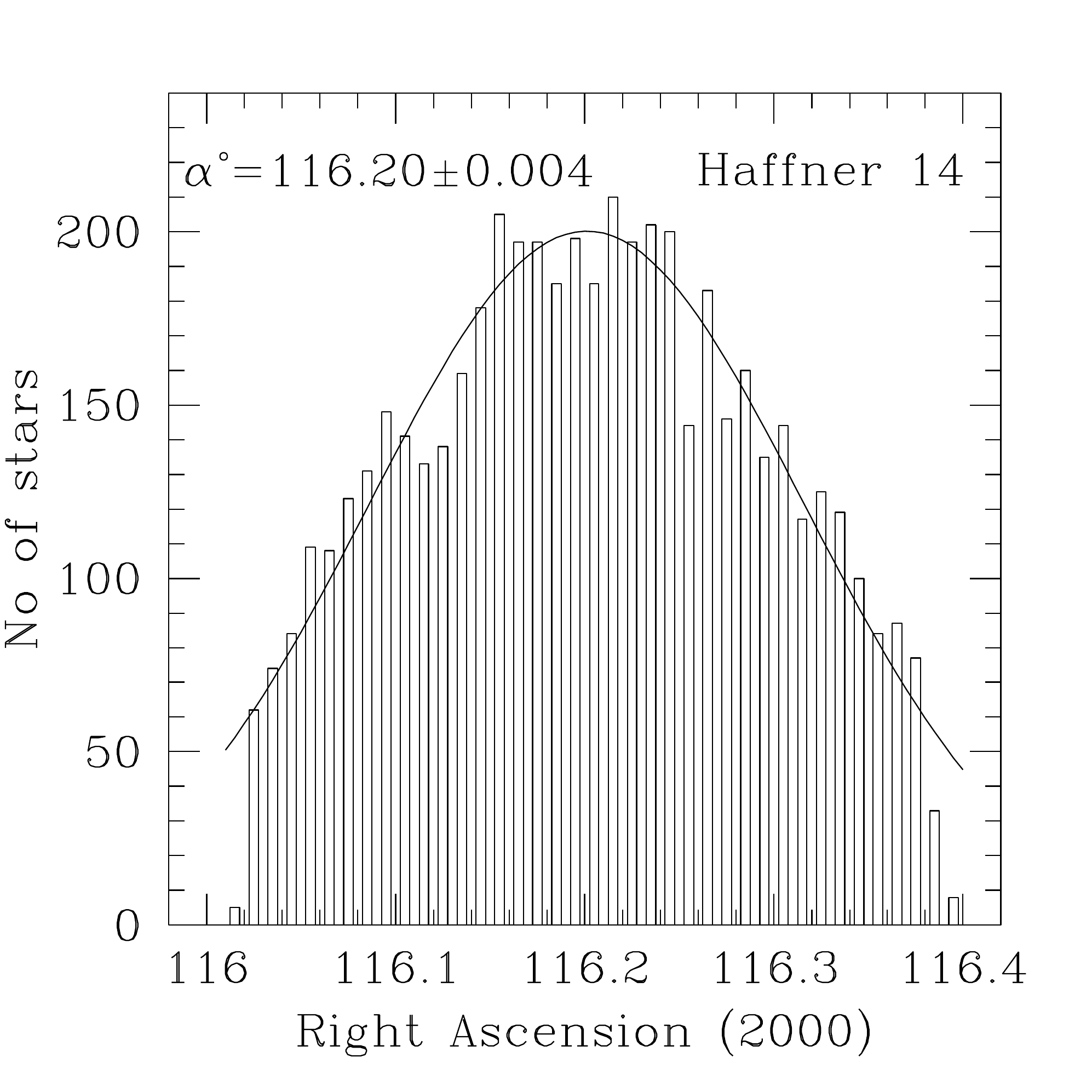}
\includegraphics[width=4.2cm, height=4.2cm]{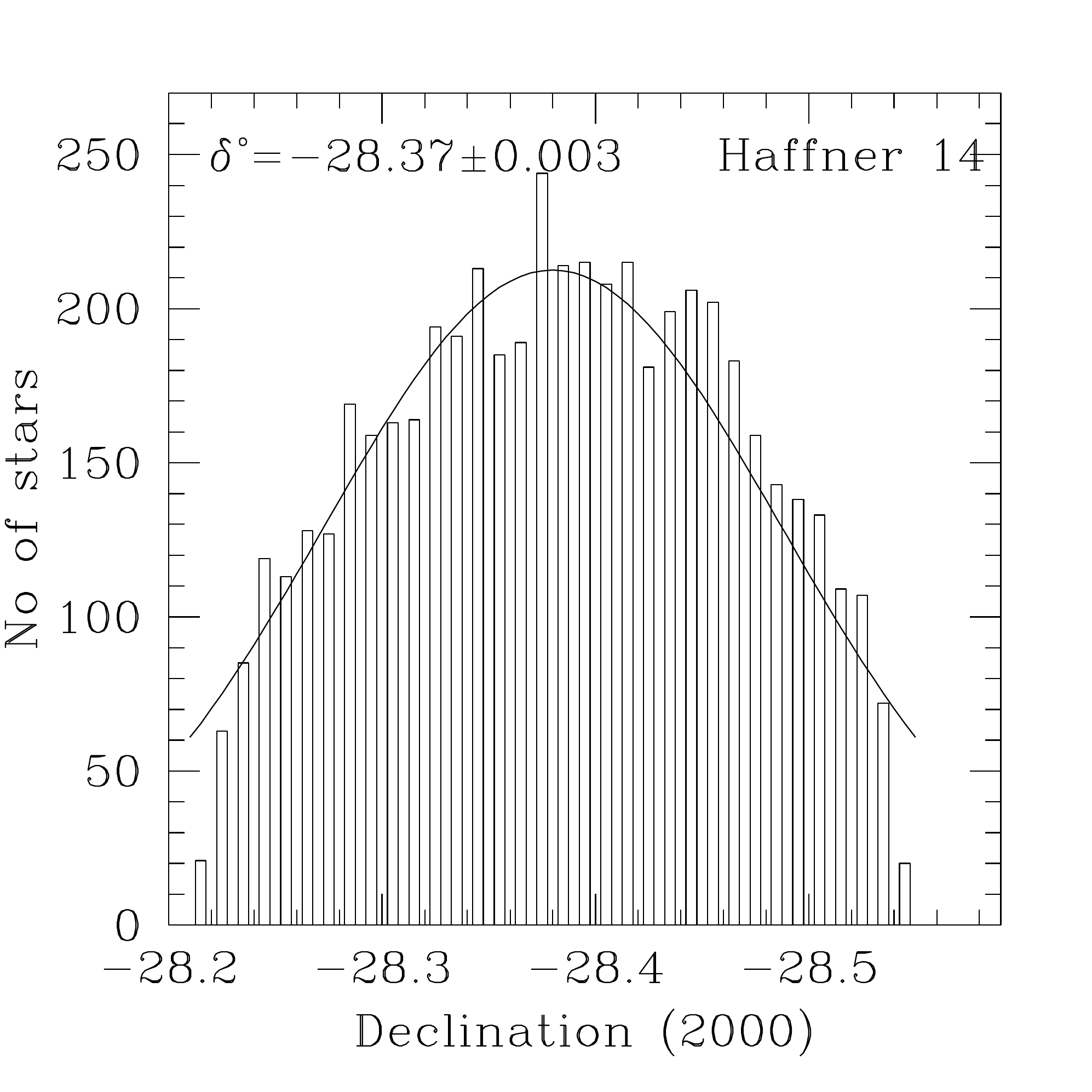}
}
\hbox{
\includegraphics[width=4.2cm, height=4.2cm]{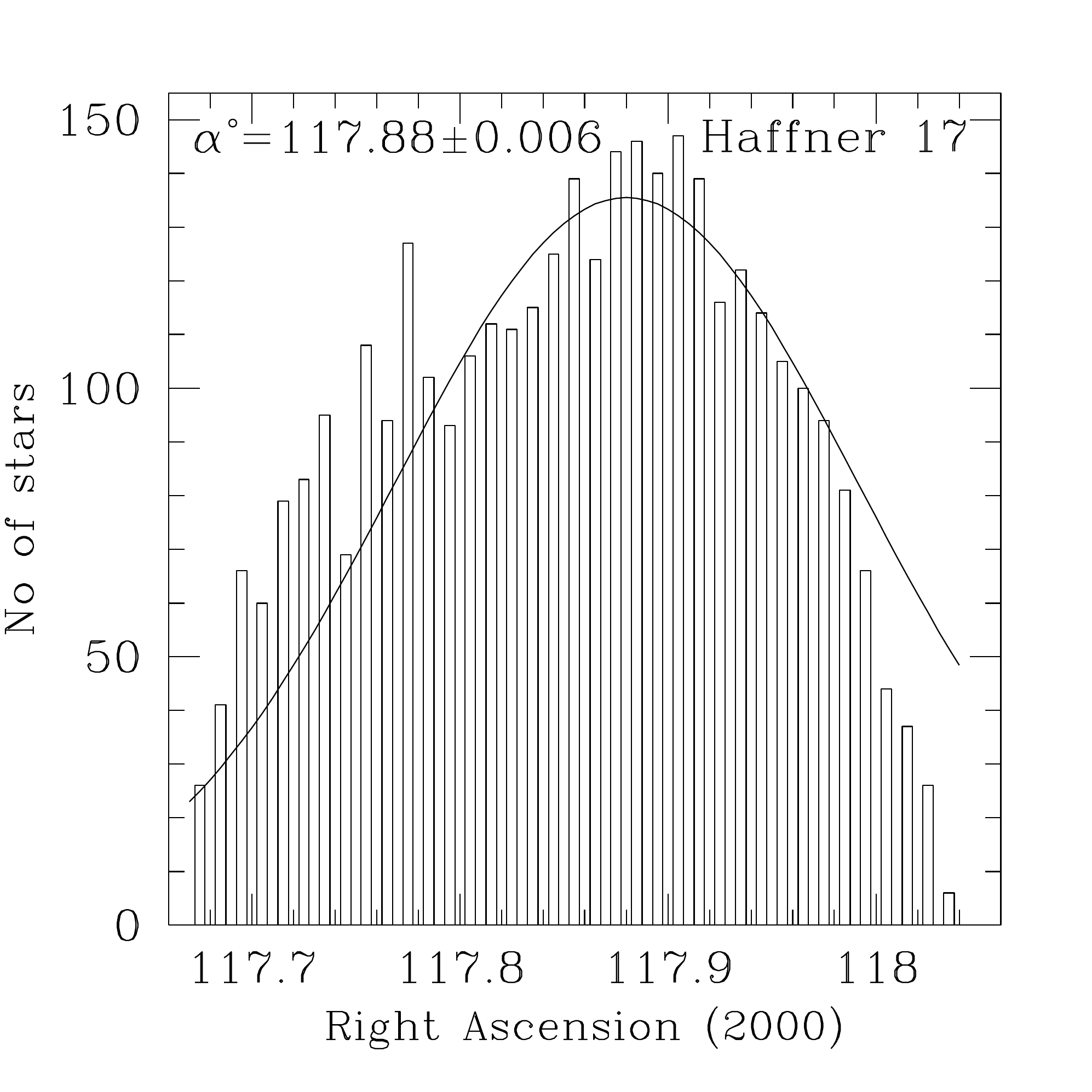}
\includegraphics[width=4.2cm, height=4.2cm]{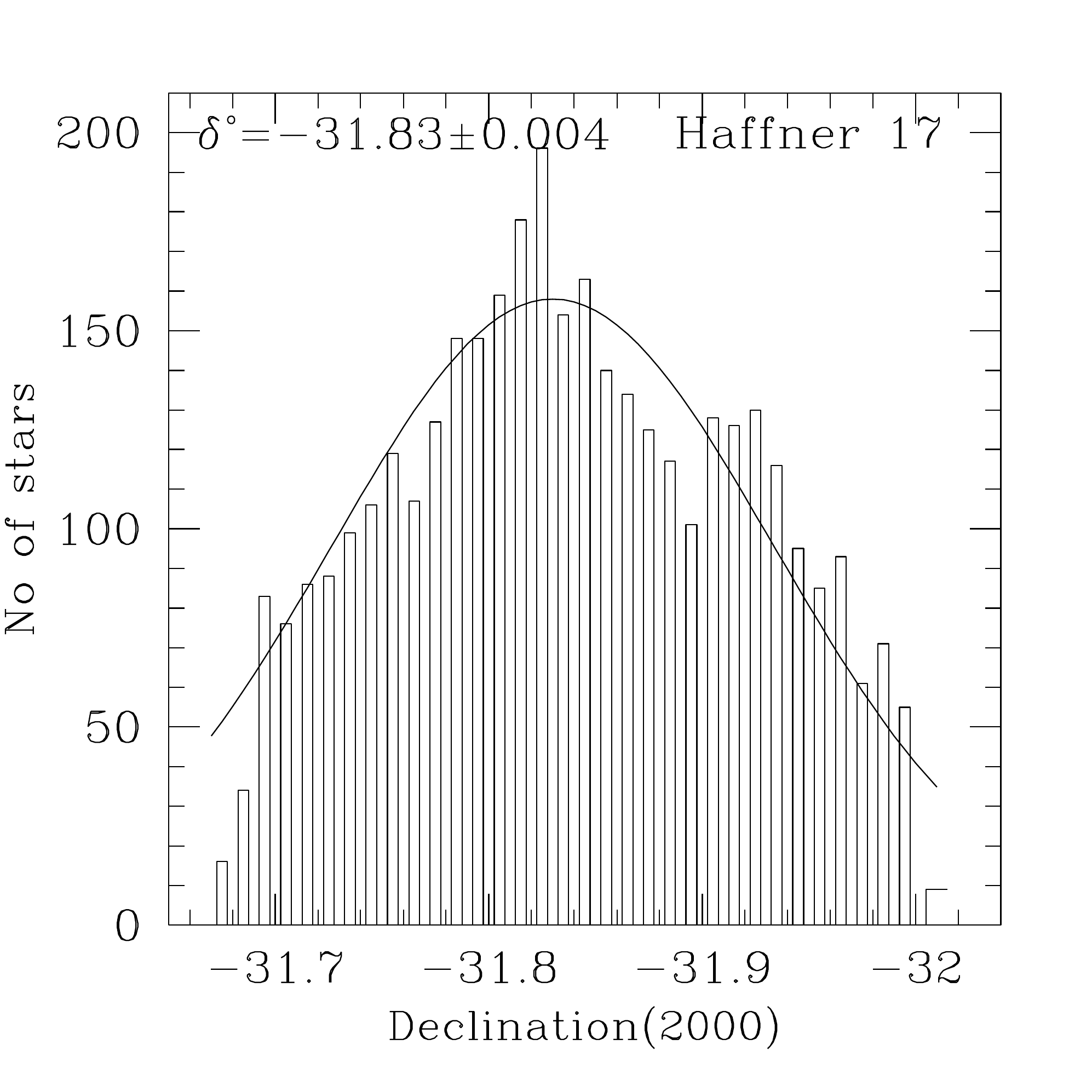}
}
\hbox{
\includegraphics[width=4.2cm, height=4.2cm]{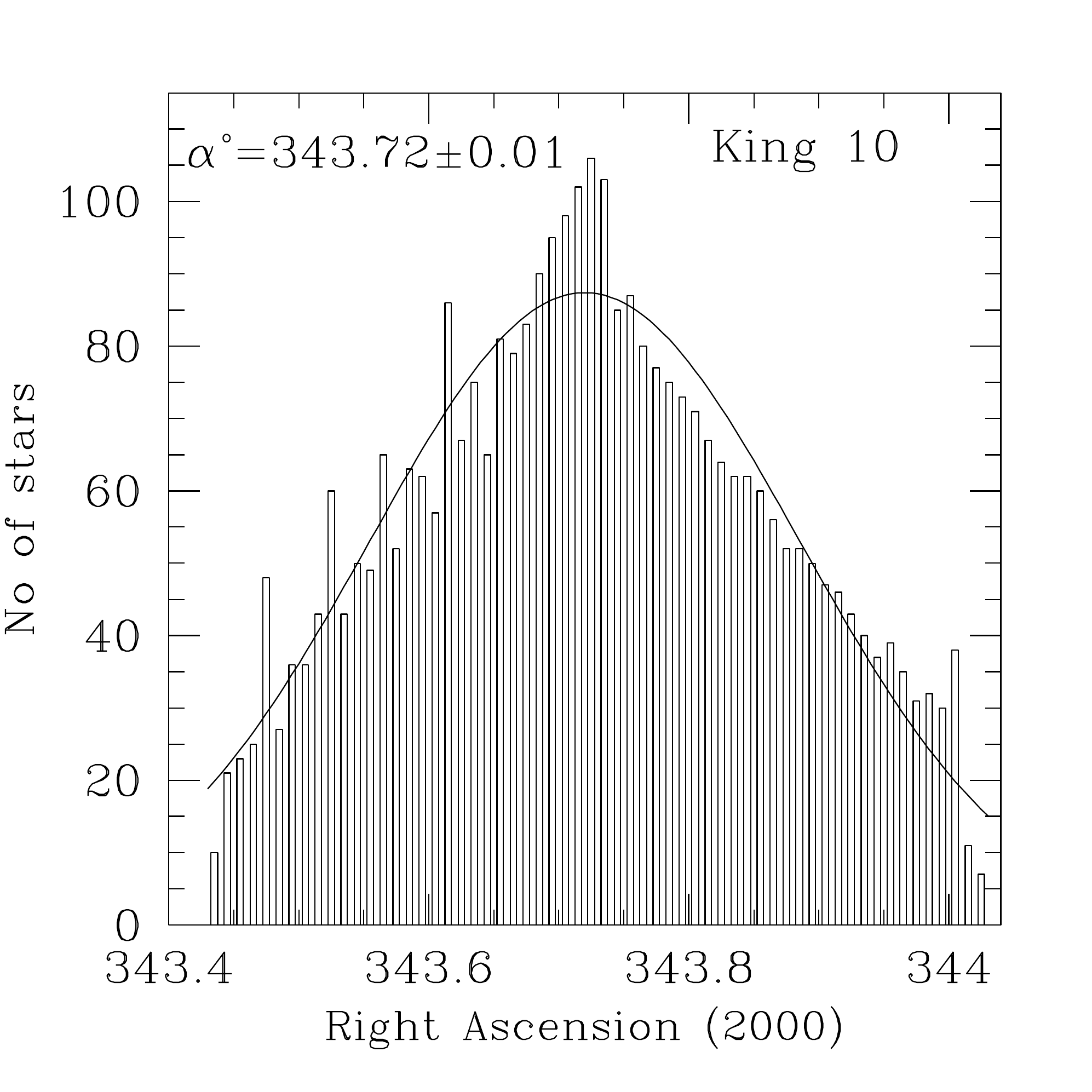}
\includegraphics[width=4.2cm, height=4.2cm]{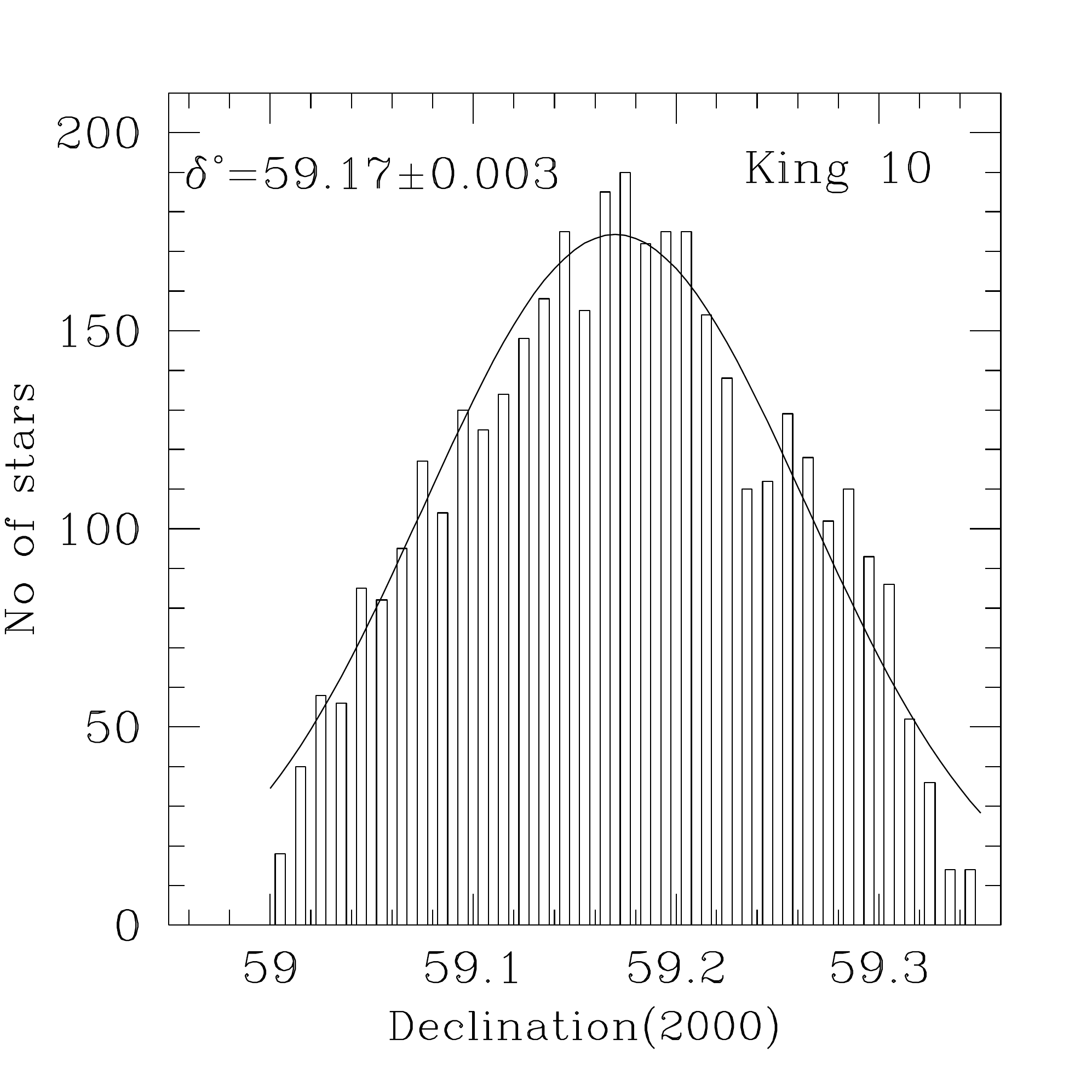}
}
\caption{Profiles of stellar counts across clusters Czernik 14, Haffner 14, Haffner 17 and King 10 using Gaia~DR2. The Gaussian
fits have been applied. The center of symmetry about the peaks of Right Ascension and Declination is taken to be the position of
the cluster's center.}
\label{center}
\end{center}
\end{figure}
The Pan-STARRS1 survey (Hodapp et al. 2004) imaged the sky in five broad-band filters, $g$, $r$, $i$, $z$, $y$, covering
from 400 nm to 1 μm (Stubbs et al. 2010). The mean 5$\sigma$ point source limiting sensitivities in $g$, $r$, $i$, $z$, and
$y$ bands are 23.3, 23.2, 23.1, 22.3, and 21.4 mag, respectively (Chambers et al. 2016). The effective wavelengths of these
filters are 481, 617, 752, 866, and 962 nm, respectively (Schlafly et al. 2012; Tonry et al. 2012). The photometric accuracy
of this data has been demonstrated by Schlafly et al. (2012) and Magnier et al. (2013).

\section{Mean Proper motion of clusters and member selection}

Field region stars always affect the accurate measurements of cluster fundamental parameters. Proper motion plays the most
important role to separate non members from cluster's area. To select possible cluster members of Czernik 14, Haffner 14,
Haffner 17 and King 10, we applied Gaia~DR2 proper motion data and parallax data. 

PMs, $\mu_{\alpha} cos{\delta}$ and $\mu_{\delta}$ are plotted as VPDs in the bottom panels of Fig \ref{pm_dist} and  Fig \ref{pm_dista}.
The top row panels in each cluster show the corresponding $J$ versus $(J-H)$ CMDs and $G$ versus $(G_{BP}-G_{RP})$. The left panel in the
CMDs shows all stars present in the cluster's area, while the middle and right panels show the probable cluster members and
non member stars respectively. A circle of 0.5, 0.4, 0.4 and 0.5  $mas~ yr^{-1}$ around the center of the member stars distribution
in the VPDs characterize our membership criteria. The picked radius is an agreement between losing cluster members with poor
PMs and the involvement of non member stars. The CMDs of the probable members are shown in the middle row panels in each clusters
CMDs as shown in Fig \ref{pm_dist} and Fig \ref{pm_dista}. The main sequence of the cluster is separated out. These stars have a
PM error of $\le 1 ~ mas~ yr^{-1}$.

For the precise estimation of mean proper motion, we deal with only probable cluster members based on clusters VPDs and 
CMDs as shown in Fig \ref{pm_dist1}. By fitting the Gaussian function into the constructed histograms provides mean proper
motion in the directions of RA and DEC, as shown in Fig \ref{pm_dist1}. To centered the Gaussian, we focused on the 
peak of the distribution. In this way, we found the mean-proper motion in RA and DEC directions as $-0.42\pm0.02$ and
$-0.38\pm0.01$ $mas yr^{-1}$ for Czernik 14, $-1.82\pm0.009$ and $1.73\pm0.008$ $mas yr^{-1}$ for Haffner 14, $-1.17\pm0.007$
and $1.88\pm0.006$ $mas yr^{-1}$ for Haffner 17 and $-2.75\pm0.008$ and $-2.04\pm0.006$ $mas yr^{-1}$ for King 10. The estimated
values of mean proper motions for each cluster is in fair agreement with the values given by Cantat-Gaudin (2018).
Cantat-Gaudin catalogue (2018) reports the membership probabilities of these OCs. We have matched our probable members with this
catalogue and selected stars having a probability higher than $40\%$ for each cluster and used to derive fundamental parameters
of the clusters.

\section{Clusters Structure, extinction law and fundamental parameters evaluation}

\subsection{ Center estimation:}

To understand cluster properties, the elementary step is to find clusters central coordinates. In the previous studies, 
the center has been determined just by the visual inspection (Becker \& Fenkart 1971; Romanishim \& Angel 1980).
In this paper, we applied the star-count method using the stars selected from proper motion.
The histograms are constructed for the clusters in RA and DEC directions by using {bf most probable cluster
members, selected in above section} as shown in Fig \ref{center}. The Gaussian 
curve-fitting is performed to the star counts profiles in RA and DEC directions. Using this method, the coordinates of the center are
found to be $\alpha^{\circ} = 49.25\pm0.008$ deg $(3^{h} 17^{m} 00^{s})$ and $\delta^{\circ} = 58.59\pm0.009$ deg 
$(58^{\circ} 35^{\prime} 24^{\prime\prime})$ for Czernik 14, $\alpha^{\circ} = 116.20\pm0.004$ deg $(7^{h} 44^{m} 48^{s})$
and $\delta^{\circ} = -28.37\pm0.003$ deg $ (-28^{\circ} 22^{\prime} 12^{\prime\prime})$ for Haffner 14, 
$\alpha^{\circ} = 117.88\pm0.006$ deg $(7^{h} 51^{m} 31^{s})$ and $\delta^{\circ} = -31.83\pm0.004$ deg $ 
(-31^{\circ} 49^{\prime} 48^{\prime\prime})$ for Haffner 17 and $\alpha^{\circ} = 343.72\pm0.01$ deg $(22^{h} 54^{m} 53^{s})$
and $\delta^{\circ} = 59.17\pm0.003$ deg $ (59^{\circ} 10^{\prime} 12^{\prime\prime})$ for King 10 . These  estimated values
are in good agreement with the values given by Dias et al. (2002). Our derived values of the center of cluster are also matched
well with Cantat-Gaudin (2018) catalogue within uncertainty.

\subsection{Cluster radius and radial stellar surface density}

\begin{figure}
\includegraphics[width=8.5cm, height=8.5cm]{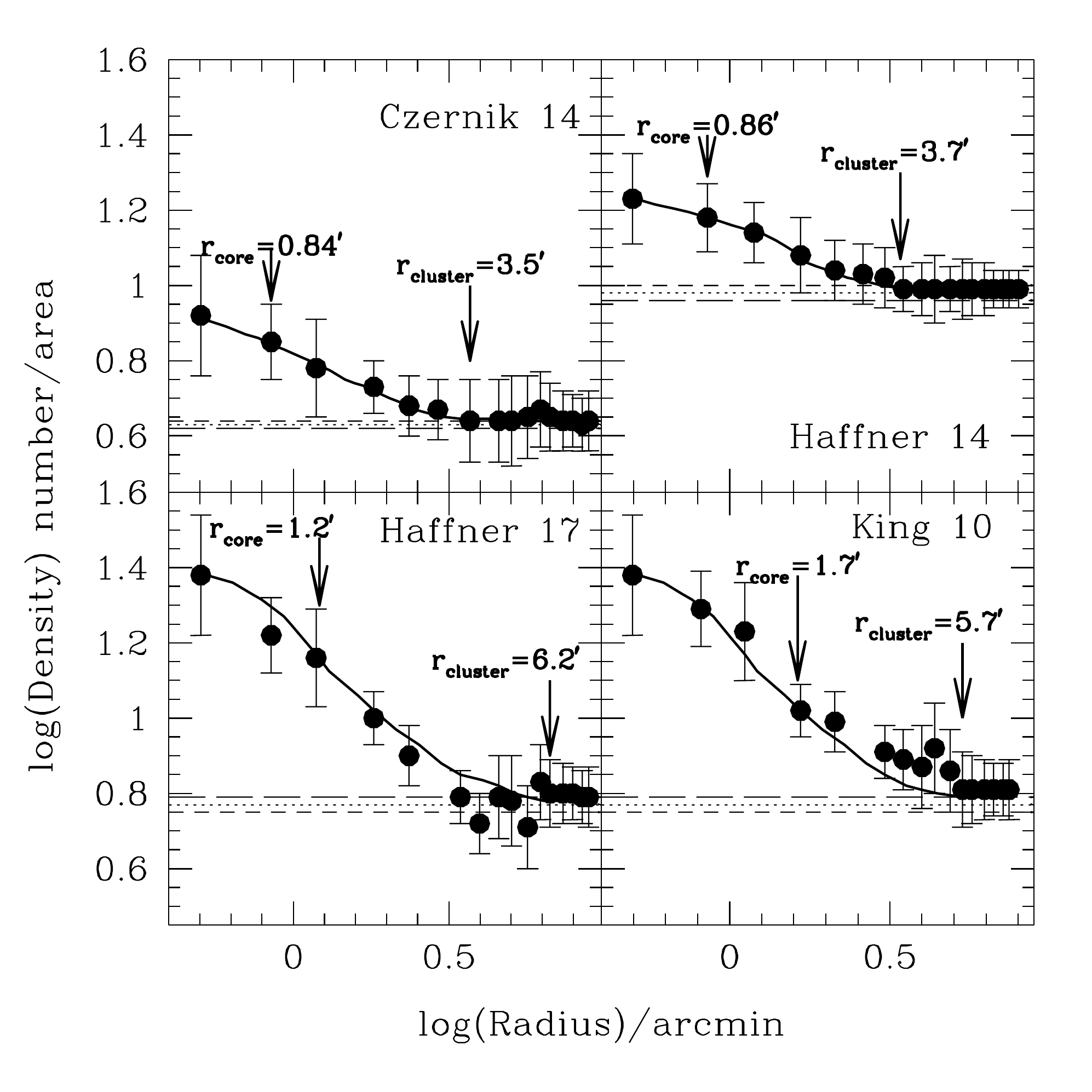}
\caption{Surface density distribution of the clusters Czernik 14, Haffner 14, Haffner 17 and King 10 using Gaia~DR2 $G$ band
data. Errors are determined from sampling  statistics (=$\frac{1}{\sqrt{N}}$ where $N$ is the number of cluster members
used in the density estimation at that point). The smooth line represents the fitted profile of King (1962)  whereas
the dotted line shows the background density level. Long and short dash lines represent the errors in background
density.}
\label{dens}
\end{figure}

Estimation of cluster radius is one of the most important fundamental properties. We construct a radial density profile (RDP) 
for open clusters Czernik 14, Haffner 14, Haffner 17 and King 10 using stars with $G\le19$ mag. We drew many 
concentric rings around the cluster center with an equal increment in radius. The number density, $\rho_{i}$, in the $i^{th}$ zone is determined
by using the formula of $\rho_{i}$ = $\frac{N_{i}}{A_{i}}$, where $N_{i}$ is the number of cluster stars and $A_{i}$ is the area of the
$i^{th}$ zone. The radii of the clusters are calculated on the basis
of the visual inspection of RDPs. The radii at which each distribution flattens are considered as cluster radius. 
The error in the background density level is shown with dotted lines in Fig \ref{dens}. RDP becomes flat at 
$r\sim$ 3.5$^{\prime}$ (log(radius)=0.54), 3.7$^{\prime}$ (log(radius)=0.57), 6.2$^{\prime}$ (log(radius)=0.79) and
5.7$^{\prime}$ (log(radius)=0.76)  for the clusters Czernik 14, Haffner 14, Haffner 17 and King 10. After this point, cluster 
stars merged with non-member stars, which is seen in Fig \ref{dens}. Therefore, we considered 3.5$^{\prime}$, 3.7$^{\prime}$,
6.2$^{\prime}$ and 5.7$^{\prime}$ as the cluster radius. The observed radial density profile was fitted using 
King (1962) profile:\\

$f(r) = f_{b}+\frac{f_{0}}{1+(r/r_{c})^2}$\\

where $f_{b}$, $f_{0}$ and $r_c$ are background density, central star density and the core radius of the cluster respectively. The
errors bar are calculated using the statistics Poisson error in each shell as $P_{err}=\frac{1}{\sqrt{N}}$. By fitting the King model
to the cluster density profiles, we estimate the structural parameters of each cluster. Using Gaia~DR2 photometry, the structural 
parameters for the clusters are obtained as, $f_{b}$=4.27 $star/arcmin^{2}$, $f_{0}$=9.77 $star/arcmin^{2}$ and $r_{c}$=0.84 arcmin 
for Czernik 14, $f_{b}$=9.68 $star/arcmin^{2}$, $f_{0}$=18.89 $star/arcmin^{2}$ and $r_{c}$=0.86 arcmin for Haffner 14, 
$f_{b}$=5.90 $star/arcmin^{2}$, $f_{0}$=32.33 $star/arcmin^{2}$ and $r_{c}$=1.2 arcmin for Haffner 17 and $f_{b}$=5.57 $star/arcmin^{2}$, 
$f_{0}$=24.46 $star/arcmin^{2}$ and $r_{c}$=1.7 arcmin for King 10.

\begin{table*}
\centering
\caption{Structural parameters of the clusters under study. Background and central
density are in the unit of stars per arcmin$^{2}$. Core radius ($r_c$) and tidal radius ($R_t$) are
in arcmin and pc. 
}
\vspace{0.5cm} 
\begin{center}
\small
\begin{tabular}{cccccccccc}
\hline\hline
Name & $f_{0}$ &$f_{b}$& $r_{c}$&$r_{c}$&$R_{t}$& $R_{t}$&$\delta_{c}$&$r_{lim}$&$c$ \\
&&& arcmin & parsec & arcmin & parsec & & arcmin
\\
Czernik 14  & $9.77$&$4.27$&$0.84$&$0.71$&$11.51\pm0.49$&$9.71\pm0.50$&$3.3$&$4.5$&$0.73$ \\
Haffner 14  & $18.89$&$9.68$&$0.86$&$1.20$&$9.21\pm0.50$&$12.86\pm0.70$&$2.9$&$4.1$&$0.67$ \\
Haffner 17  & $32.33$&$5.90$&$1.2$&$1.3$&$12.52\pm0.57$&$13.11\pm0.60$&$6.3$&$6.8$&$0.75$ \\
King 10     & $24.46$&$5.57$&$1.7$&$1.9$&$12.13\pm0.45$&$13.41\pm0.50$&$5.4$&$6.2$&$0.56$ \\
\hline
\end{tabular}
\label{stru_para}
\end{center}
\end{table*}

Limiting radius ($r_{lim}$) of each cluster is calculated by comparing $f(r)$ to a background density level, $f_{b}$, defined
as\\

$f_{b}=f_{bg}+3\sigma_{bg}$\\

where $\sigma_{bg}$ is uncertainty of $f_{bg}$. Therefore, $r_{lim}$ is calculated according to the following formula
(Bukowiecki et al. 2011)\\

$r_{lim}=r_{c}\sqrt(\frac{f_{0}}{3\sigma_{bg}}-1)$\\

The value of limiting radius is found to be 4.5, 4.1, 6.8, and 6.2 arcmin for Czernik 14, Haffner 14, Haffner 17 and King 10, 
respectively. $r_{c}$ and $r_{lim}$ are used to determine the value of concentration parameter $c = log (\frac{r_{lim}} {r_{c}})$
(Peterson \& King, 1975) as 0.73, 0.67, 0.75 and 0.56 for the 
clusters Czernik 14, Haffner 14, Haffner 17 and King 10, respectively. Maciejewski \& Niedzielski (2007) suggested that
$r_{lim}$ may vary for particular clusters from 2$r_{c}$ to 7$r_{c}$. In this study, all clusters show a good agreement with
Maciejewski \& Niedzielski (2007).

The density contrast parameter ($\delta_{c} = 1 +\frac{f_{0}}{f_{b}}$) is calculated for all the clusters under study using 
member stars selected from proper motion data. Current evaluation of $\delta_{c}$ (3.3, 2.9, 6.3 and 5.4 for Czernik 14, Haffner 14, 
Haffner 17 and King 10, respectively) are lower than the values ($7\le \delta_{c}\le 23$) given by Bonatto \& Bica (2009). 
This estimation of $\delta_{c}$ indicates that all clusters are sparse.

\begin{figure*}
\begin{center}
\centering
\hbox{
\includegraphics[width=8.5cm, height=8.5cm]{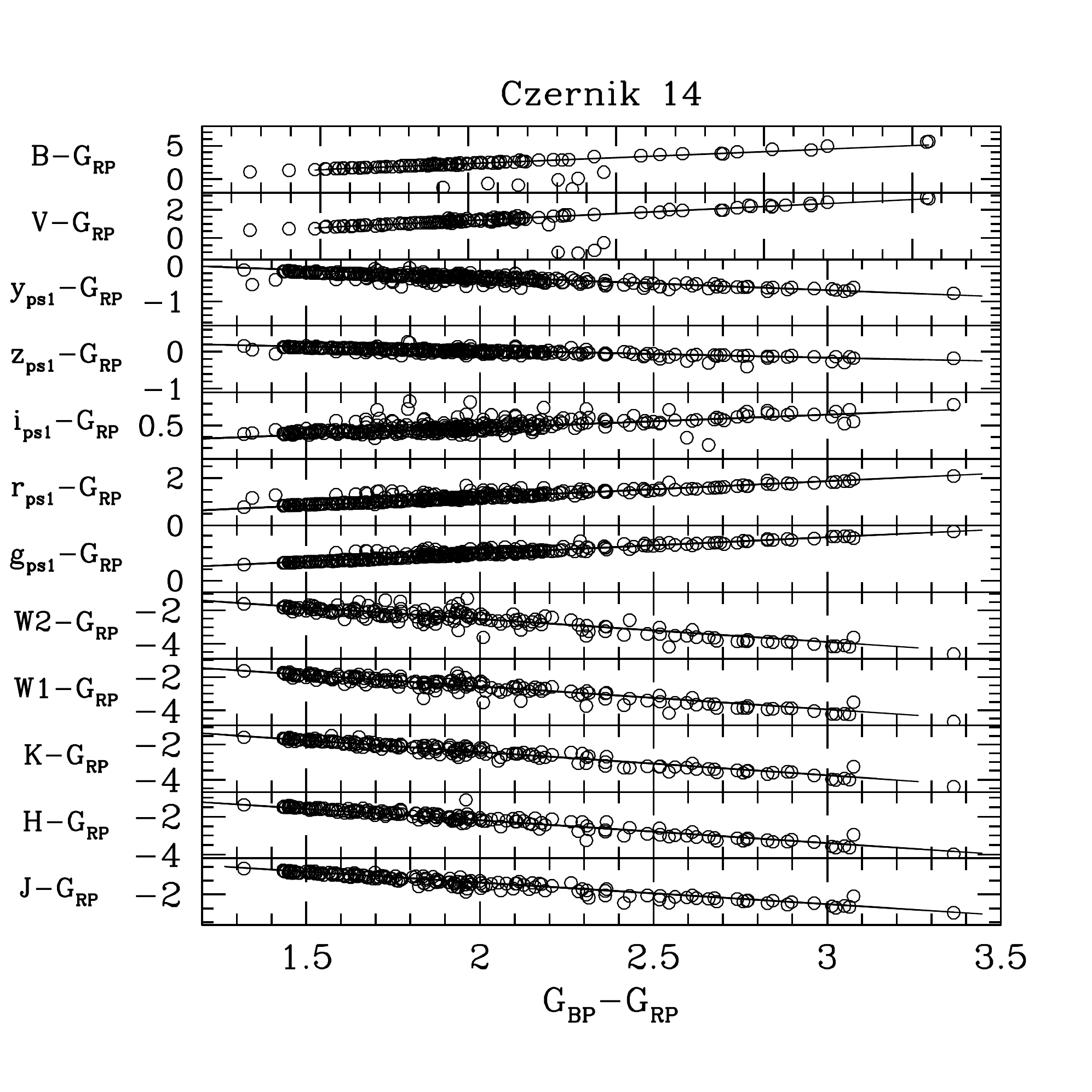}
\includegraphics[width=8.5cm, height=8.5cm]{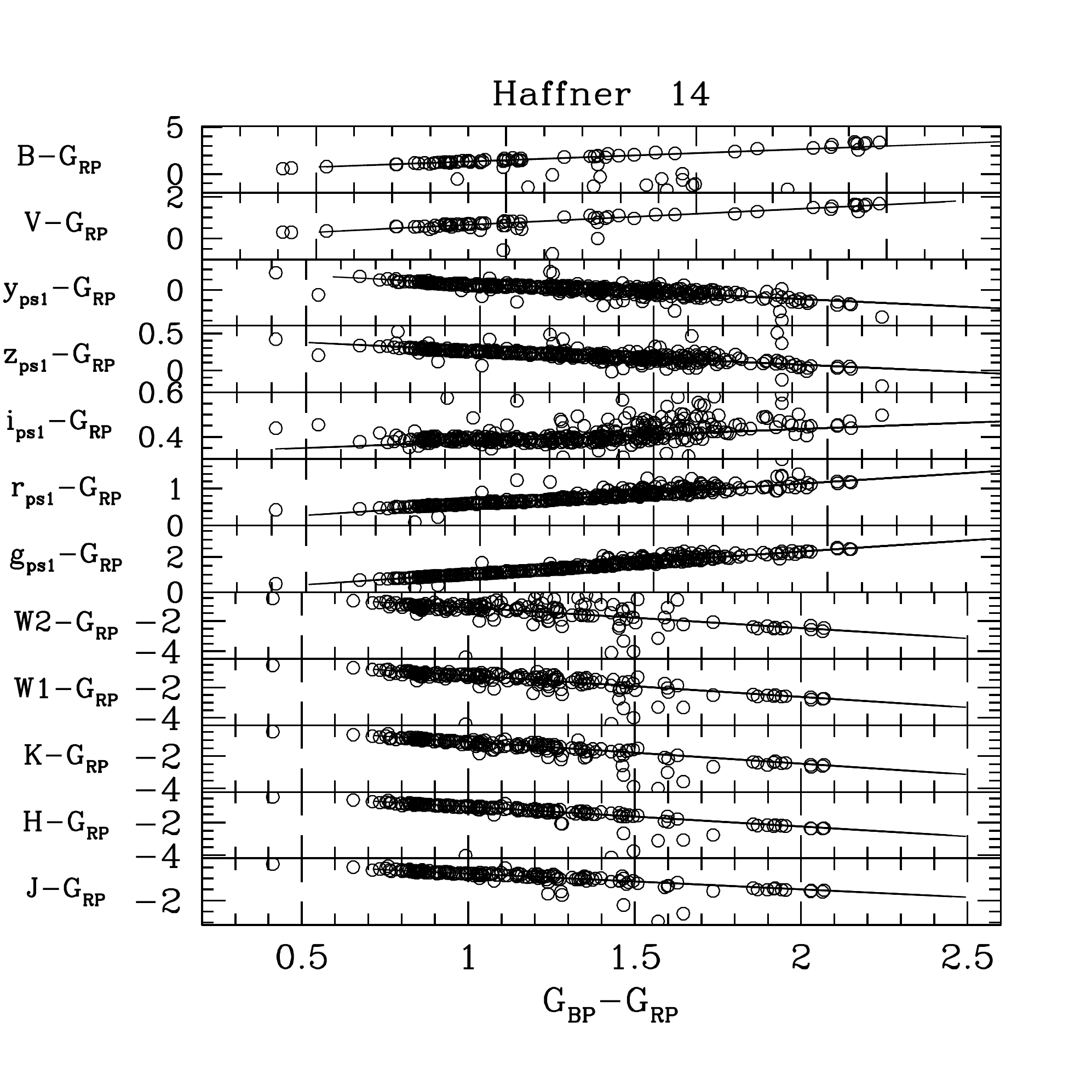}
}
\hbox{
\includegraphics[width=8.5cm, height=8.5cm]{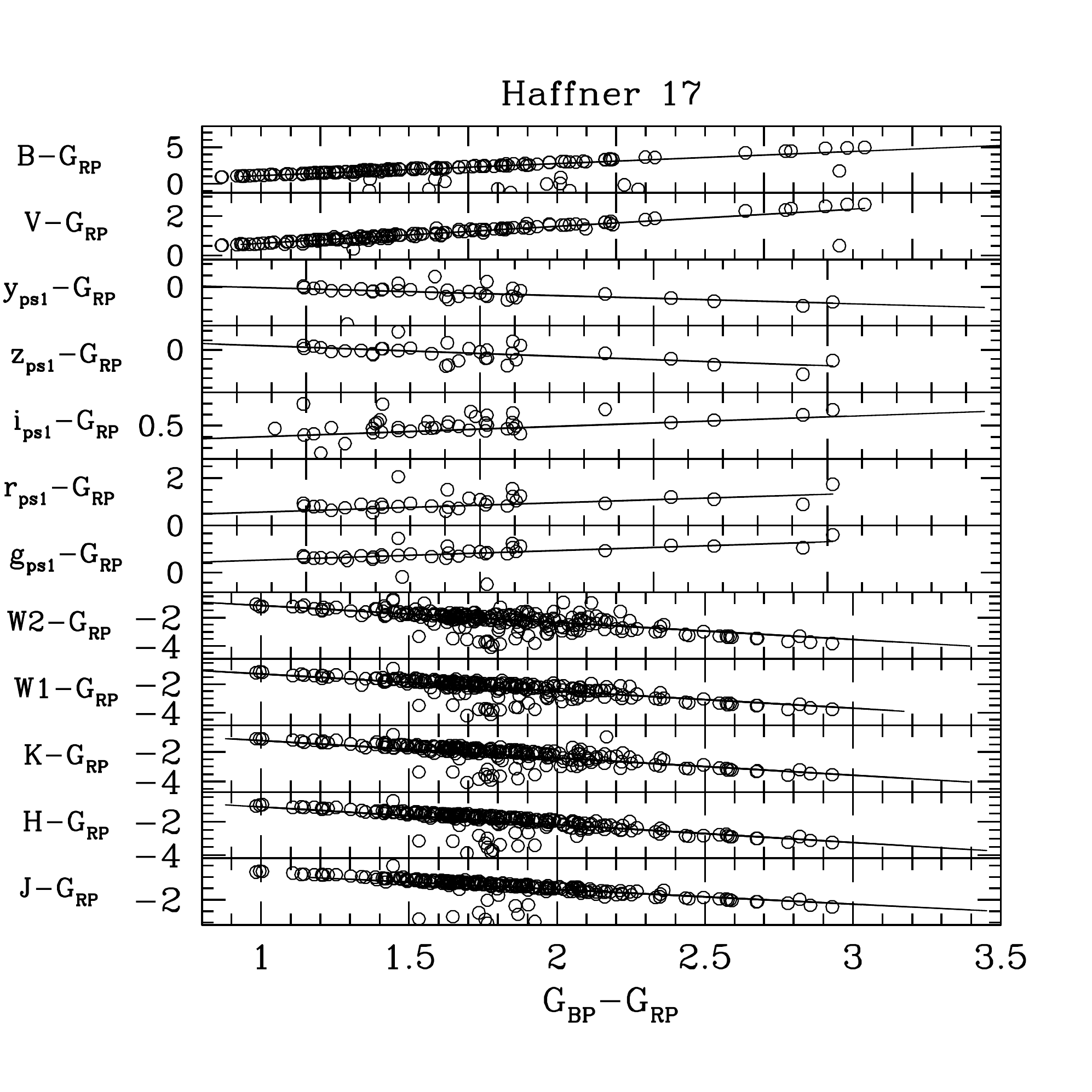}
\includegraphics[width=8.5cm, height=8.5cm]{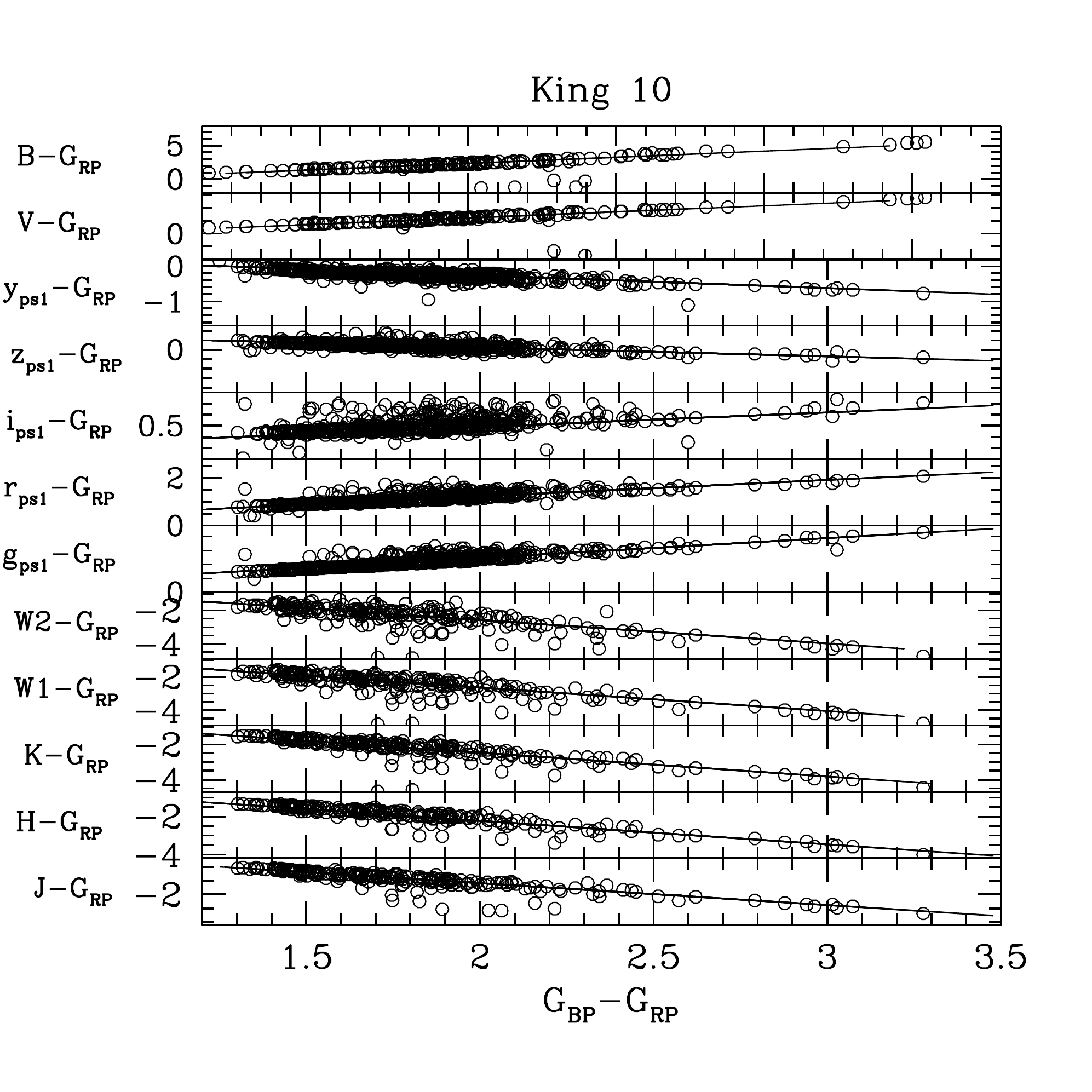}
}              

\caption{The $(\lambda-G_{RP})/(G_{BP}-G_{RP})$ TCDs using the stars selected from VPDs of clusters Czernik 14, Haffner 14,
Haffner 17 and King 10. The continuous lines represent the slope determined through least-squares linear fit.
}
\label{cc_gaia}
\end{center}
\end{figure*}

The tidal radius of clusters are normally influenced by the effects of Galactic tidal fields and later internal relaxation
and dynamical evolution of clusters (Allen \& Martos 1988). The tidal radius is determined as follows.

The Galactic mass $M_{G}$ inside a Galactocentric radius $R_{G}$ is given by (Genzel \& Townes, 1987),\\

$M_{G}=2\times10^{8} M_{\odot} (\frac{R_{G}} {30 pc})^{1.2}$\\

Estimated values of Galactic mass inside the Galactocentric radius (see Sec. 4.5) are found as $2.2\times10^{11} M_{\odot}$, 
$2.8\times10^{11} M_{\odot}$, $2.5\times10^{11} M_{\odot}$ and $2.1\times10^{11} M_{\odot}$ for the clusters 
Czernik 14, Haffner 14, Haffner 17 and King 10 respectively. Present estimation of Galactic mass are close to the value 
(2.9$\pm0.4\times10^{11} M_{\odot}$) determined by Gibbons, Belokurov \& Evans (2014) within 50 kpc radius of the Galaxy.

Kim et al. (2000) has specified the clusters tidal radius $R_{t}$ as, \\

$R_{t}=(\frac{M_{c}} {2M_{G}})^{1/3}\times R_{G}$\\

\begin{table*}
\centering
\caption{Multi-band colour excess ratios in the direction of clusters Czernik 14, Haffner 14, Haffner 17 and King 10.
}
\vspace{0.5cm}
\begin{center}
\small
\begin{tabular}{cccccc}
\hline\hline
Band $(\lambda)$ & Effective wavelength  &&   $\frac{\lambda-G_{RP}}{G_{BP}-G_{RP}}$ \\
&              (nm) &  Czernik 14 &  Haffner 14 & Haffner 17 & King 10
\\
\hline\hline
Johnson~ B        &445              &$1.78\pm0.02$ & $1.50\pm0.01$ &$1.65\pm0.02$ &$1.82\pm0.01$\\
Johnson~ V        &551              &$0.99\pm0.01$ & $0.88\pm0.01$ &$0.87\pm0.02$ &$0.95\pm0.02$\\
Pan-STARRS~ g     &481              &$1.43\pm0.02$ & $1.33\pm0.03$ &$1.14\pm0.03$ &$1.42\pm0.02$\\
Pan-STARRS~ r     &617              &$0.68\pm0.03$ & $0.60\pm0.02$ &$0.45\pm0.03$ &$0.69\pm0.02$\\
Pan-STARRS~ i     &752              &$0.11\pm0.03$ & $0.06\pm0.04$ &$0.11\pm0.04$ &$0.13\pm0.04$\\
Pan-STARRS~ z     &866              &$-0.20\pm0.04$ & $-0.21\pm0.03$ &$-0.26\pm0.03$ &$-0.19\pm0.03$\\
Pan-STARRS~ y     &962              &$-0.37\pm0.04$ & $-0.32\pm0.03$ &$-0.41\pm0.04$ &$-0.36\pm0.04$\\
2MASS~ J          &1234.5           &$-0.78\pm0.03$ & $-0.71\pm0.04$ &$-0.64\pm0.03$ &$-0.79\pm0.04$\\
2MASS~ H          &1639.3           &$-1.20\pm0.04$ & $-1.12\pm0.04$ &$-1.07\pm0.05$ &$-1.24\pm0.04$\\
2MASS~ K          &2175.7           &$-1.33\pm0.06$ & $-1.31\pm0.07$ &$-1.17\pm0.06$ &$-1.36\pm0.05$\\
WISE ~W1          &3317.2           &$-1.38\pm0.07$ & $-1.27\pm0.08$ &$-1.21\pm0.06$ &$-1.41\pm0.07$\\
WISE~ W2          &4550.1           &$-1.43\pm0.08$ & $-1.12\pm0.09$ &$-1.14\pm0.07$ &$-1.41\pm0.08$\\
\hline
\end{tabular}
\label{gaia_slope}
\end{center}
\end{table*}
where $R_{t}$ and  $M_{c}$ indicate the cluster's tidal radius and total mass (see Sect.~8), respectively. The estimated
values of tidal radius are $9.71\pm0.5$, $12.86\pm0.7$, $13.11\pm0.6$ and $13.41\pm0.5$ pc for Czernik 14, Haffner 14, Haffner 17
and King 10, respectively and listed in Table~\ref{stru_para}. 

\subsection{Optical to mid-infrared extinction law}

In this section, we combined multi-wavelength photometric data with Gaia astrometry to check the extinction law from
optical to mid-infrared region for clusters under study. The resultant $(\lambda-G_{RP})/(G_{BP}-G_{RP})$ two colour diagrams
(TCDs) are shown in Fig \ref{cc_gaia} for all the clusters. Here, $\lambda$ denotes the filters other than $G_{RP}$.
All stars showing in Fig \ref{cc_gaia} are probable cluster members. A linear fit to the data points is performed and slopes
are listed in Table \ref{gaia_slope}. The estimated values of slopes are in good agreement with the value given by Shu Wang
and Xiaodian Chen (2019). We estimated $\frac{A_{V}}{E(B-V)}$ as $\sim$ 3.1 for all the clusters under study. This indicates
that reddening law is normal towards the clusters under study.

\subsection{Interstellar reddening from 2MASS colours}

\begin{figure}
\begin{center}
\centering
\includegraphics[width=8.5cm, height=8.5cm]{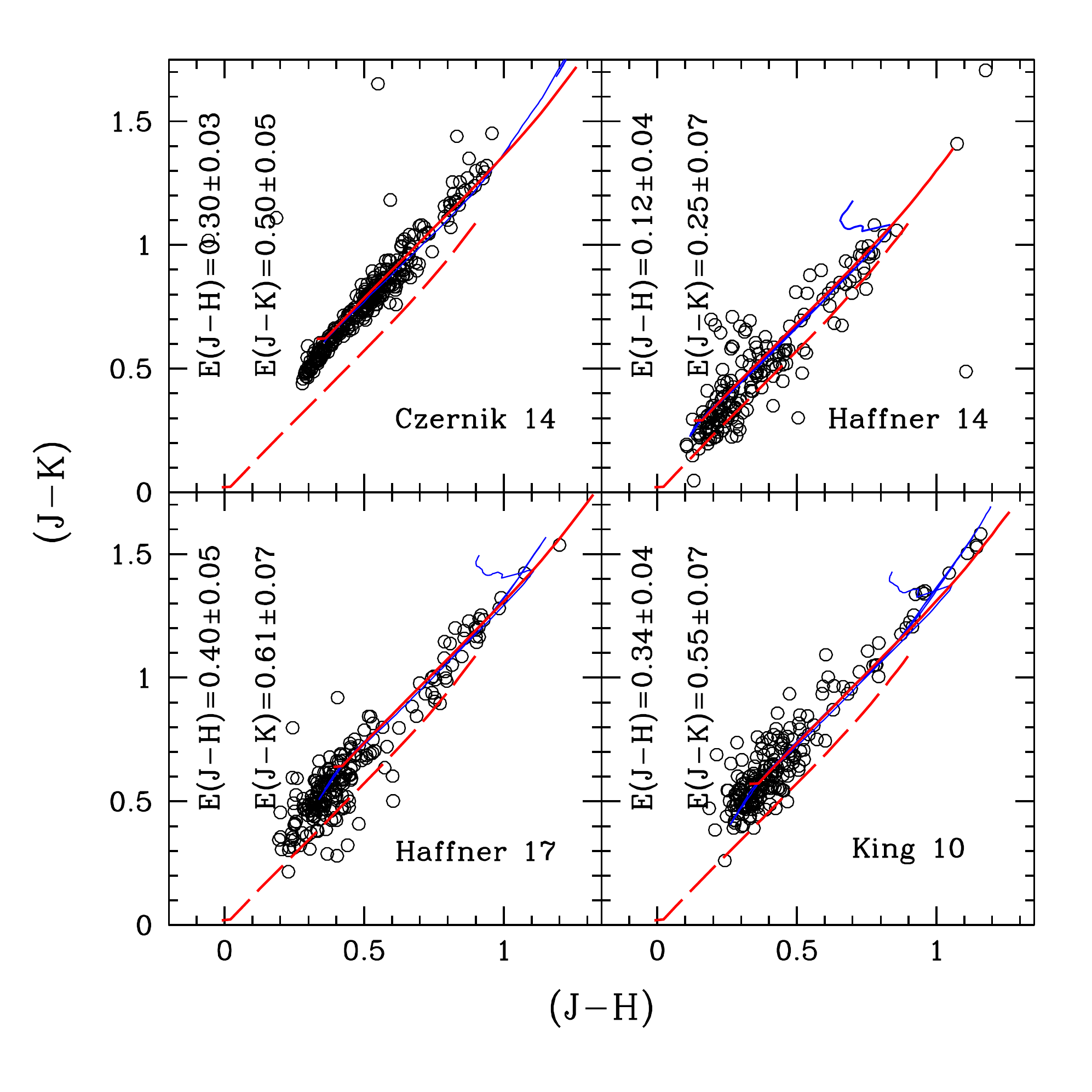}
\caption{The colour-colour diagrams (CCDs) for clusters Czernik 14, Haffner 14, Haffner 17 and King 10 using probable
cluster members. In CCDs, the red solid line is the ZAMS taken from Caldwell et al. (1993) while the red dotted lines
are the same ZAMS shifted by the values as described in the text. The blue line is the theoretical isochrones of 
log(age)=8.75, 8.50, 7.95 and 7.65 for clusters Czernik 14, Haffner 14, Haffner 17 and King 10, respectively.}
\label{cc}
\end{center}
\end{figure}

To estimate the cluster reddening in the near-IR region, we used $(J-H)$ versus $(J-K)$ colour-colour diagrams
as shown in Fig \ref{cc}. Stars plotted in this figure are the probable cluster members described in Sec. 3. The solid
line is cluster's zero age main sequence (ZAMS) taken from Caldwell et al. (1993). The ZAMS shown by the dotted
line is displaced by $E(J-H) = 0.30\pm0.03$ mag and $E(J-K) = 0.50\pm0.05$ mag for Czernik 14, $0.12\pm0.04$ and $0.25\pm0.07$ mag
for Haffner 14, $E(J-H) = 0.40\pm0.05$ mag and $E(J-K) = 0.61\pm0.07$ mag for Haffner 17 and $E(J-H) = 0.34\pm0.04$ mag and
$E(J-K) = 0.55\pm0.07$ mag for King 10. The solid line in this figure is theoretical isochrone taken from Marigo et al. (2017)
of log(age)=8.75, 8.50, 7.95 and 7.65 for the clusters Czernik 14, Haffner 14, Haffner 17 and King 10, respectively. The ratio 
of $E(J-H)$ and $E(J-K)$ shows a good agreement with the normal value 0.55 proposed by 
Cardelli et al. (1989). Using $E(J-H)$ and $E(J-K)$, we have calculated the interstellar reddenings ($E(B-V)$) as, 0.96, 0.38,
1.29 and 1.09 for the clusters Czernik 14, Haffner 14, Haffner 17 and King 10, respectively. Our derived value of $E(B-V)$ is 
higher than Tadross (2014) for Czernik 14. The Present estimate of $E(B-V)$ is reliable than the value given by Tadross (2014)
because it is based on the cluster members selected using proper motion data.  

\subsection{Age, distance and Galactocentric coordinates}

\begin{figure*}
\begin{center}
\centering
\hbox{
\hspace{1cm}\includegraphics[width=8.5cm, height=8.5cm]{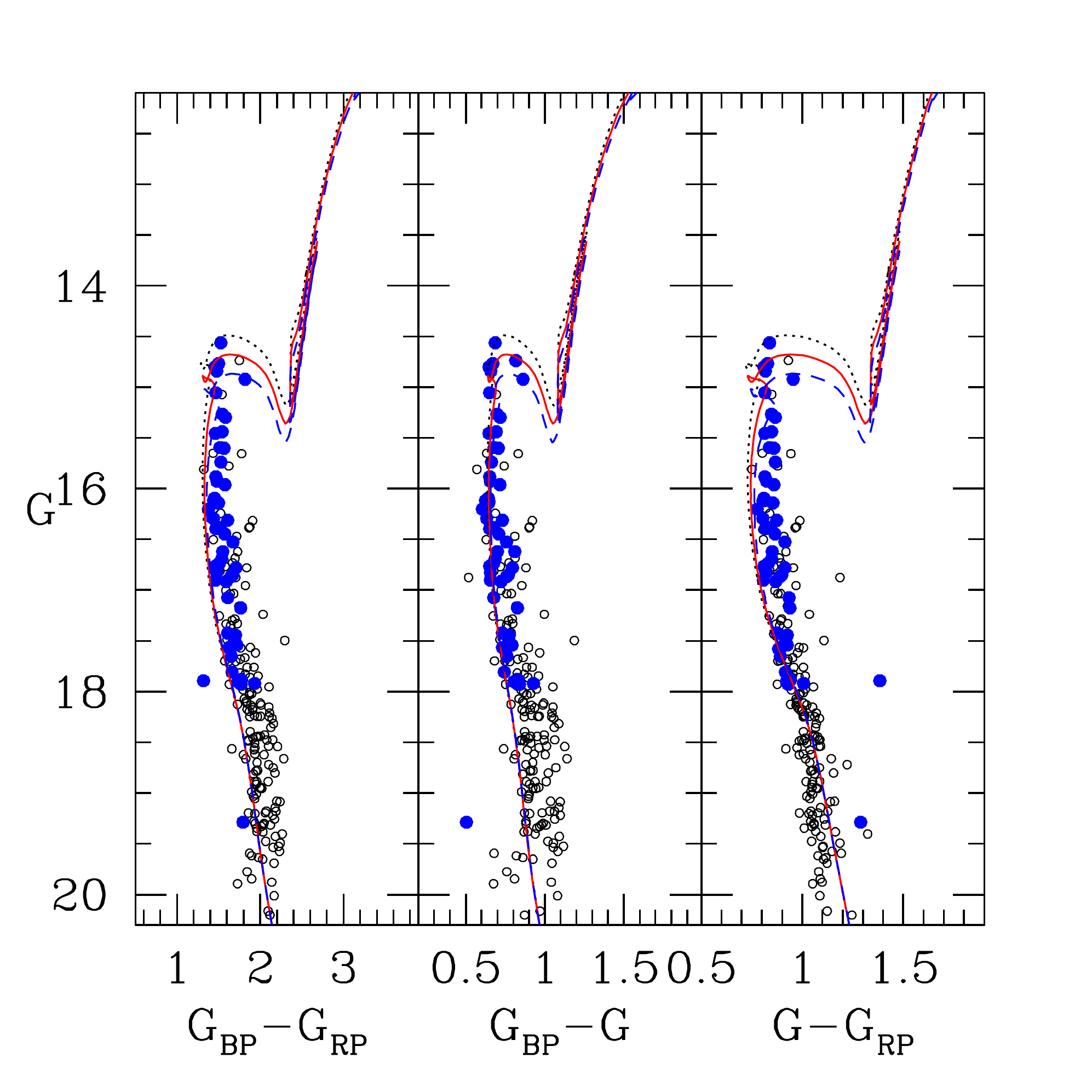}
\hspace{-1cm}\includegraphics[width=8.5cm, height=8.5cm]{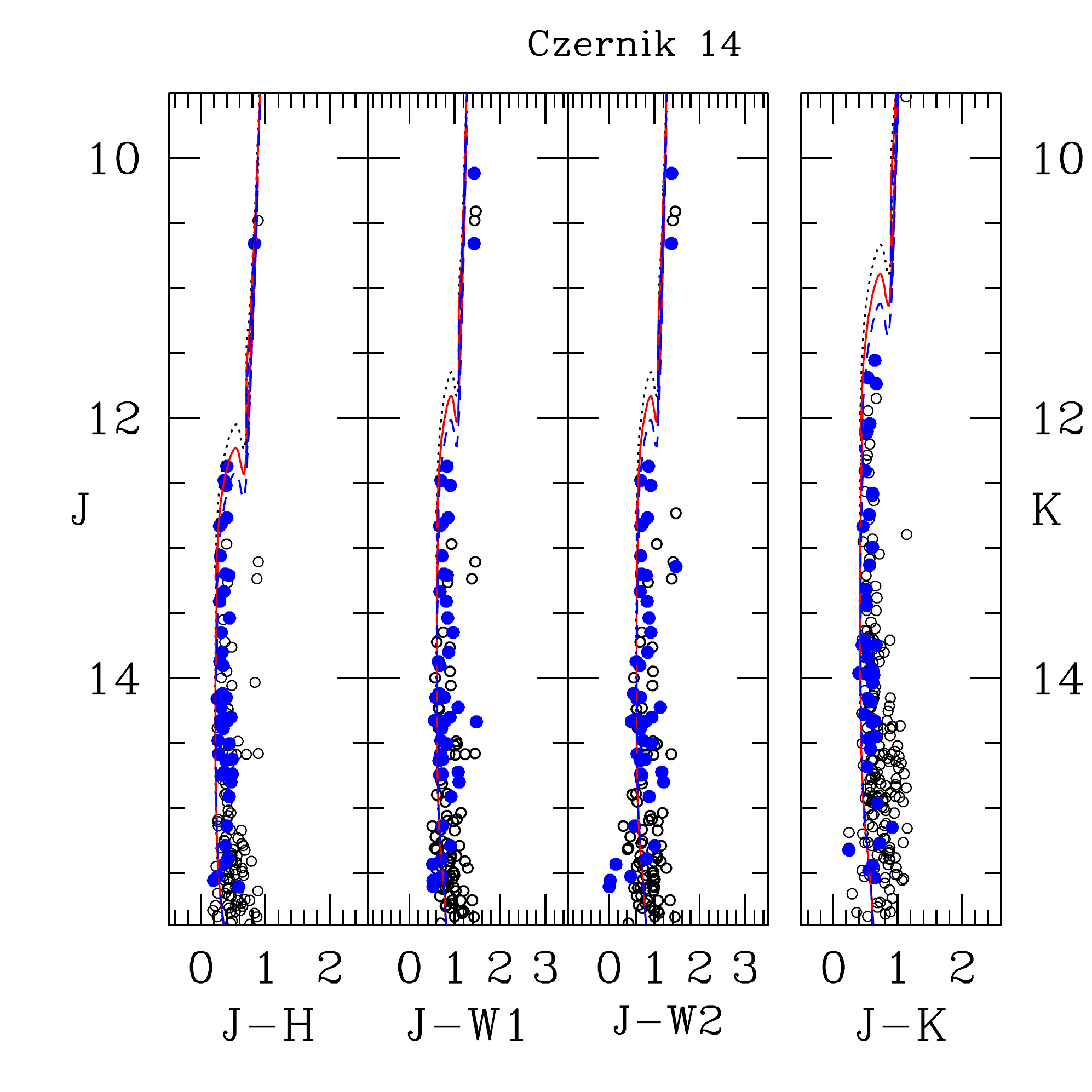}
}
\hbox{
\hspace{1cm}\includegraphics[width=8.5cm, height=8.5cm]{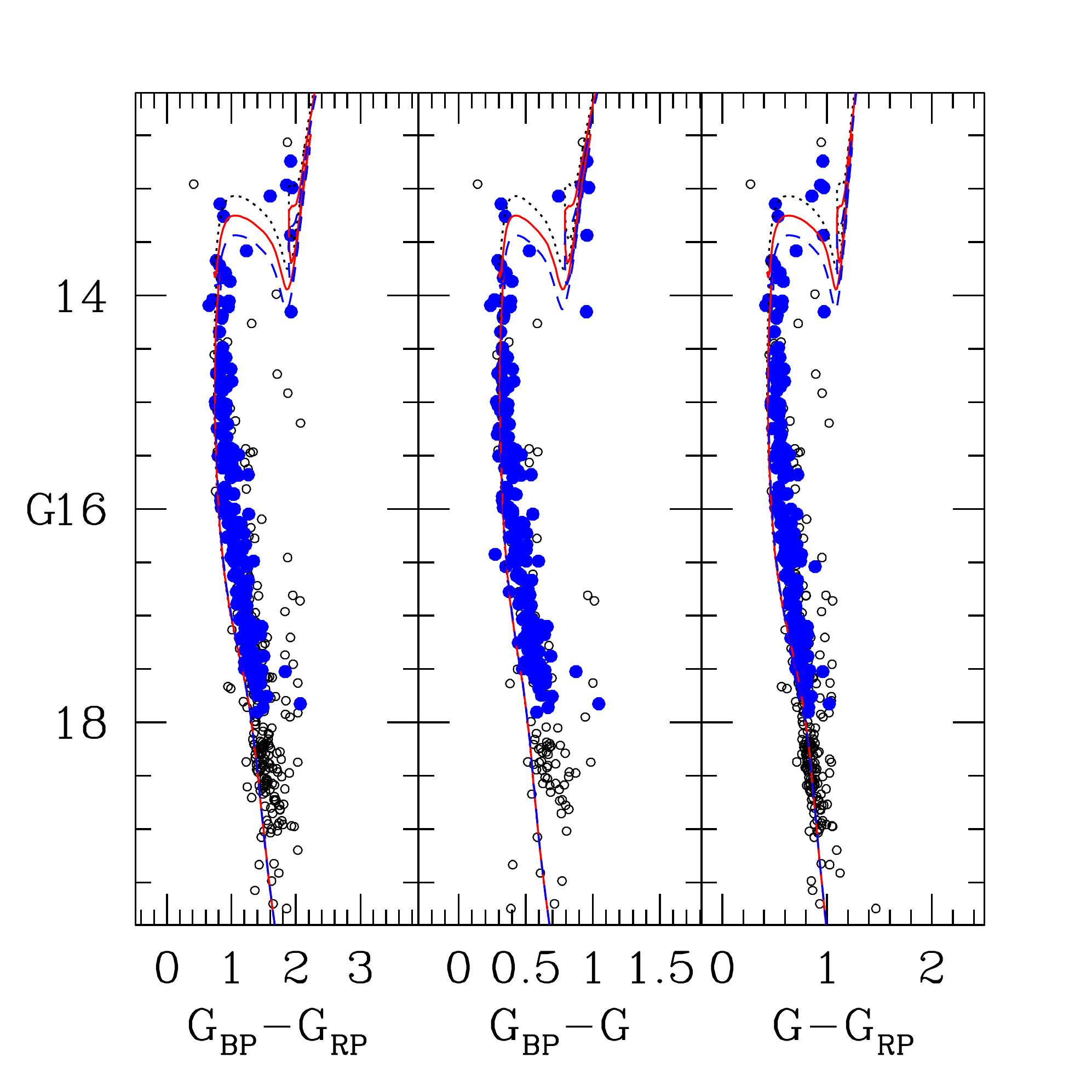}
\hspace{-1cm}\includegraphics[width=8.5cm, height=8.5cm]{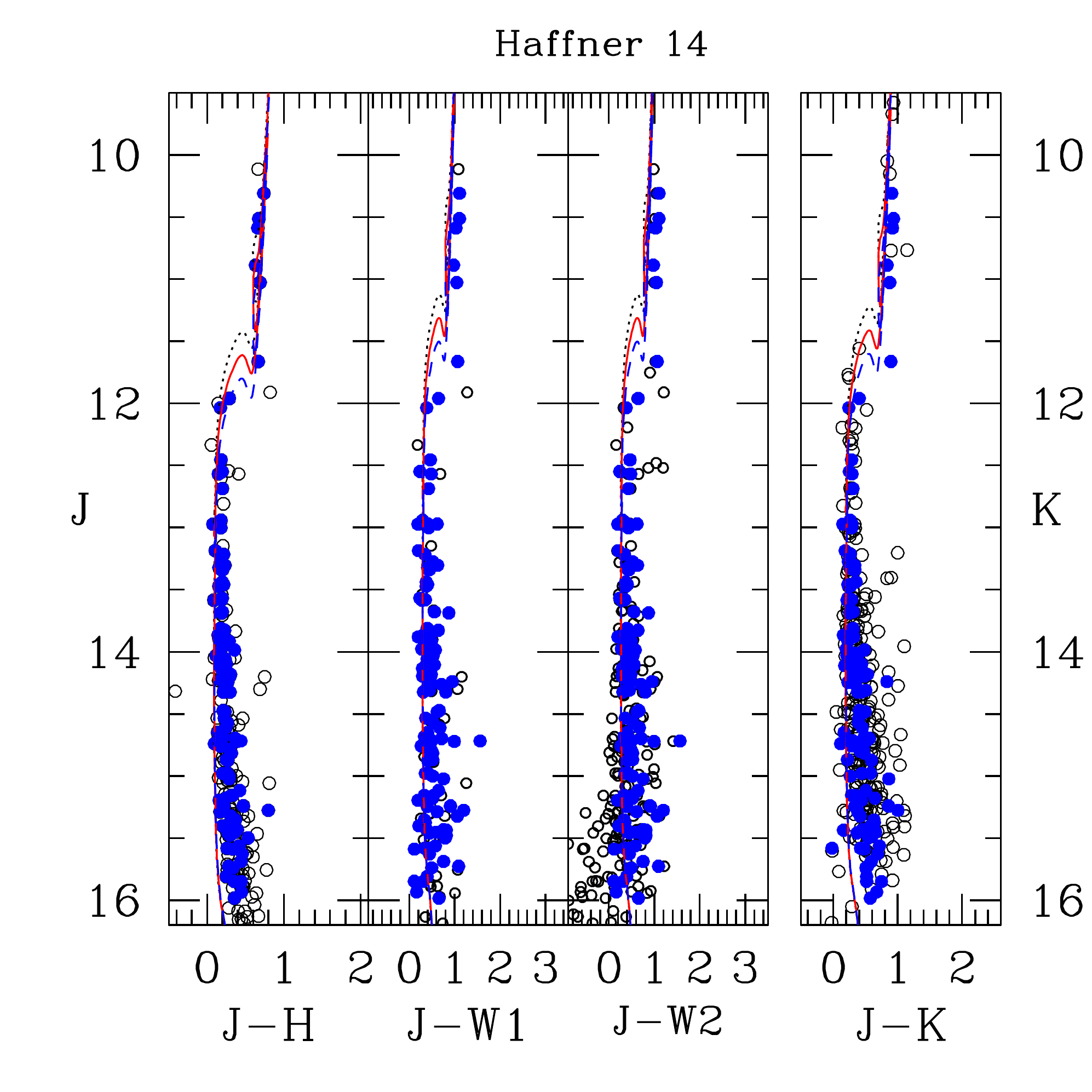}
}
\caption{The $G, (G_{BP}-G_{RP})$, $G, (G_{BP}-G)$, $G, (G-G_{RP})$, $J, (J-H)$, $J, (J-W1)$, $J, (J-W2)$ and $K, (J-K)$ colour-magnitude
diagrams of open star cluster Czernik 14 and Haffner 14. Black open circles are probable cluster members as selected
from VPDs. Blue dots are most likely members with membership probability greater than 40 $\%$. The curves are the isochrones of 
(log(age)=8.70, 8.75 and 8.80) for Czernik 14 and (log(age)=8.45, 8.50 and 8.55) for Haffner 14. All these isochrones are taken
from Marigo et al. (2017) for solar metallicity.}
\label{cmd}
\end{center}
\end{figure*}

\begin{figure*}
\begin{center}
\centering
\hbox{
\hspace{1cm}\includegraphics[width=8.5cm, height=8.5cm]{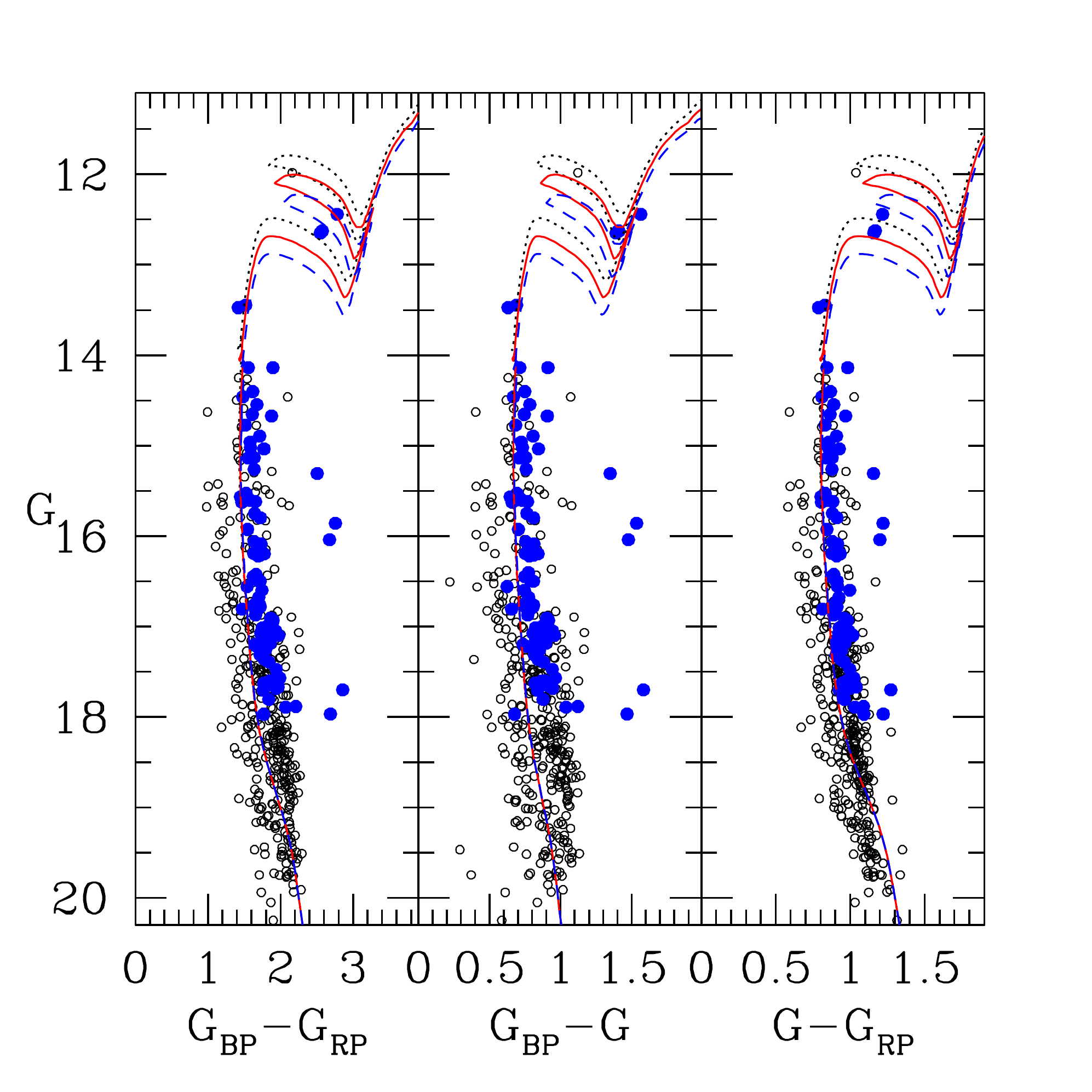}
\hspace{-1cm}\includegraphics[width=8.5cm, height=8.5cm]{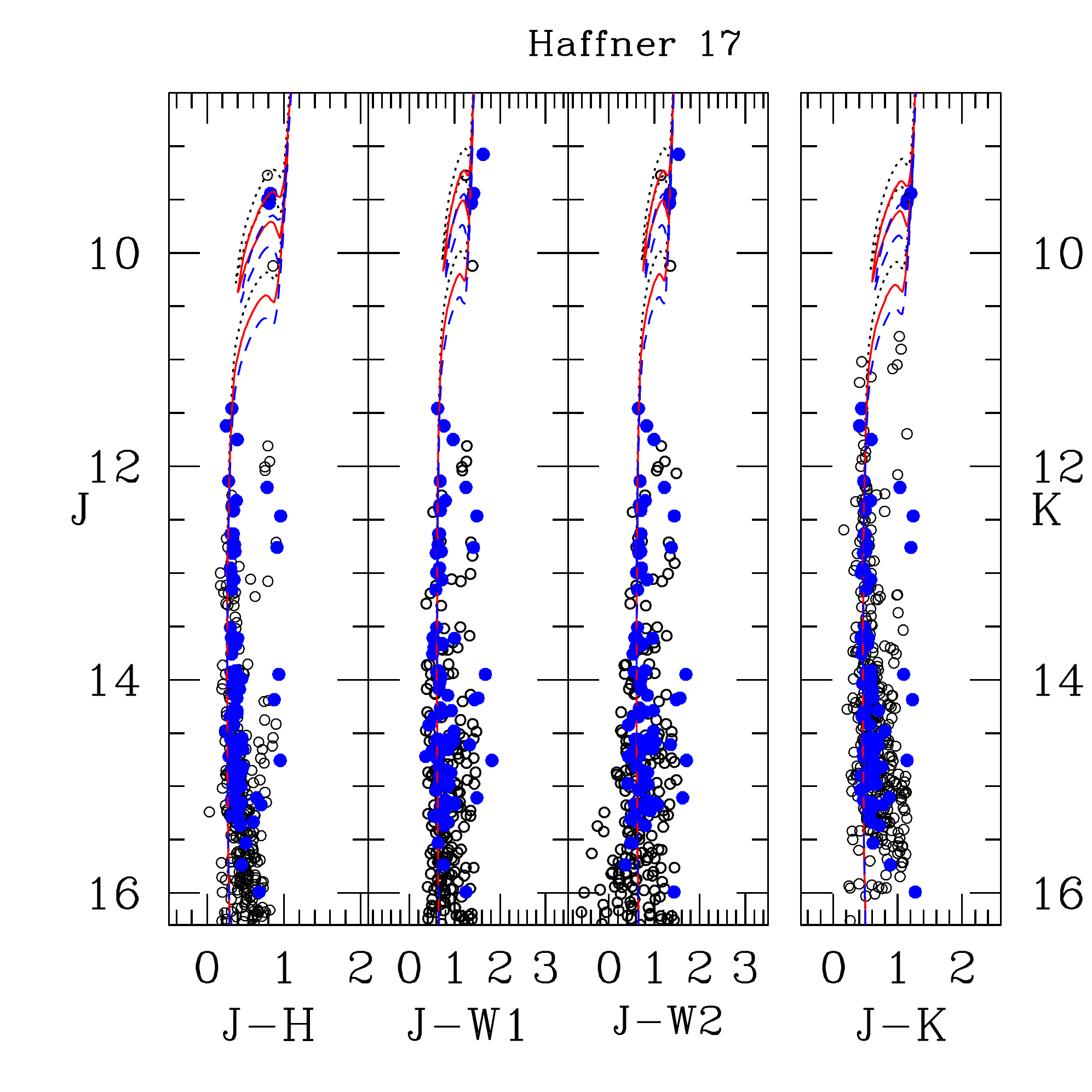}
}
\hbox{
\hspace{1cm}\includegraphics[width=8.5cm, height=8.5cm]{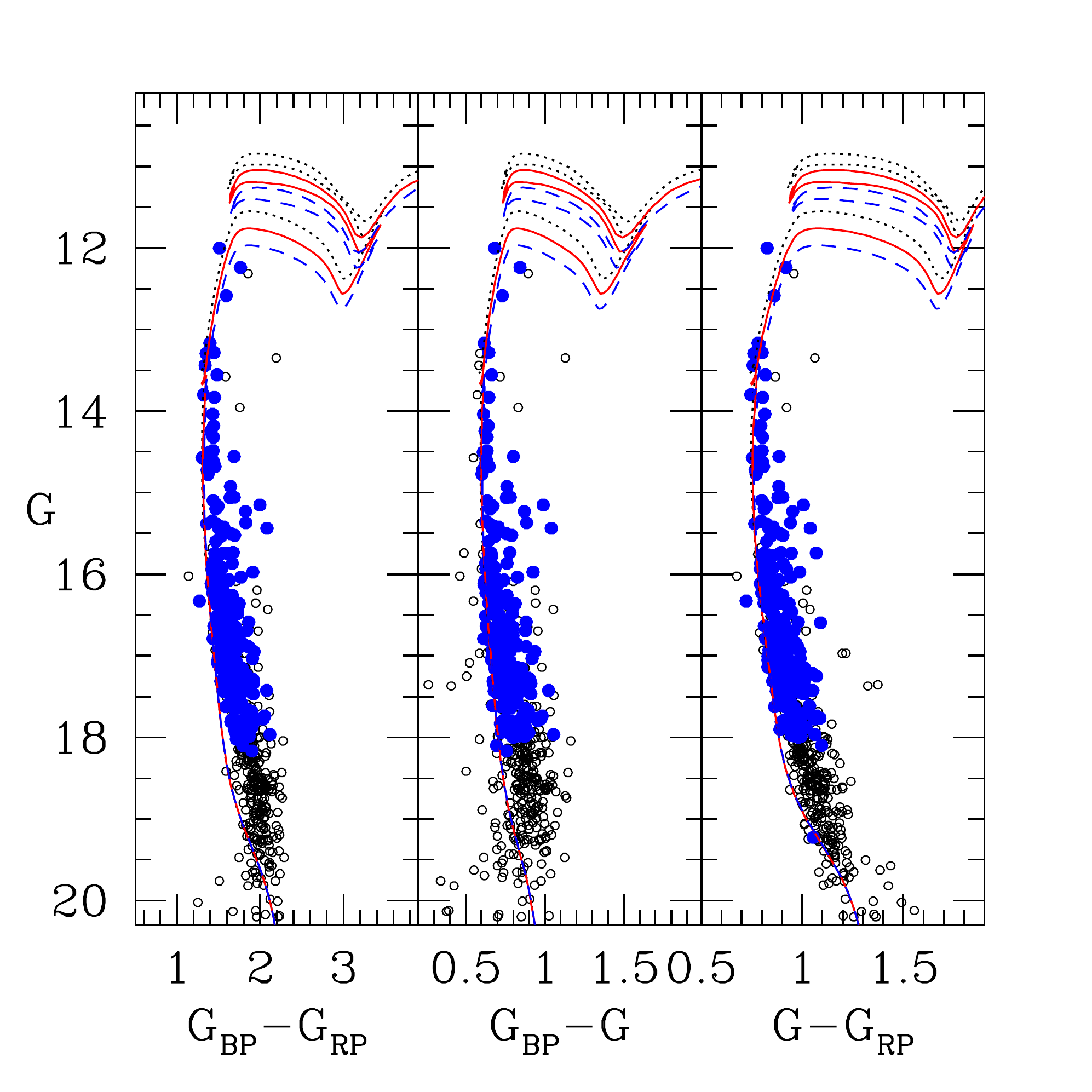}
\hspace{-1cm}\includegraphics[width=8.5cm, height=8.5cm]{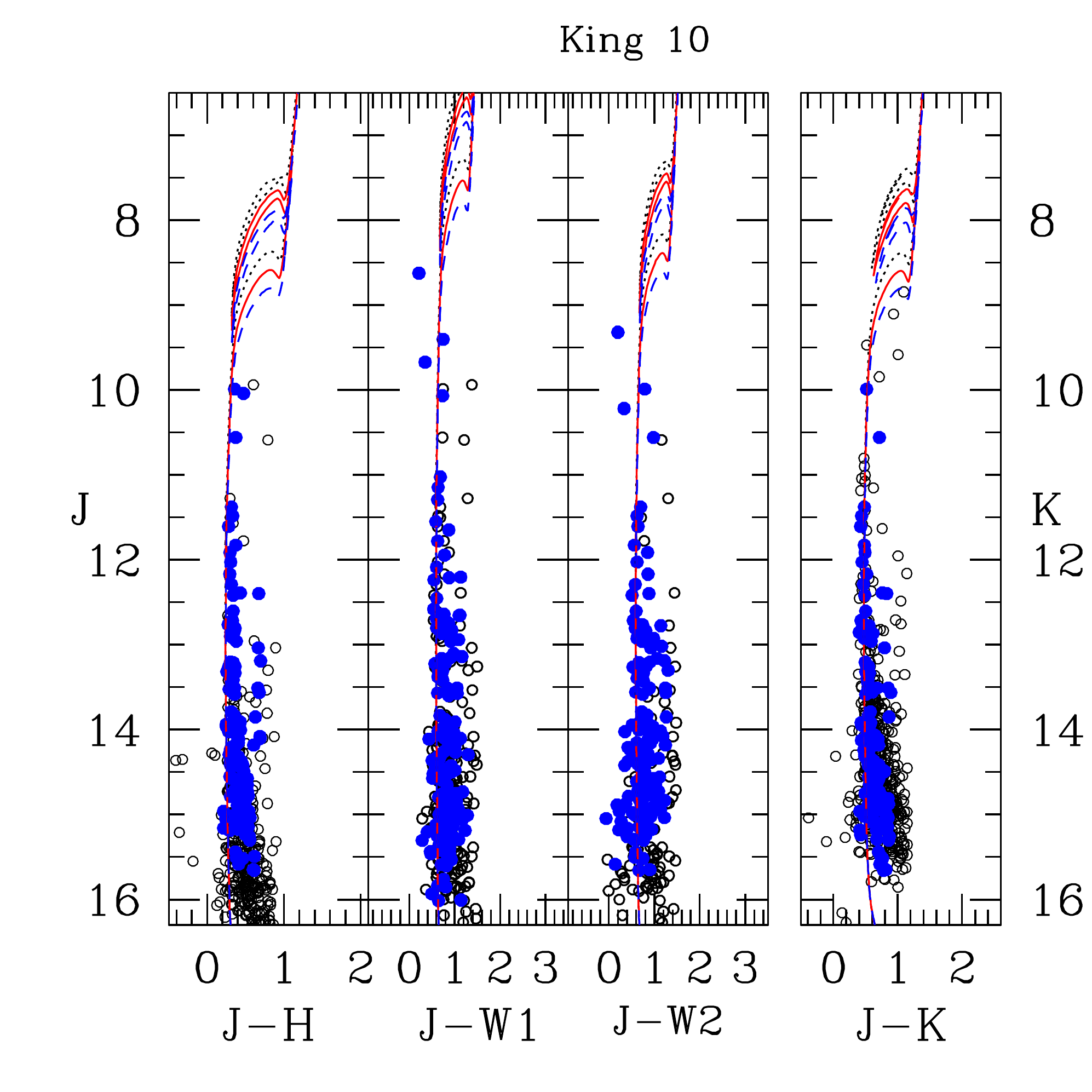}
}
\caption{Same as Fig \ref{cmd} for clusters Haffner 17 and King 10. The curves are the isochrones of (log(age)=7.90, 7.95 and 8.00) for
Haffner 17 and (log(age)=7.60, 7.65 and 7.70) for King 10.}
\label{cmd1}
\end{center}
\end{figure*}

%The average value of metallicity ([Fe/H]) have been estimated as $-0.02\pm0.01$, $-0.08\pm0.01$, $-0.06\pm0.01$ and $-0.05\pm0.02$ for
%clusters Czernik 14, Haffner 14, Haffner 17 and King 10 respectively using Gaia DR2 data with TOPCAT tool. {\bf Isochrones of the estimated metallicity
%with three different log(ages) to clusters CMDs are plotted. We considered
%only our estimated metallicities.} We used the relationship given by 
%Mowlavi et al. (2012) to transform [Fe/H] into the mass fraction $Z$. We found the value of $Z$ as $\sim0.01$ for all clusters.
The essential fundamental parameters (age, distance, and reddening) are estimated by fitting the solar metallicity ($Z=0.019$) isochrones
of Marigo et al. (2017) to all the CMDs $(G, G_{BP}-G_{RP}), (G, G_{BP}-G), (G, G-G_{RP}) (J, J-H), (J, J-W1), (J, J-W2) \& (K, J-K)$ 
as shown in Fig \ref{cmd} and Fig \ref{cmd1}. To reduce the field star contamination we have used only probable cluster members based
on clusters VPDs. The ratios 
$\frac{A_{J}}{A_{V}}$=0.276 and $\frac{A_{H}}{A_{V}}$=0.176 are taken from Schlegel et al. (1998), while the ratio 
$\frac{A_{K_{s}}}{A_{V}}$=0.118 was derived from Dutra et al. (2002). The Gaia~DR2 absorption ratios $\frac{A_{G}}{A_{V}}$=0.859,
$\frac{A_{G_{BP}}}{A_{V}}$=1.068 and $\frac{A_{G_{RP}}}{A_{V}}$=0.652 are taken from Hendy (2018).

The galactocentric coordinates of the clusters $X$ (directed towards the galactic center in the Galactic disc), $Y$ (directed towards
the Galactic rotation) and distance from the galactic plane $Z$ (directed towards Galactic north pole) can be estimated using
clusters' distances, longitude and latitude. The estimated Galactocentric coordinates are given in the corresponding paragraph 
of the clusters.

The estimation of the main fundamental parameters for the clusters are given below:

{\bf Czernik 14:} We superimposed theoretical isochrones of different age (log(age)=8.70,8.75 and 8.80) in all the 
CMDs for the cluster Czernik 14, shown in Fig \ref{cmd}. The overall fit is favorable for log(age)=8.75 (middle isochrone) to
the brighter cluster members. The estimated apparent distance modulus ($(m-M)=15.10\pm0.2$ mag) provides a distance $2.9\pm0.20$ kpc from
the Sun. Present estimate of distance is very close to the value 3.0 kpc derived by Cantat-Gaudin (2018).
The Galactocentric coordinates are derived as $X$=$2.32$~kpc, $Y$=$10.24$~kpc and $Z$=$0.04$~kpc. The Galactocentric distance of the cluster
is calculated to be $10.49\pm0.5$~kpc. The value of $Z$ indicates that Czernik 14 is above $\sim$40 pc from Galactic plane.\\

{\bf Haffner 14:} In the CMDs of Haffner 14, the isochrones of different age (log(age)=8.45, 8.50 and 8.55), are over plotted
in Fig \ref{cmd}. A satisfactory fitting of isochrones provides an age of $320\pm35$ Myr for this object. The inferred apparent distance
modulus $(m-M)=13.80\pm0.3$ mag provides a heliocentric distance as $4.8\pm0.2$ kpc. Our estimated value of distance is slightly
higher than the value 3.9 kpc derived by Cantat-Gaudin (2018), but very close to the value derived by us using parallax. The
Galactocentric distance is determined as $12.57\pm0.9$ kpc, which is calculated by considering 8.5 kpc as the distance of the
Sun to the Galactic center. The Galactocentric coordinates are determined as $X$=$-3.05$~kpc, $Y$=$12.20$~kpc and $Z$=$-0.15$~kpc.
This cluster is $\sim$150 pc below the Galactic plane.

{\bf Haffner 17:} Isochrones of different age (log(age)=7.90, 7.95 and 8.00) are used for the CMDs of Haffner 17, shown in
Fig \ref{cmd1}. By the isochrone fitting, we found an age $90\pm10$ Myr. The apparent distance modulus $(m-M)=16.20\pm0.25$ mag provides a
heliocentric distance as $3.6\pm0.1$ kpc which is similar to 3.5 kpc derived by Cantat-Gaudin (2018). The Galactocentric distance
is determined as $11.40\pm0.7$ kpc. The Galactocentric coordinates are determined as $X$=$-2.45$~kpc, $Y$=$11.13$~kpc and $Z$=$-0.14$~kpc.
This cluster is $\sim$140 pc below the Galactic plane.

{\bf King 10:} We superimposed isochrones of different age (log(age)=7.60, 7.65 and 7.70) to CMDs, shown in Fig \ref{cmd1}.
This provides an age of $45\pm5$ Myr which is similar to the value 50 Myr derived by Mohan et al. (1992). The inferred apparent distance modulus
$(m-M)=16.70\pm0.3$ mag provides a heliocentric distance as $3.8\pm0.1$ kpc which is not much different with 3.5 kpc derived by
Cantat-Gaudin (2018). The Galactocentric distance is calculated as $9.76\pm0.4$ kpc. The Galactocentric coordinates are determined as
$X$=$3.76$~kpc, $Y$=$9.10$~kpc and $Z$=$-0.02$~kpc. This cluster is very near to the Galactic plane.

\subsubsection{Distance of clusters from Gaia DR2 parallax}

\begin{figure}
\begin{center}
\hbox{
\includegraphics[width=4.2cm, height=4.2cm]{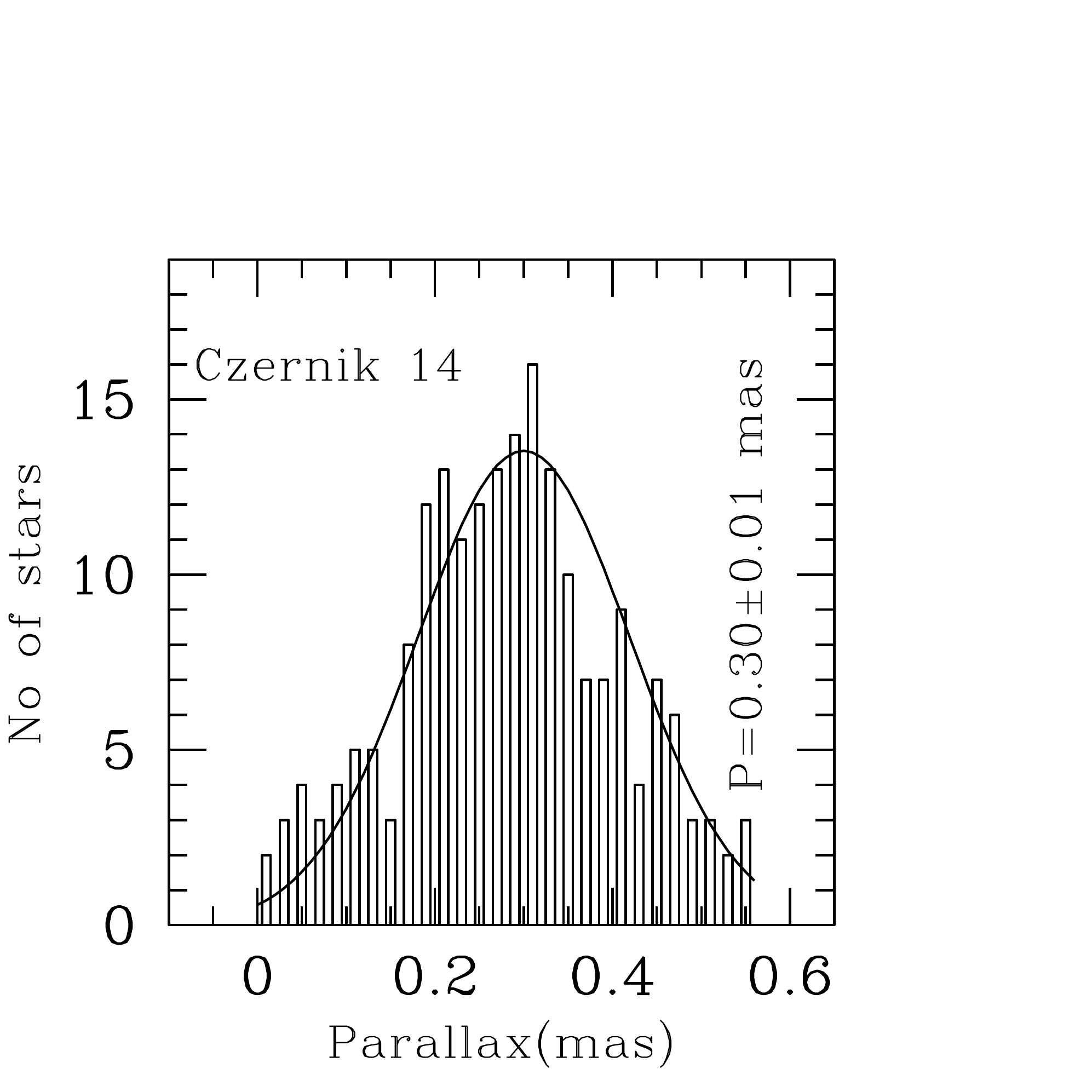}
\includegraphics[width=4.2cm, height=4.2cm]{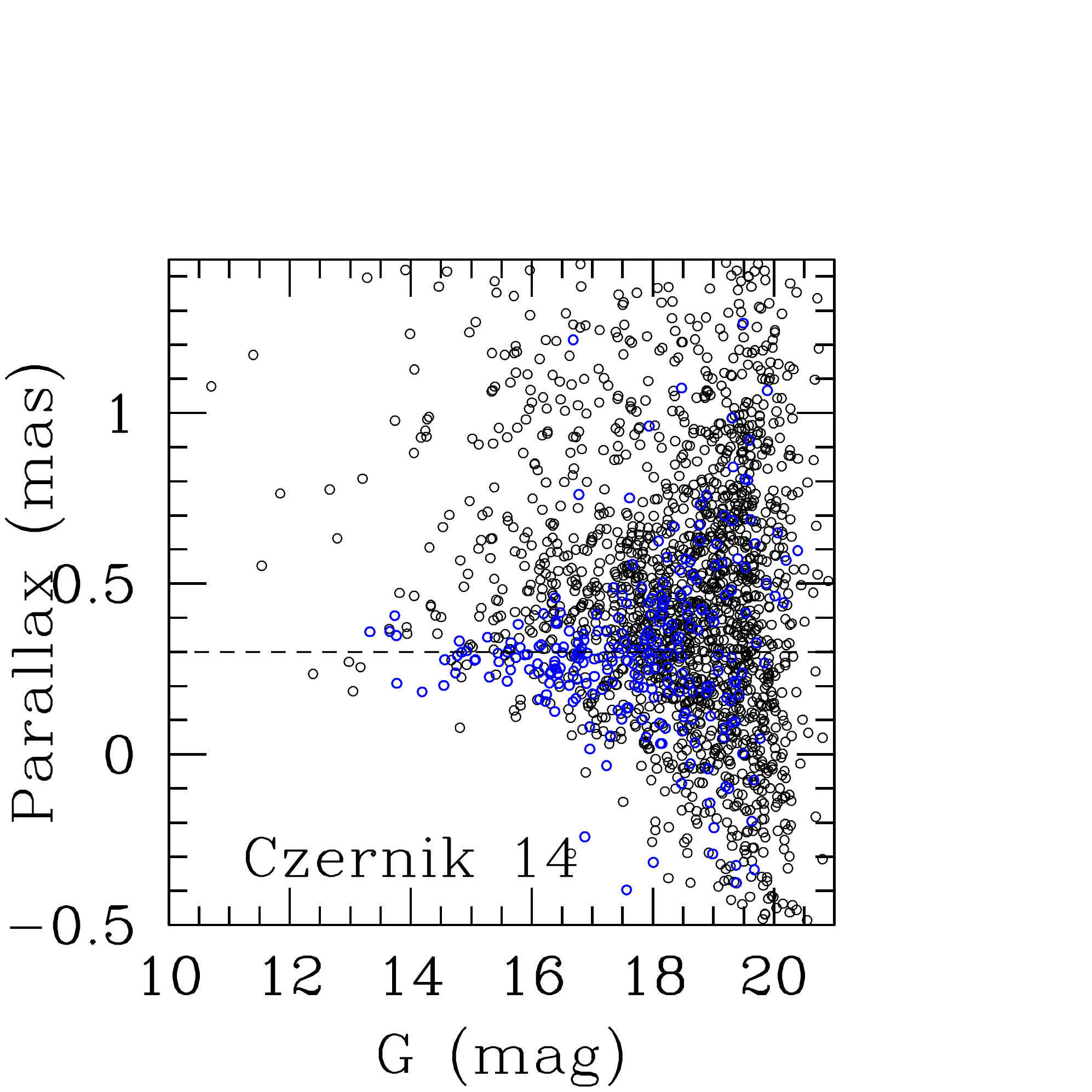}
}
\hbox{
\includegraphics[width=4.2cm, height=4.2cm]{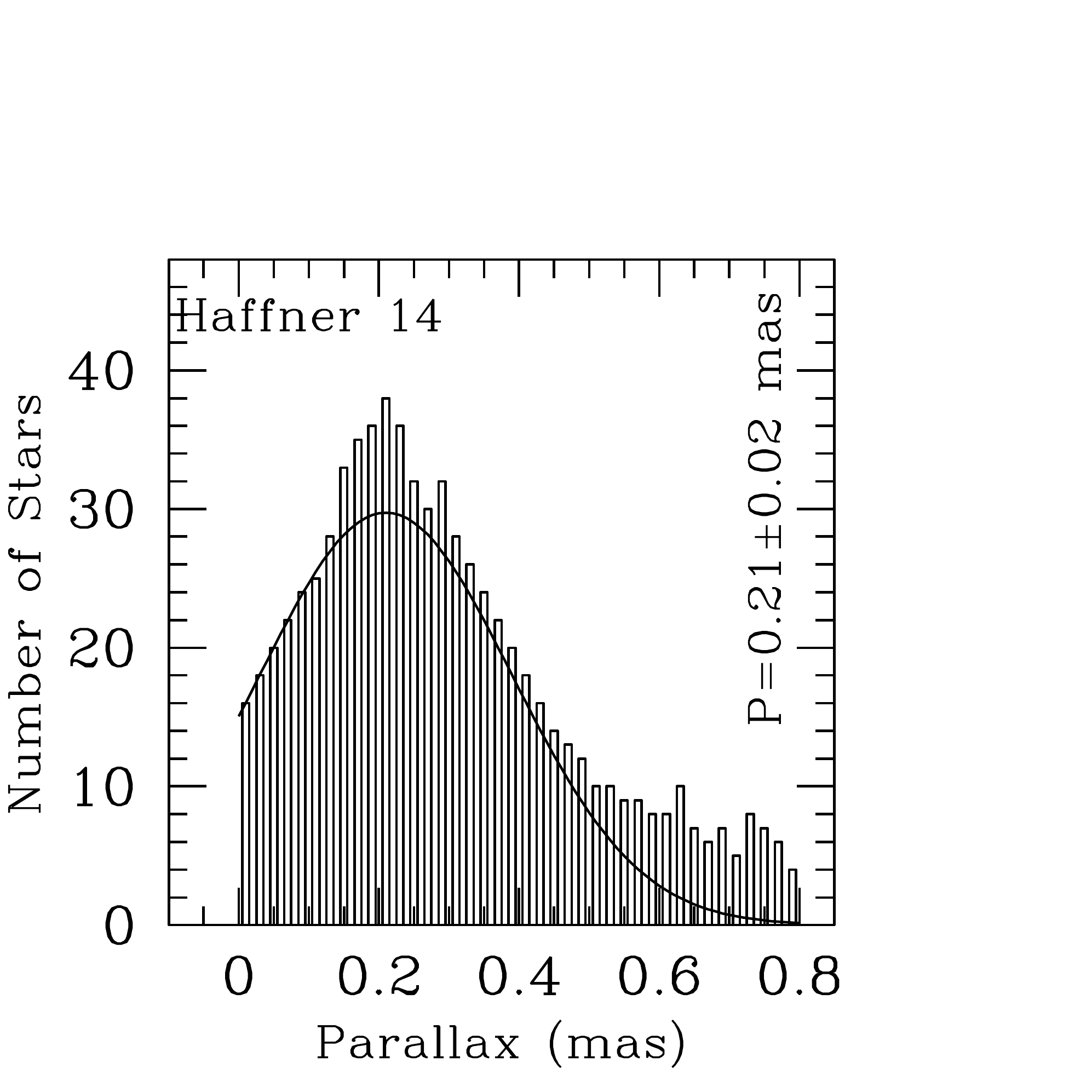}
\includegraphics[width=4.2cm, height=4.2cm]{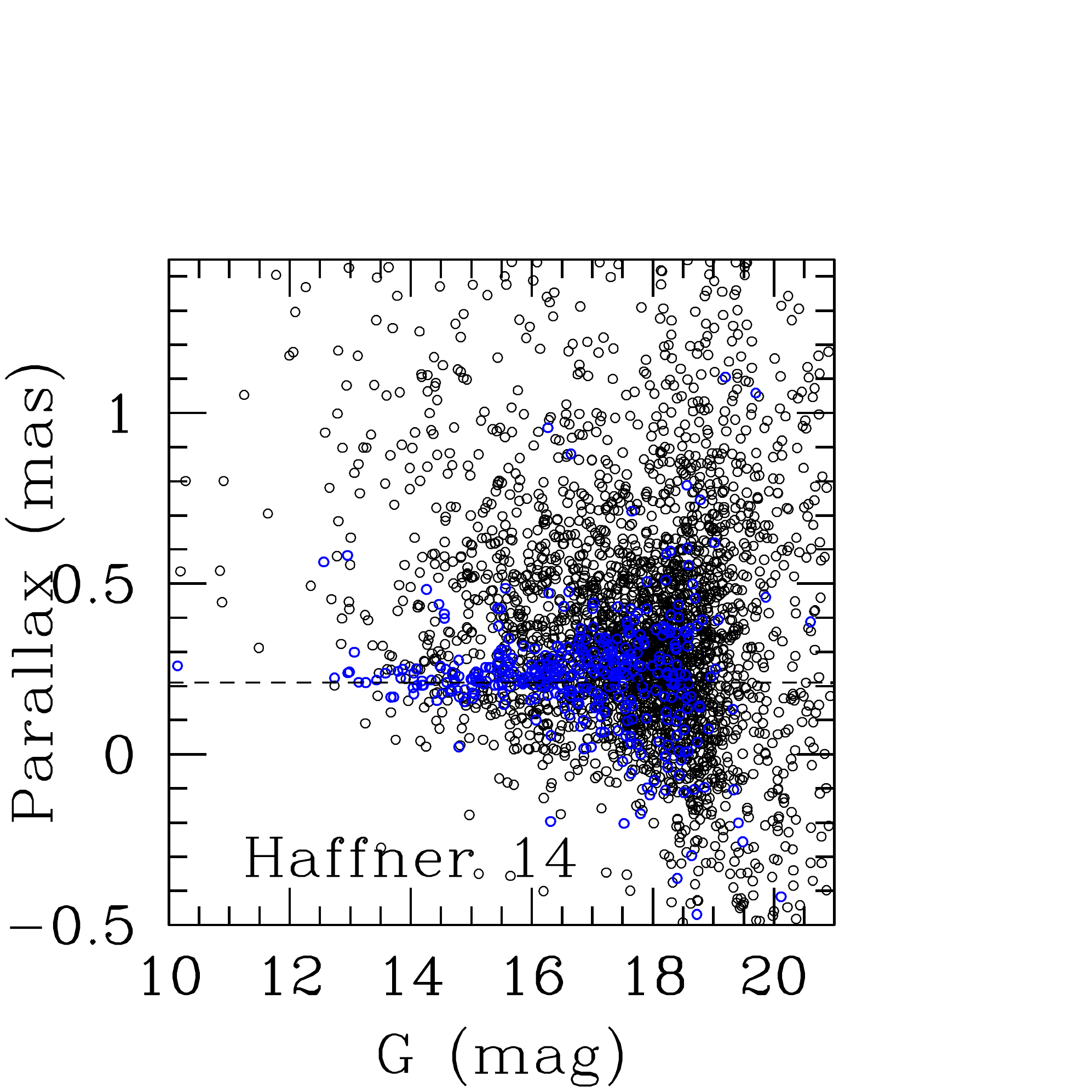}
}
\hbox{
\includegraphics[width=4.2cm, height=4.2cm]{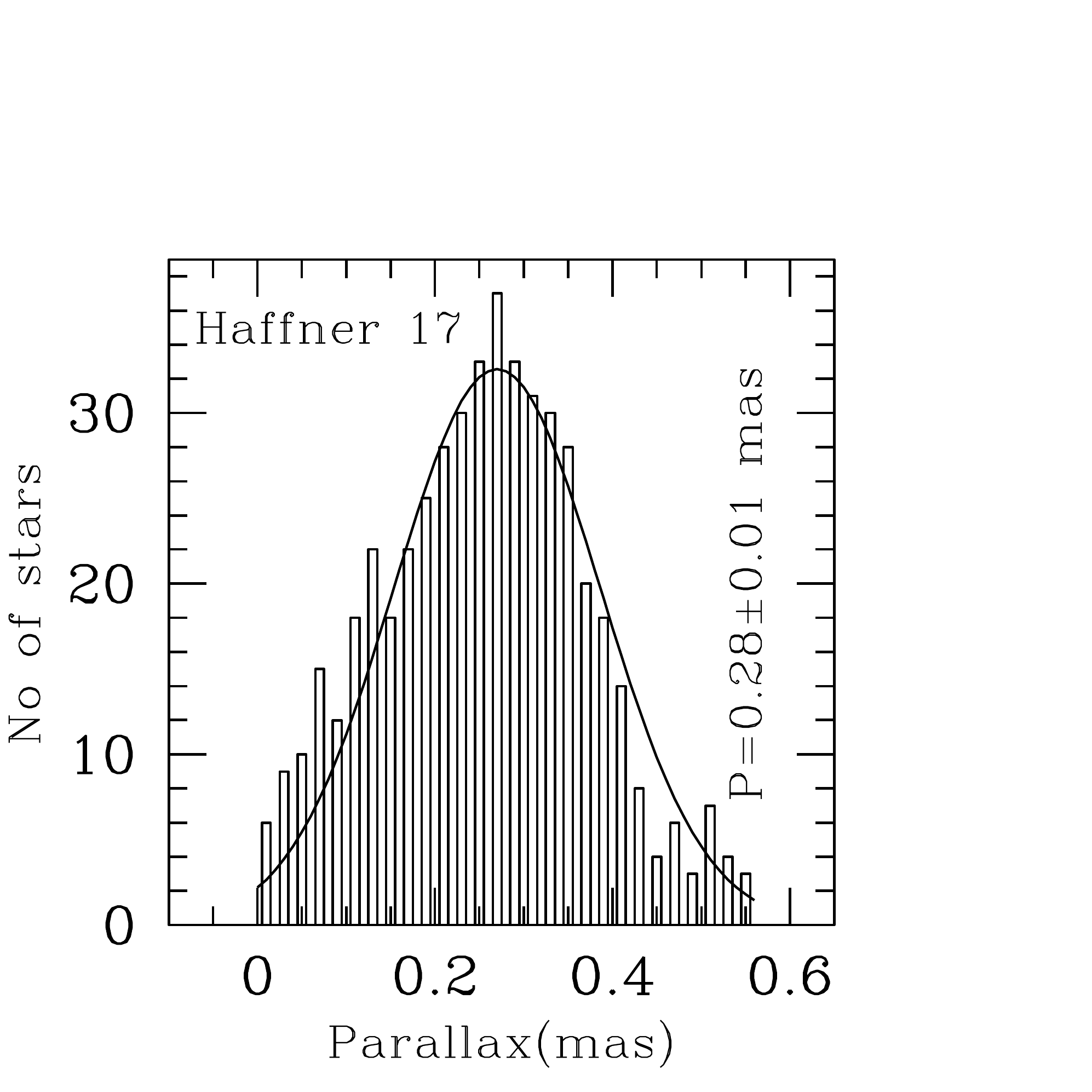}
\includegraphics[width=4.2cm, height=4.2cm]{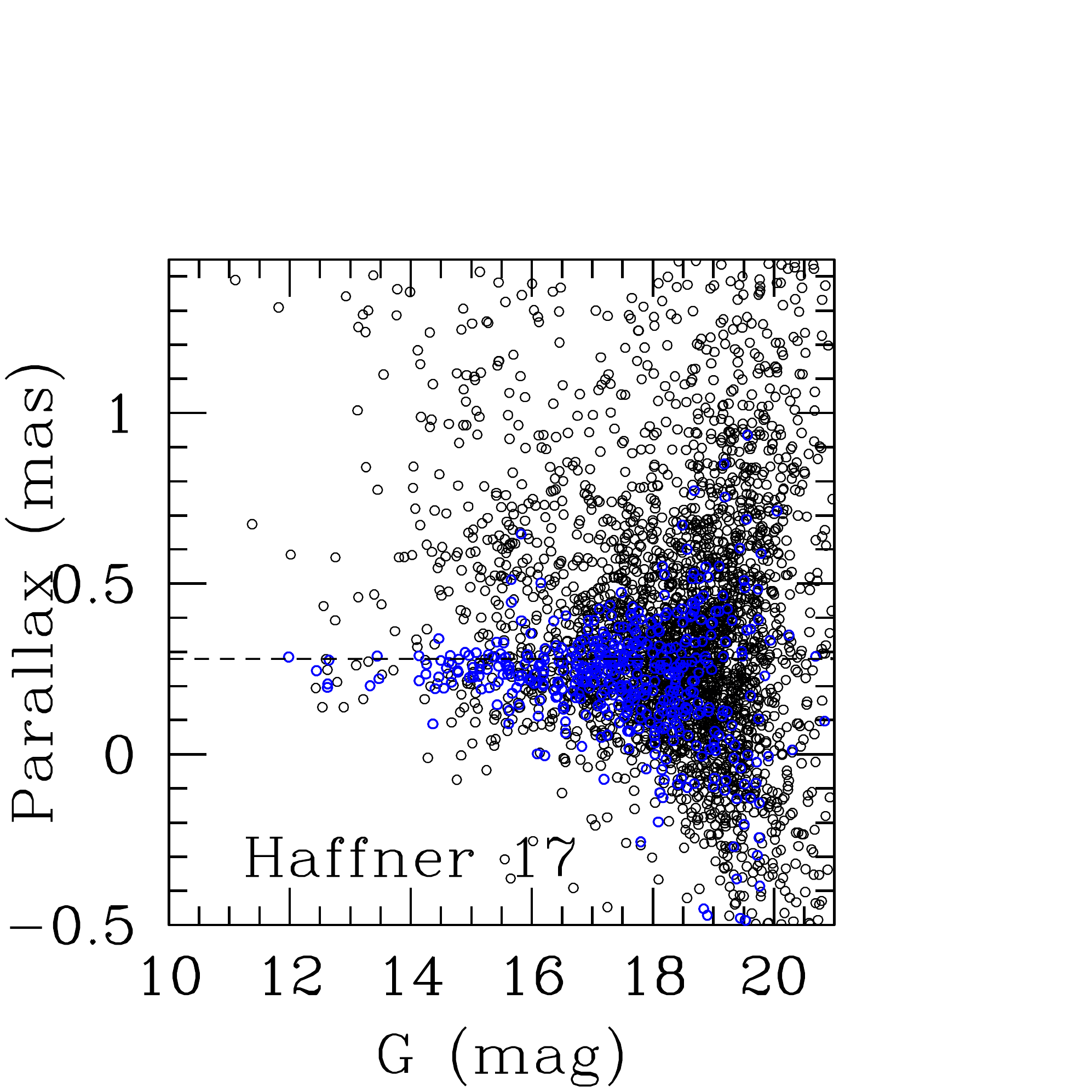}
}
\hbox{
\includegraphics[width=4.2cm, height=4.2cm]{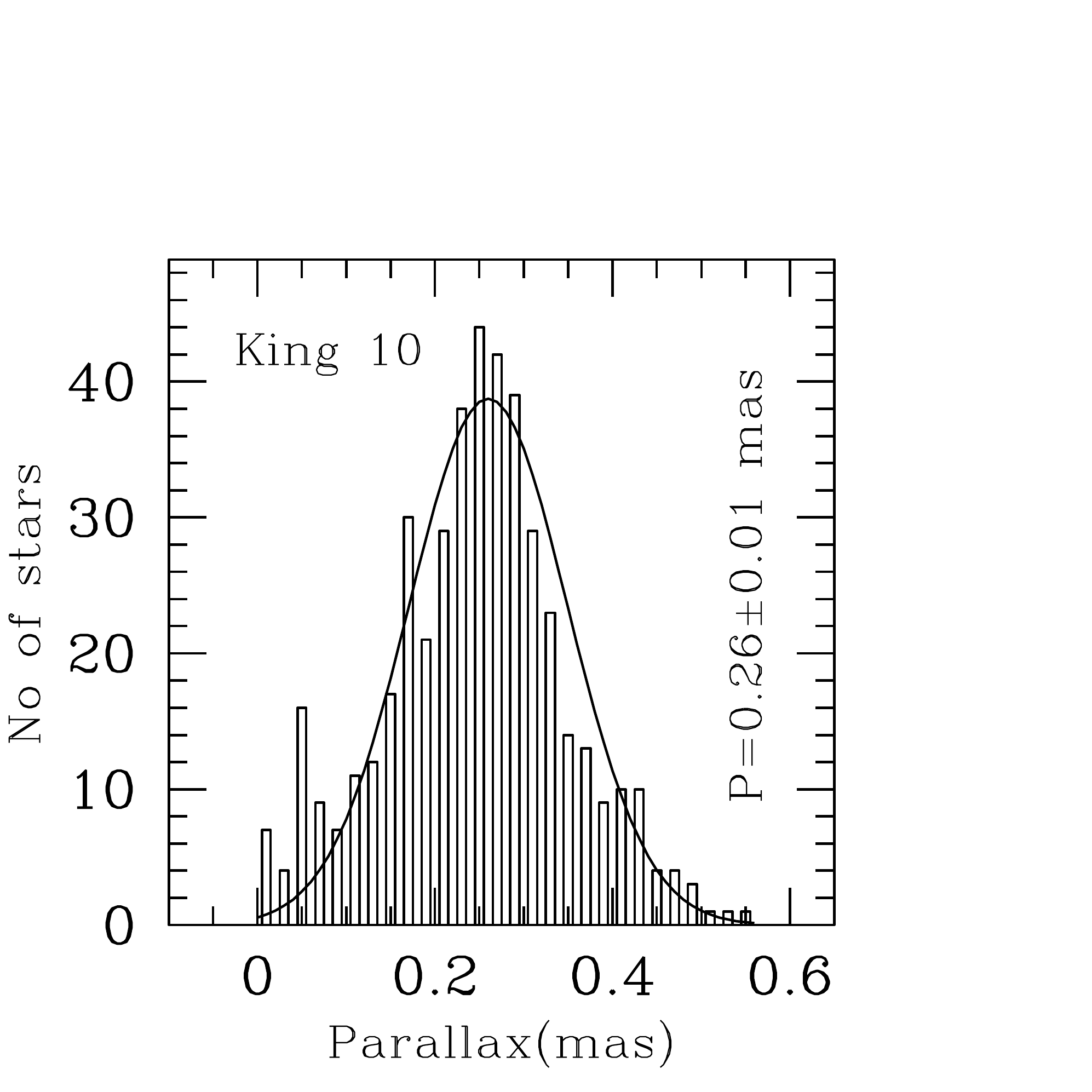}
\includegraphics[width=4.2cm, height=4.2cm]{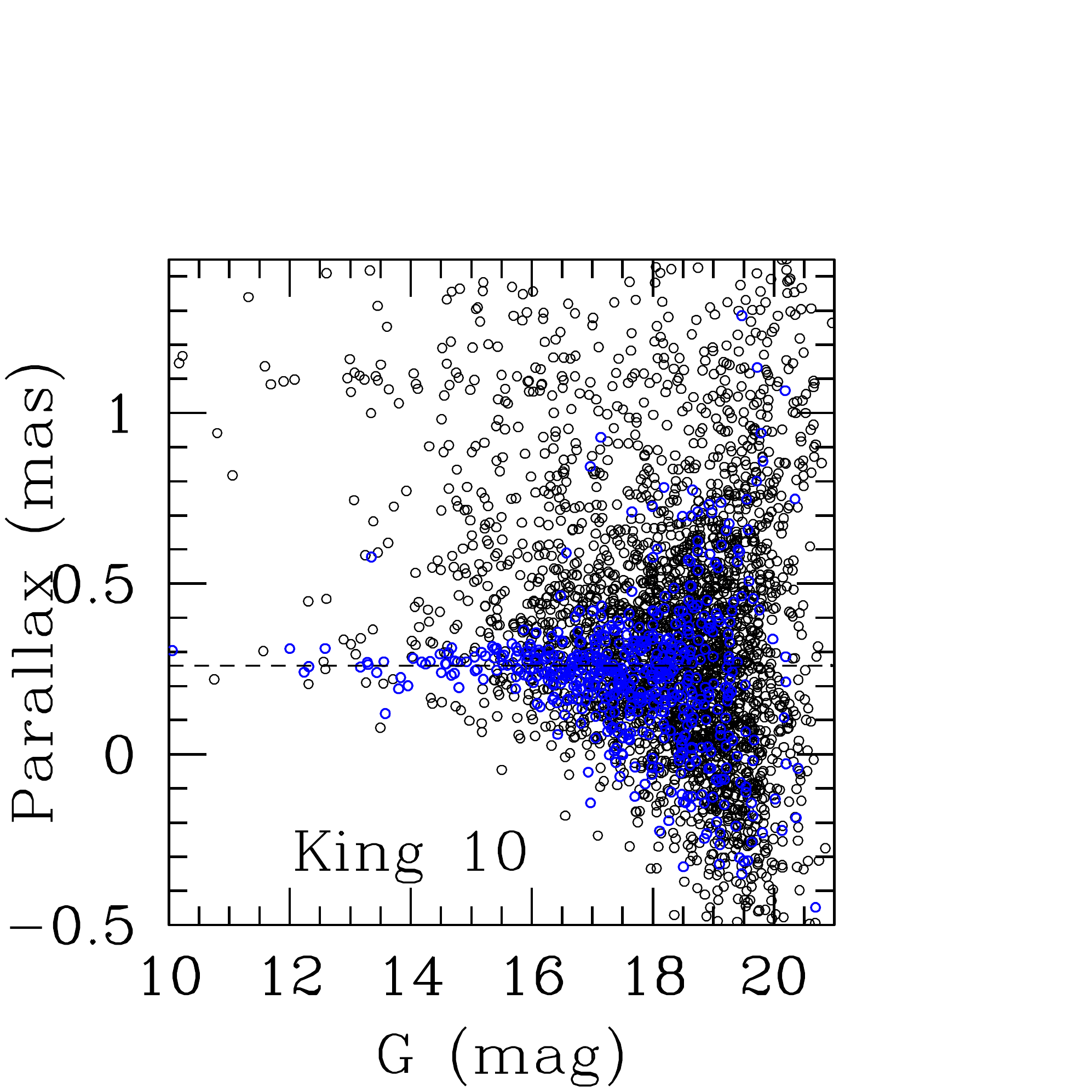}
}
\caption{(left):Histogram for parallax estimation of the clusters Czernik 14, Haffner 14, Haffner 17 and King 10 using
probable cluster members based on clusters VPDs. The Gaussian function is fitted to the central bins provides 
a mean value of parallax. (Right): G magnitude vs parallax diagrams. Black open circles are all stars while blue ones are
probable cluster members. The dashed line is the mean value of clusters parallax.}
\label{pllax}
\end{center}
\end{figure}

Gaia~DR2 has provided precise parallaxes for many stars in our sample clusters. Reliable value of distance can be 
estimated using the average value of parallax of cluster members (Lauri et al. 2018). We have corrected Gaia~DR2 parallax by
adopting zero-point offset (-0.05 mas) as given by Riess et al. (2018). The histograms of parallax for the clusters in 0.15 mas bins are shown
in Fig \ref{pllax}. In this figure, we also plotted G mag versus stars parallax. Black circles are all stars while blue denotes
the probable cluster members. The dashed line indicates the average value of parallax for each cluster. The mean parallax is 
estimated as $0.30\pm0.01$ mas, $0.21\pm0.02$ mas, $0.28\pm0.01$ mas and $0.26\pm0.01$ mas for the clusters Czernik 14, Haffner 14, Haffner 17 
and King 10 respectively and the corresponding distances are $2.9\pm0.1$ kpc, $4.8\pm0.4$ kpc, $3.6\pm0.1$ kpc and $3.8\pm0.1$ kpc.
These values of cluster distance are in good agreement with the results obtained in the previous section.

\section{Dynamics of clusters}

\subsection{Luminosity and mass function}

\begin{figure}
\begin{center}
\includegraphics[width=8.5cm, height=8.5cm]{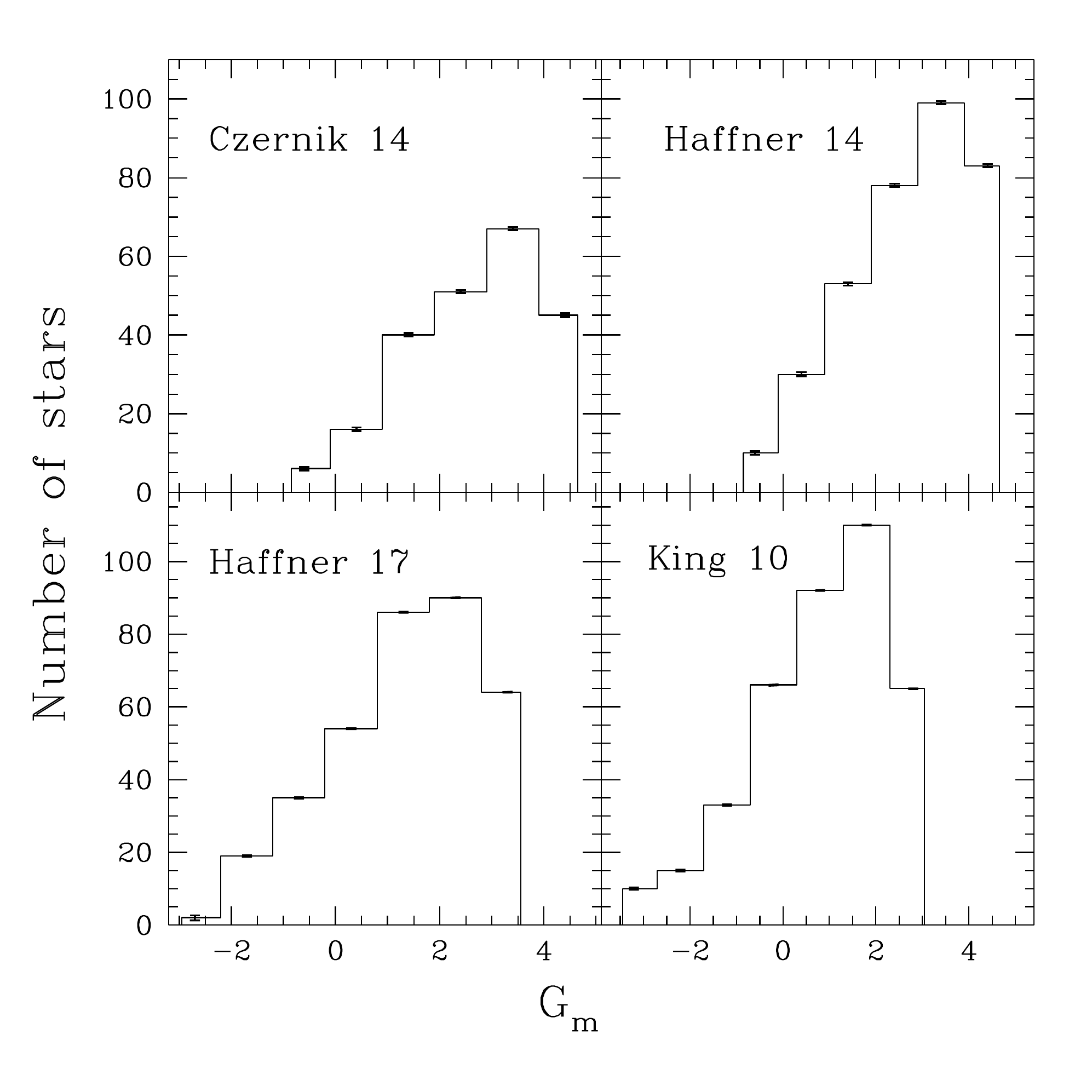}
\caption{Luminosity function of main sequence stars in the region of the clusters Czernik 14, Haffner 14, Haffner 17 and King 10.}
\label{lf}
\end{center}
\end{figure}

Generally, OCs represent hundreds of stars having similar ages and compositions but differ in masses. Luminosity 
and mass functions (LF\&MF) mainly depends on the membership information of stars.
The new data of Gaia~DR2 are used to get reliable members using
proper motions and parallaxes. The LF is the total number of member stars in different magnitude bins. To estimate the LF, we 
converted the apparent $G$ magnitudes of main sequence stars into absolute one using the distance and reddening
estimated in the present analysis. A histogram was constructed with 1.0-mag intervals and shown in Fig \ref{lf} for all the clusters 
under study. This figure shows that the LF continues to rise up to $M_{G}\sim$ 3.4, 3.4, 2.5 and 2.0 mag for clusters Czernik 14, 
Haffner 14, Haffner 17 and King 10, respectively.

\begin{figure}
\begin{center}
\includegraphics[width=8.5cm, height=8.5cm]{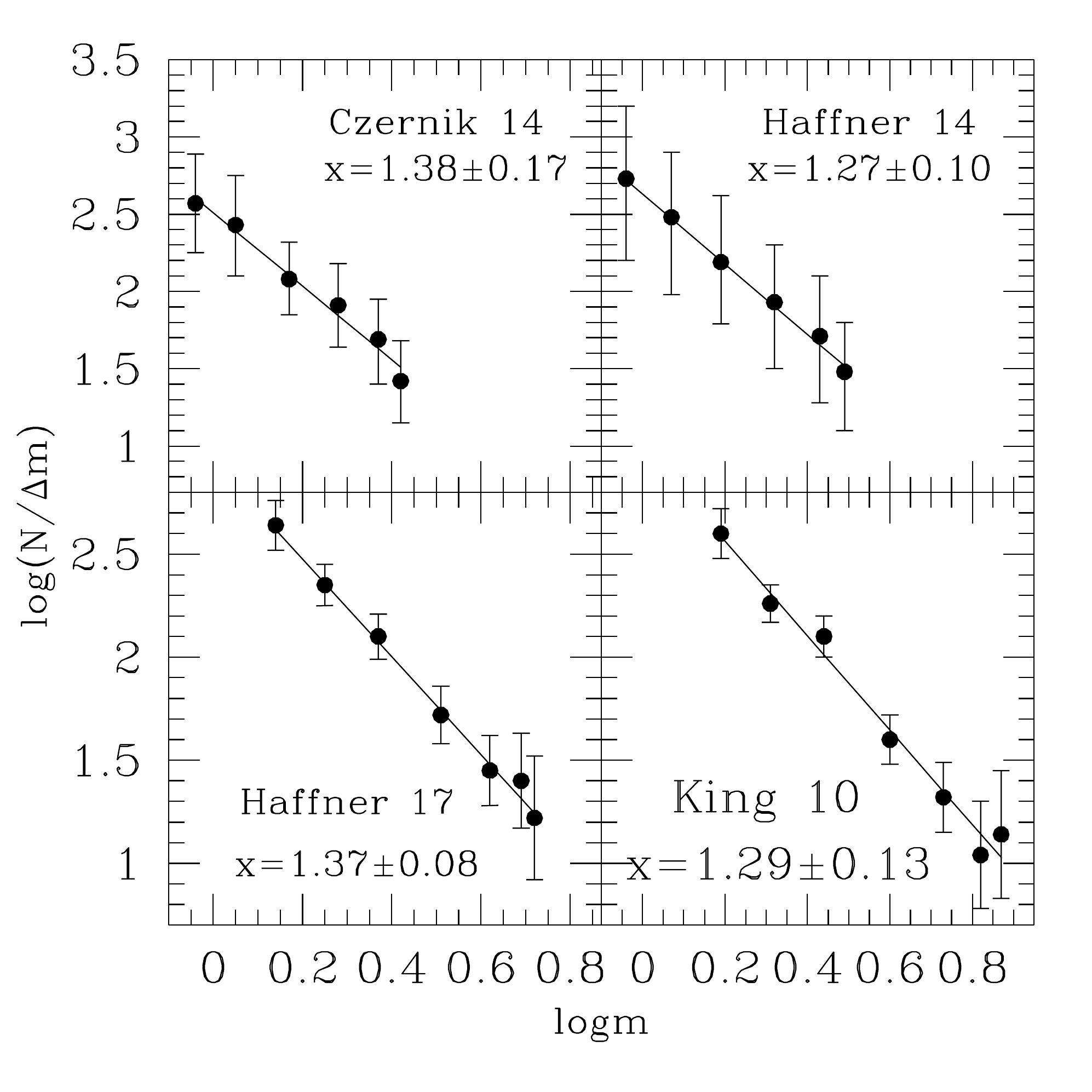}
\caption{Mass function for the clusters Czernik 14, Haffner 14, Haffner 17 and King 10 derived using probable cluster members 
and Marigo et al. (2017) isochrones. The error bars represent $\frac{1}{\sqrt{N}}$.} 
\label{mass}
\end{center}
\end{figure}

To transform the LF to MF, we used the theoretical isochrones of Marigo et al. (2017). Absolute magnitude bins are converted to 
mass bins. The resulting mass function created for the clusters is shown in Fig. \ref{mass}. It can be described by a power-law 
given by,\\

$\log\frac{dN}{dM}=-(1+x)\log(M)$+constant\\

Where $dN$ is the number of stars in a mass bin $dM$ with central mass $M$ and $x$ is mass function slope. 
To derive the mass function, we only considered stars more massive than
1 $M_{\odot} (G \sim 19^{th}$ mag). This is because Gaia data ($G$ mag) is not complete below $\sim$ 1 $M_{\odot}$.
(Arenou et al. (2018)). The initial mass
function for massive stars ($\ge$ 1 $M_{\odot}$) has been well studied and established by Salpeter (1955), where $x$=1.35.
This form of Salpeter shows that the number of stars in each mass range decreases rapidly with the increasing mass. Our derived
values of the MF slope, $x=1.38\pm0.17, 1.27\pm0.10, 1.37\pm0.08$ and $1.29\pm0.13$ for clusters Czernik 14, Haffner 14,
Haffner 17 and King 10, respectively, are in good agreement with the Salpeters initial mass function slope within error. We have an estimated
total mass of the clusters using the above mass function slope. The mass range, mass function slope and the total mass estimated in the present 
analysis are listed in the Table \ref{massf_tab}.

%We have also evaluated the total mass of clusters considering the above mass function slopes within the mass range
%1.0~-~2.8 $M_{\odot}$ for Czernik 14, 1.0~-~3.5 $M_{\odot}$ for Haffner 14, 1.2~-~5.2 $M_{\odot}$ for Haffner 17 and 
%1.4~-~7.2 $M_{\odot}$ for King 10. Total mass was estimated as $\sim$348 $M_{\odot}$, $\sim$595 $M_{\odot}$, $\sim$763 $M_{\odot}$ and
%$\sim$1088 $M_{\odot}$ for the clusters Czernik 14, Haffner 14, Haffner 17 and King 10 respectively.

\begin{table}
\centering
\caption{The main mass function parameters in clusters. 
}
\vspace{0.5cm} 
\begin{center}
\small
\begin{tabular}{lcccc}
\hline\hline
Object & Mass range & MF slope & Total mass & Mean mass \\
&      $M_{\odot}$ &  & $M_{\odot}$ & $M_{\odot}$
\\
Czernik 14  & $1.0-2.8$&$1.38\pm0.17$&$348$&$1.55$ \\
Haffner 14  & $1.0-3.5$&$1.27\pm0.10$&$595$&$1.68$ \\
Haffner 17  & $1.2-5.2$&$1.37\pm0.08$&$763$&$2.18$ \\
King 10     & $1.4-7.2$&$1.29\pm0.13$&$1088$&$2.75$ \\
\hline
\end{tabular}
\label{massf_tab}
\end{center}
\end{table}

\begin{table}
\centering
\caption{Distribution of stars in different mass ranges along with the percentage of confidence level in mass-segregation effect for the 
clusters. 
}
\vspace{0.5cm} 
\begin{center}
\small
\begin{tabular}{lcc}
\hline\hline
Object & Mass ranges & Confidence level\\
&      $M_{\odot}$ & $\%$
\\
Czernik 14  & $3.8-2.2, 2.2-1.4, 1.4-1.0$& 91 \\
Haffner 14  & $3.3-2.4, 2.4-1.3, 1.3-0.8$& 88 \\
Haffner 17  & $5.2-3.7, 3.7-1.9, 1.9-1.2$& 75 \\
King 10     & $7.2-4.7, 4.7-2.3, 2.3-1.4$& 77 \\
\hline
\end{tabular}
\label{masss_tab}
\end{center}
\end{table}
\subsection{Mass-segregation }

There are plenty of works related to mass-segregation of clusters in the literature (e.g. Sagar et al. 1988,
Hillenbrand \& Hartmann 1998, Lada \& Lada 1991, Campbell et al. 1992, Pandey et al. 1992, Brandl et al. 1996,
Meylan 2000, Bisht et al. 2019). Under this process, massive stars tend to move towards the core region and faint
stars generally move towards the halo region of the cluster. For mass segregation study, we considered probable
cluster members as selected in section 3. Cluster members are divided into three mass ranges as shown in
Table \ref{masss_tab} for the clusters under study.
%3.8$\le\frac{M}{M_{\odot}}\le$~2.2,
%2.2$\le\frac{M}{M_{\odot}}\le$~1.4 and 1.4$\le\frac{M}{M_{\odot}}\le$~1.0 for Czernik 14, 3.3$\le\frac{M}{M_{\odot}}\le$~2.4,
%2.4$\le\frac{M}{M_{\odot}}\le$~1.3 and 1.3$\le\frac{M}{M_{\odot}}\le$~0.8 for Haffner 14, 5.2$\le\frac{M}{M_{\odot}}\le$~3.7,
%3.7$\le\frac{M}{M_{\odot}}\le$~1.9 and 1.9$\le\frac{M}{M_{\odot}}\le$~1.2 for Haffner 17 and 7.2$\le\frac{M}{M_{\odot}}\le$~4.7,
%4.7$\le\frac{M}{M_{\odot}}\le$~2.3 and 2.3$\le\frac{M}{M_{\odot}}\le$~1.4 for King 10. 

\begin{figure}
\begin{center}
\includegraphics[width=8.5cm, height=8.5cm]{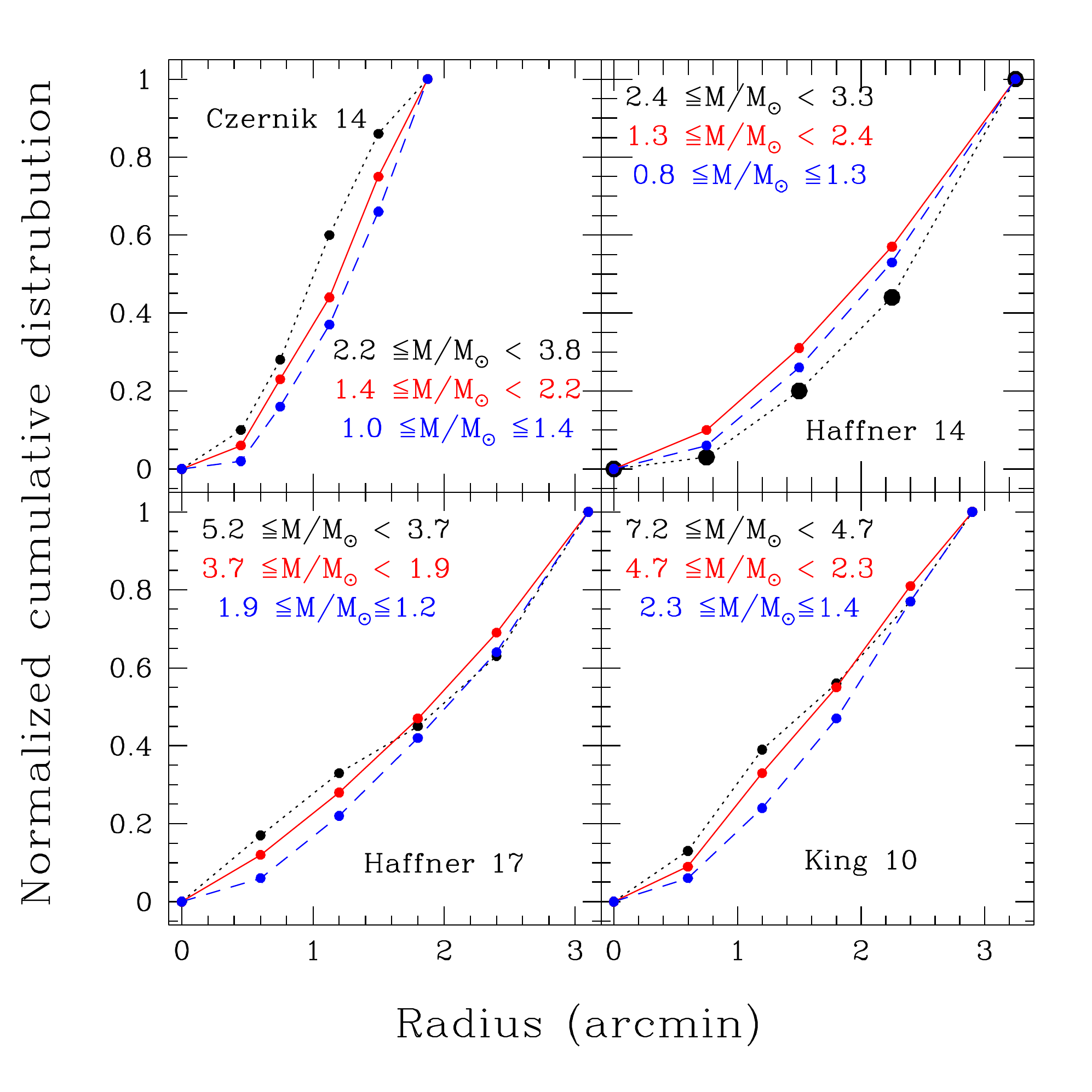}
\caption{The cumulative radial distribution for Czernik 14, Haffner 14, Haffner 17 and King 10 in various mass-ranges.}
\label{mass_seg}
\end{center}
\end{figure}

We present cumulative radial stellar distribution of main sequence stars for three different mass ranges as shown in 
Fig \ref{mass_seg}. This diagram indicates that the cluster stars exhibit a mass-segregation in the sense that
bright stars appear to be more centrally concentrated than the low mass members. We also performed the Kolmogrov-Smirnov
test $(K-S)$ to check the statistical significance of mass segregation. Using this test, we found the confidence level
of mass-segregation as 91 $\%$, 88 $\%$, 75 $\%$  and 77 $\%$ for the clusters Czernik 14, Haffner 14, Haffner 17
and King 10 respectively.

The possible cause of the mass-segregation is not fixed and it changes from one cluster to another. The possible reason
for mass-segregation may be dynamical evolution or imprint of star formation or both in a particular cluster. Over the lifetime
of clusters, encounters between its member stars gradually lead to an increased degree of energy equipartition throughout the clusters.
The most important result of this process is that the bright stars gradually sink towards the cluster center and transfer their kinetic
energy to the more numerous lower-mass stars, thus leading to mass segregation. To understand the reason for mass-segregation
in the clusters, we calculated relaxation time ($T_{E}$). $T_{E}$ is defined as the time in which the 
stellar velocity distribution becomes Maxwellian and expressed by the following formula:\\

$T_{E}=\frac{8.9\times10^5\sqrt{N}\times{R_{h}}^{3/2}}{\sqrt{m}\times log(0.4N)}$\\

\begin{figure}
\begin{center}
\hbox{
\includegraphics[width=4.2cm, height=4.2cm]{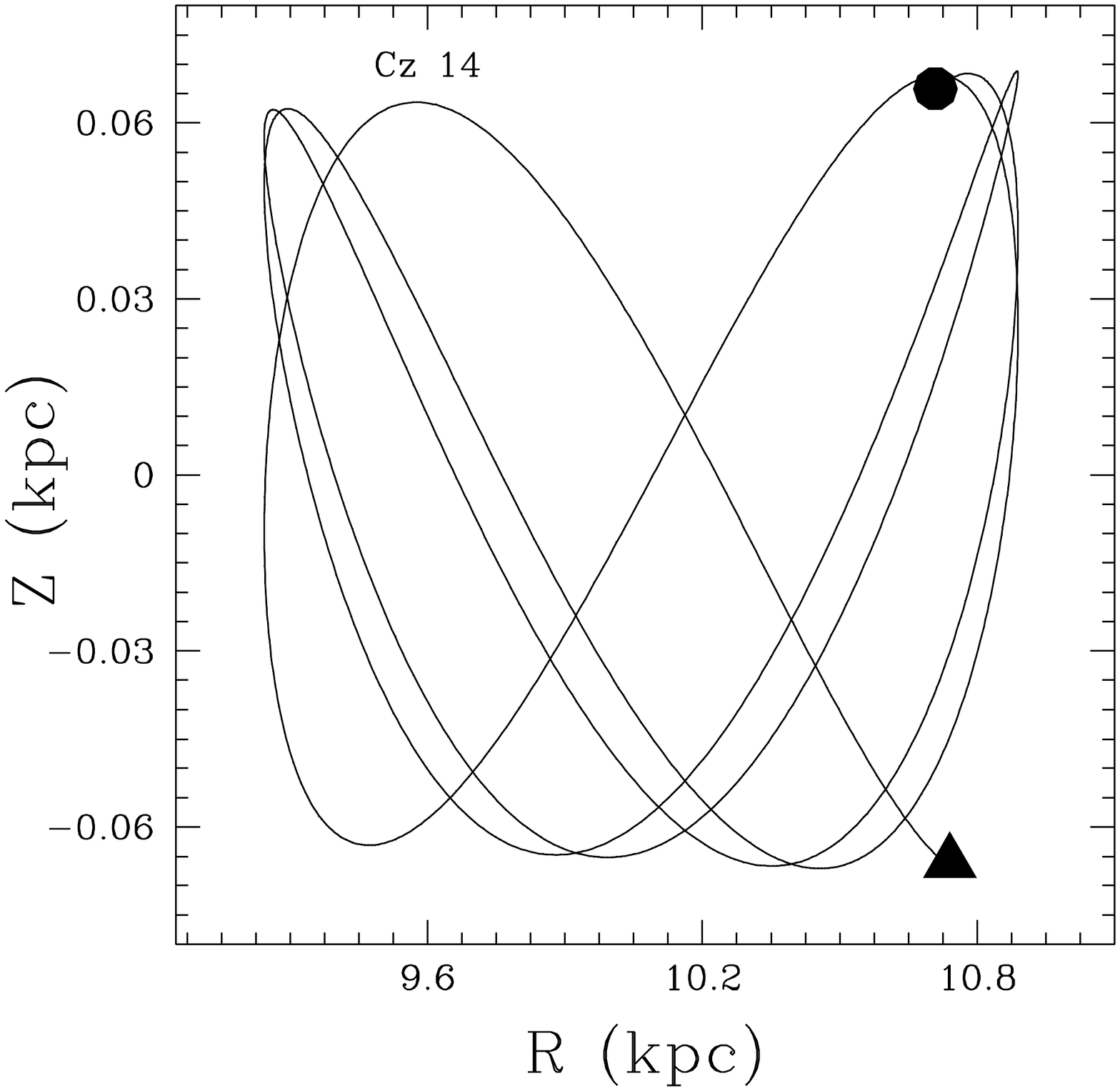}
\includegraphics[width=4.2cm, height=4.2cm]{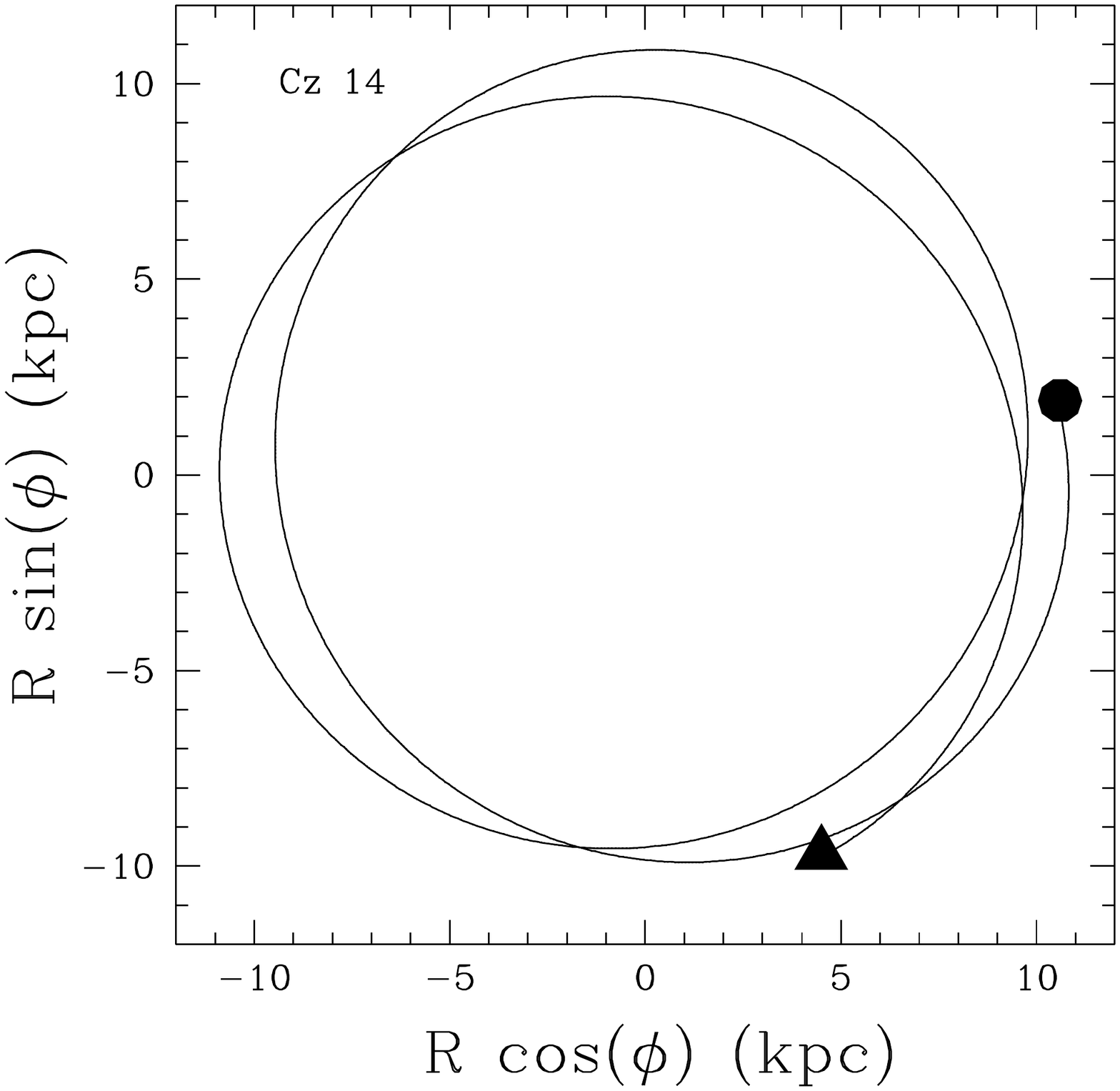}
}
\hbox{
\includegraphics[width=4.2cm, height=4.2cm]{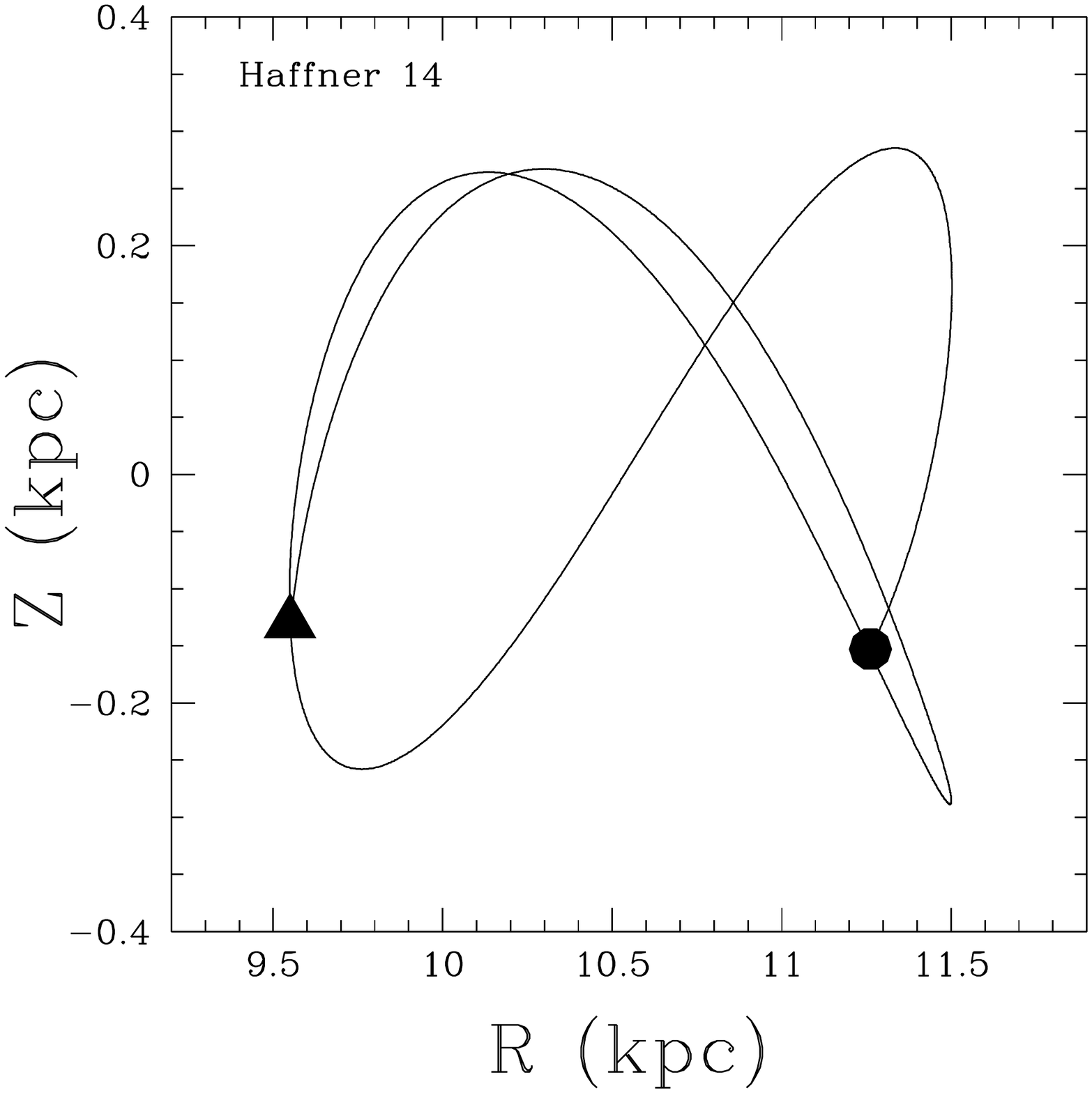}
\includegraphics[width=4.2cm, height=4.2cm]{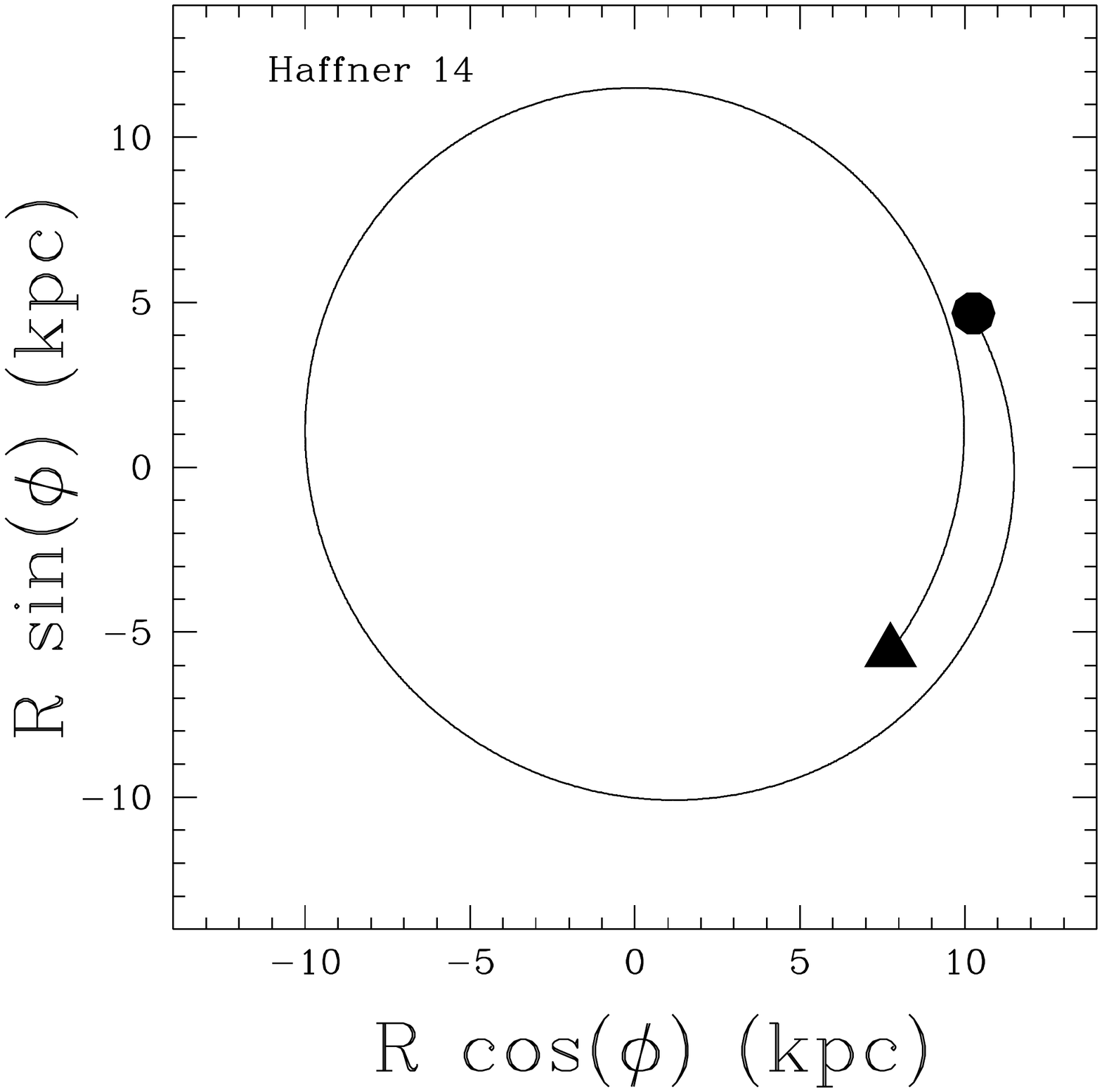}
}
\hbox{
\includegraphics[width=4.2cm, height=4.2cm]{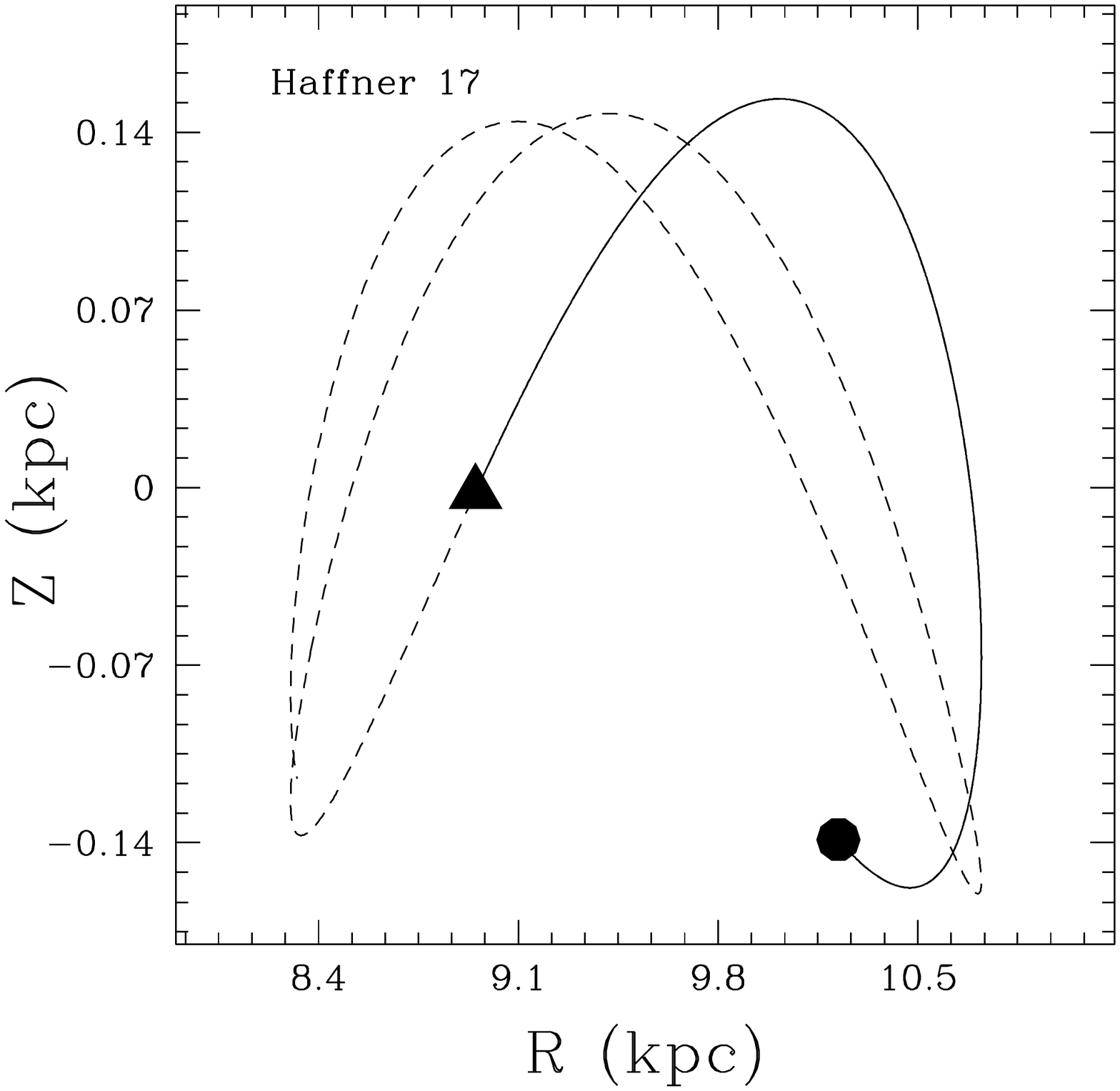}
\includegraphics[width=4.2cm, height=4.2cm]{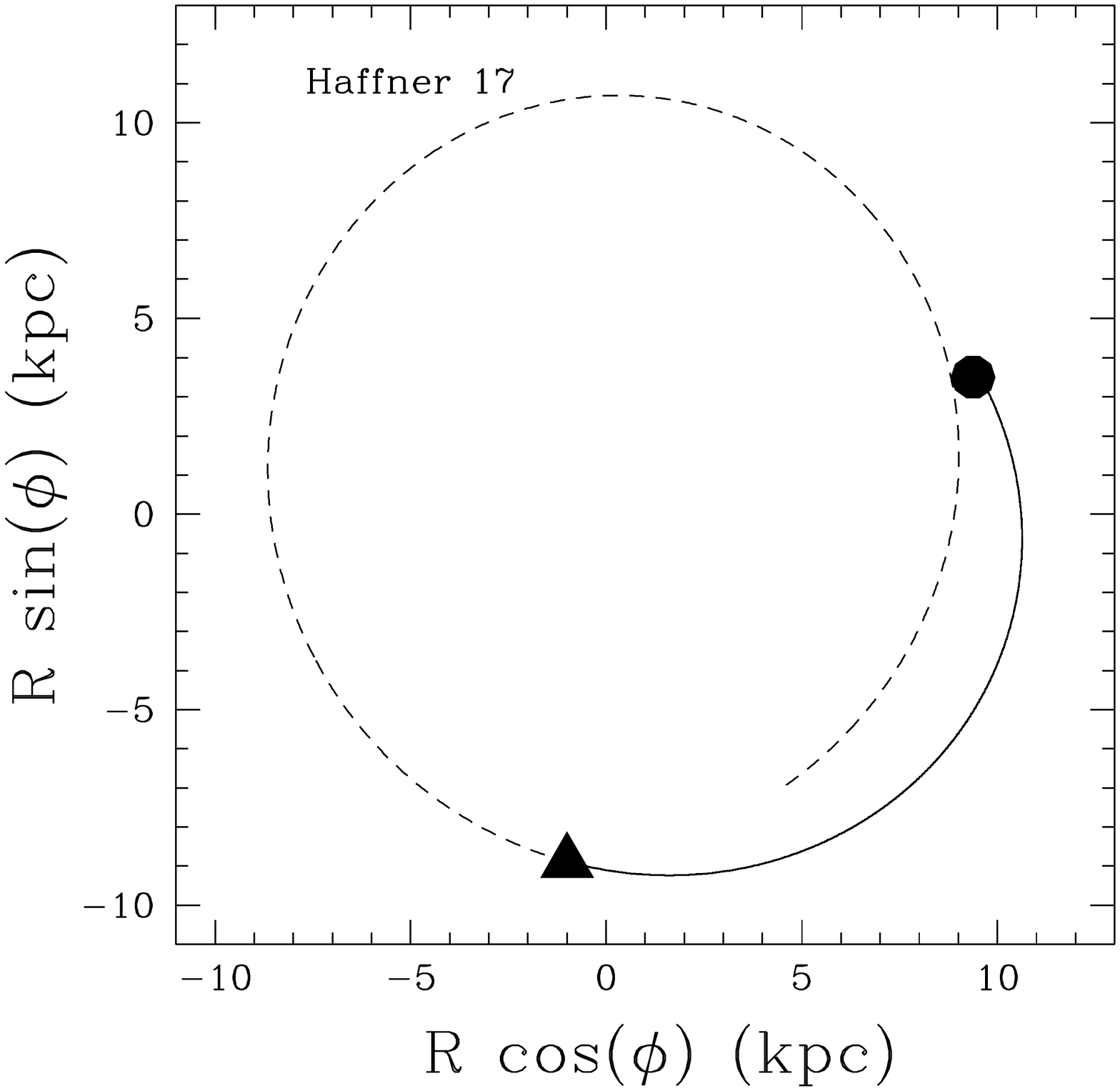}
}
\hbox{
\includegraphics[width=4.2cm, height=4.2cm]{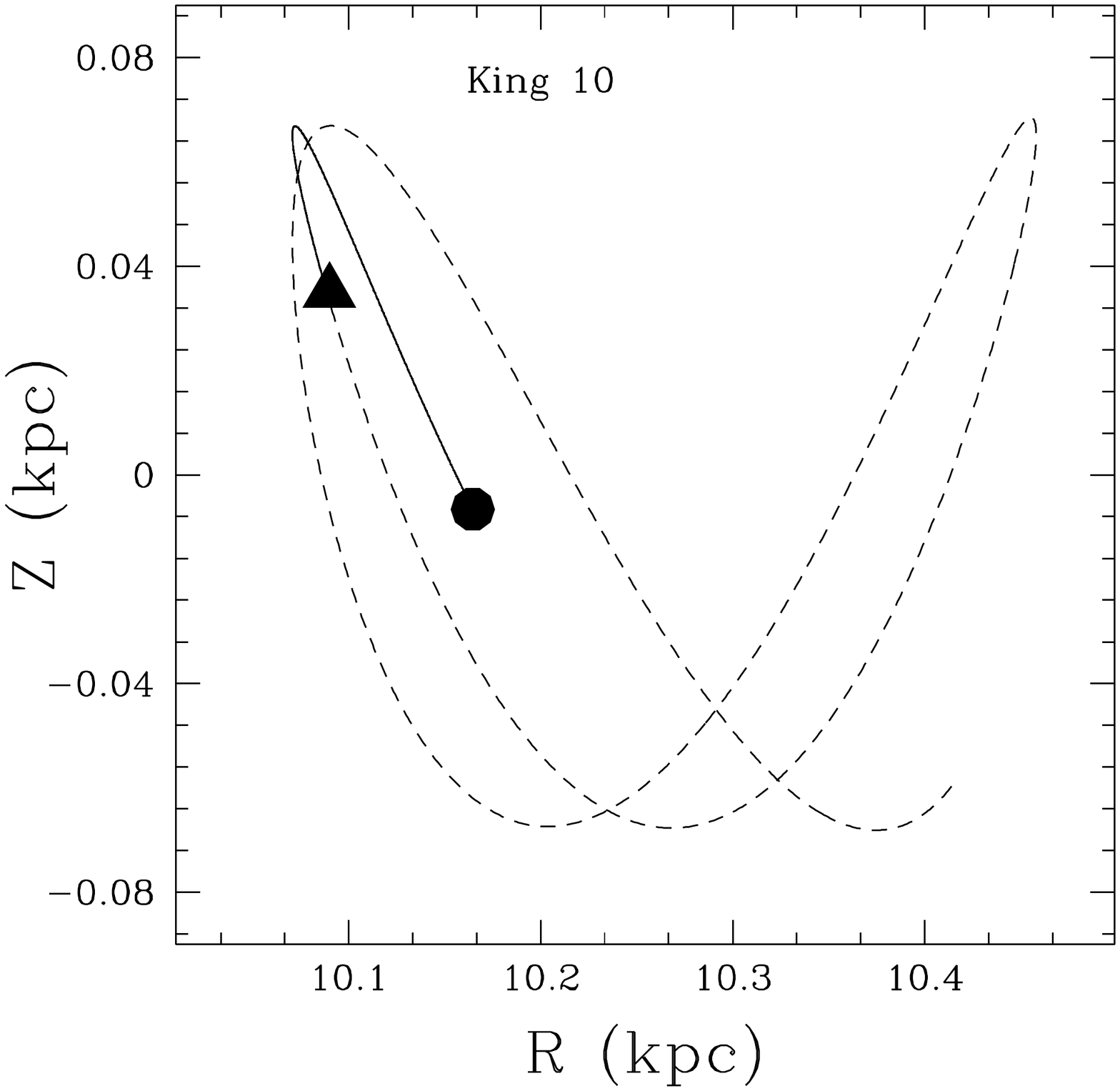}
\includegraphics[width=4.2cm, height=4.2cm]{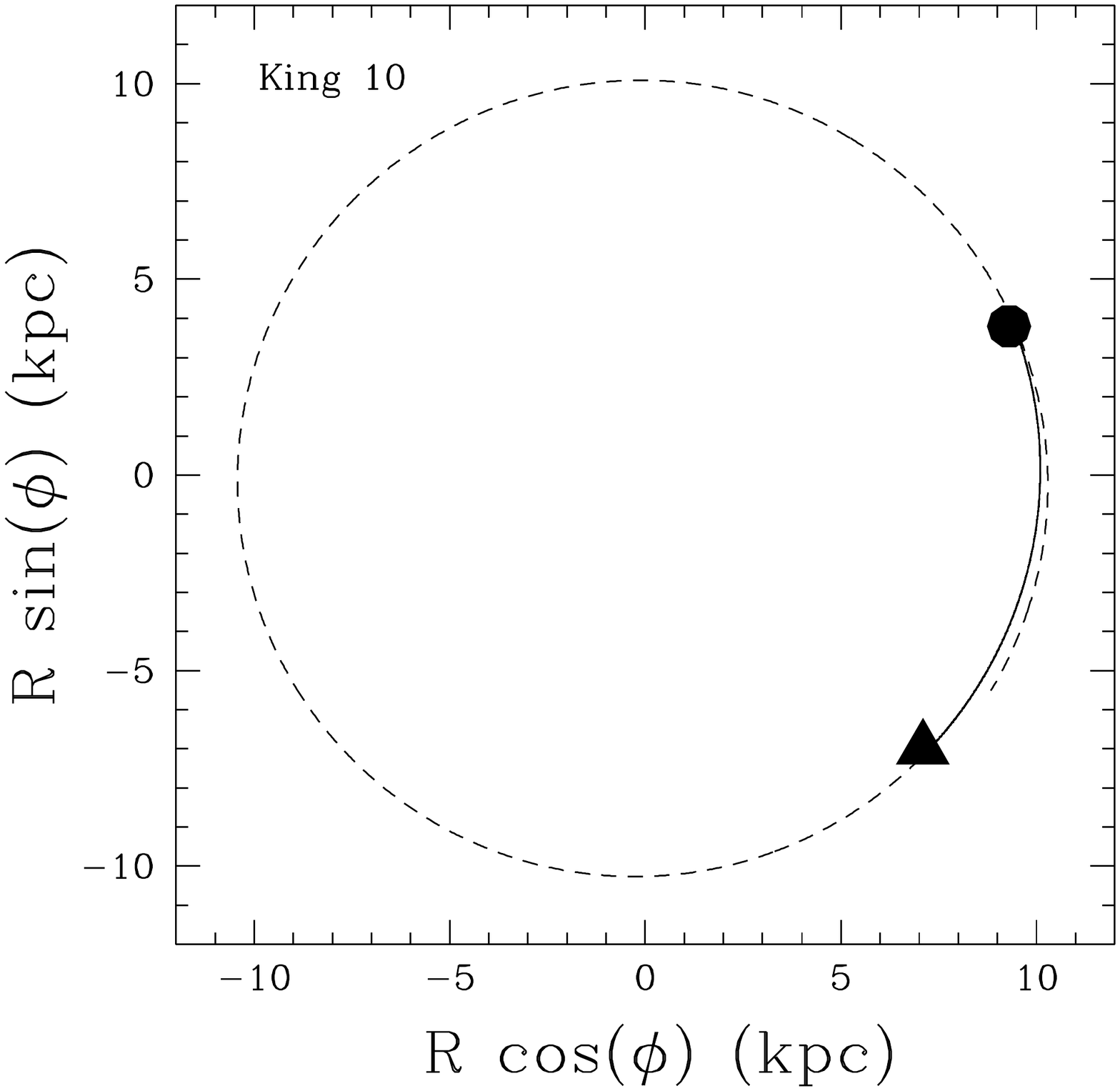}
}
\caption{Galactic orbits of the clusters Czernik 14, Haffner 14, Haffner 17 and King 10 estimated with the Galactic potential 
model described in text in the time interval of the age of each cluster. For Haffner 17 and King 10, the dotted line represents cluster
orbits for a time interval of 300 Myr. The left panel shows a side view and the right panel shows a top
view of the cluster's orbit. The filled triangle and filled circle denotes birth and present day position of clusters in the 
Galaxy.}
\label{orbit}
\end{center}
\end{figure}

where $N$ is probable cluster members, $R_{h}$ is the cluster half mass radius expressed in parsec and $m$
is the average mass of the cluster members (Spitzer \& Hart 1971) in the solar unit. We identified 225, 353, 350 and 395
stars as probable cluster members for Czernik 14, Haffner 14, Haffner 17 and King 10 respectively. 
The value of $<m>$ is estimated as 1.55 $M_{\odot}$, 1.68 $M_{\odot}$, 2.18 $M_{\odot}$ and 2.75 $M_{\odot}$
for these clusters, respectively. The value of $R_{h}$ is assumed to be equal to half of the cluster extent.
Using the above formula of dynamical relaxation time, we estimated the value of $T_{E}$ as 9.8, 20.4, 30,
27.2 Myr for Czernik 14, Haffner 14, Haffner 17 and King 10, respectively.

By using the cluster's age and dynamical relaxation time, we estimated the dynamical evolution parameter
($\tau$) of the clusters using the following relation:\\

$\tau=\frac{age}{T_{E}}$

Our estimated values of relaxation time are found to be lower than the age of the clusters under study. This gives $\tau\ge1.0$. Hence,
we conclude that Czernik 14, Haffner 14,  Haffner 17 and King 10 are dynamically relaxed open clusters.

%%__________________________________________________________________________
%%__________________________________________________________________________

\section{Orbits of the clusters}

\subsection{Galactic potential model} \label{inic}

Galactic orbits are helpful to study the dynamical properties of 
stars, clusters and galaxies. We derived the Galactic orbits of 
the clusters under study using the Galactic potential models.
We adopted the approach given by Allen \& Santillan (1991) for Galactic potentials. According to their model, the mass of Galaxy is described
by three components: spherical central bulge, massive spherical halo and disc.
Recently Bajkova \& Bobylev (2016) and Bobylev et. al (2017) refined the parameters of Galactic potential models with the help of
new observational data for a distance R $\sim$ 0 to 200 kpc.
These potentials
are given as    \\

$ \Phi_{b}(r,z) = -\frac{M_{b}}{\sqrt{r^{2} + b_{b}^{2}}} $   \\

$ \Phi_{d}(r,z) = - \frac{M_{d}}{\sqrt{r^{2} + (a_{d} + \sqrt{z^{2} + b_{d}^{2}})^{2}}}  $ \\

$ \Phi_{h}(r,z) = - \frac{M_{h}}{a_{h}} ln(\frac{\sqrt{r^{2} + a_{h}^{2}} + a_{h}}{r}) $    \\

Where  $ \Phi_{b} $ , $ \Phi_{d} $ and $ \Phi_{h} $ are the potentials of central bulge, disc and halo of Galaxy respectively. $r$ and $z$ are
the distances of objects from Galactic center and Galactic disc respectively. The halo region potential is given by Wilkinson \& Evans (1999).
All three potentials are axis-symmetrical, time independent and analytical. Also, their spatial derivatives are continuous everywhere.

\subsection{Orbits Calculation}

The input parameters, such as central coordinates ($\alpha$ and $\delta$), mean
proper motions ($\mu_{\alpha}cos\delta$, $\mu_{\delta}$), parallax angles,
clusters age and heliocentric distances ($d_{\odot}$)
for the clusters under study have been taken from Table \ref{para}.
Radial velocity data has been used from Gaia DR2 catalogue for all the clusters. Average radial velocities of the clusters are calculated by
taking the mean for all probable cluster members. After five iterations, the average radial velocity of the clusters are found as
$-55.12\pm1.04$, $71.21\pm 1.03$, $49.97\pm 2.09$ and $-44.35\pm 1.83$ km/s for the clusters Czernik 14, Haffner 14,
Haffner 17 and King 10, respectively. Our estimated vlaue of radial velocities are in good agreement with Soubiran et al. (2018)
for clusters Czernik 14 and Haffner 14. For clusters Haffner 17 and King 10, our estimated values are
slightly higher than Soubiran et al. (2018). The Present estimate of radial velocity for Haffner 17 and King 10 is based on 24 and 
22 stars while in Soubiran et al. (2018) it is on 4 and 1 star respectively.

Since clusters are orbiting around the Galactic center, we cannot use position and velocity vectors in the equatorial
system. Therefore, we transformed them, into the Galacto-centric cylindrical coordinate system using the transformation
matrix given in Johnson \& Soderblom (1987). In this system, $(r,\phi,z)$ indicates the position of an object in Galaxy,
where $r$ is the distance from Galactic center, $\phi$ is the angle relative to Sun's position in the Galactic plane and $z$
is the distance from Galactic plane.

The right-hand coordinate system is adopted to transform equatorial velocity components into Galactic-space velocity components ($U,V,W$),
where $U$, $V$ and $W$ are radial, tangential and vertical velocities respectively. In this system the x-axis is taken positive towards
Galactic-center, y-axis is along the direction of Galactic rotation and z-axis is towards Galactic north pole. The Galactic center is
taken at ($17^{h}45^{m}32^{s}.224, -28^{\circ}56^{\prime}10^{\prime\prime}$) and North-Galactic pole is at ($12^{h}51^{m}26^{s}.282,
27^{\circ}7^{\prime}42^{\prime\prime}.01$) (Reid \& Brunthaler, 2004). To apply a correction for Standard Solar Motion and Motion of the
Local Standard of Rest (LSR), we used position coordinates of Sun as ($8.3,0,0.02$) kpc and its space-velocity components as
($11.1, 12.24, 7.25$) km/s (Schonrich et al. 2010). Transformed parameters in Galacto-centric coordinate system are listed in
Table \ref{inp}.

In orbit determination, we estimated the radial and vertical components 
of gravitational force, by differentiating total gravitational potentials with respect to $r$ and $z$. The second order derivatives 
of gravitational force describe the motion of the clusters. For orbit determination, the second order derivatives are integrated backwards
in time, which is equal to the age of clusters. Since potentials used are axis-symmetric, energy and $z$ component of angular momentum 
are conserved throughout the orbits.

\begin{table*}
\centering
\caption{Various fundamental parameters of the clusters Czernik 14, Haffner 14, Haffner 17 and King 10.}
\vspace{0.5cm}
\begin{center}
\small
\begin{tabular}{ccccc}
\hline\hline
Parameter &  Czernik 14 &  Haffner 14 & Haffner 17 & King 10 \\
\hline\hline
\\
RA(deg)                                  & $49.25\pm0.008$  & $116.20\pm0.004$ & $117.90\pm0.008$   & $343.72\pm0.005$    \\
DEC(deg)                                 & $58.59\pm0.009$  & $-28.37\pm0.003$ & $-31.81\pm0.006$   & $59.16\pm0.007$     \\
Radius(arcmin)                           & 3.5              & 3.7              & 6.2               & 5.7                  \\
Radius(parsec)                           & 2.9              & 4.5              & 6.5               & 6.3                  \\
$\mu_{\alpha}cos\delta$($mas~ yr^{-1}$)   & $-0.42\pm0.02$   & $-1.82\pm0.009$  & $-1.17\pm0.007$    & $-2.75\pm0.008$        \\
$\mu_{\delta}$($mas~ yr^{-1}$)            &  $-0.38\pm0.01$  &  $ 1.73\pm0.008$ &  $1.88\pm0.006$   &  $-2.04\pm0.006$       \\
Radial Velocity (Km/sec)         & $-55.12\pm1.04$  & $71.21\pm1.03$   & $49.97\pm2.09$   & $-44.35\pm1.83$        \\
Age(Myr)                                 &  $570\pm60$      &  $320\pm35$      &  $90\pm10$       &  $45\pm5$             \\
Metal abundance                          &  $0.019$          &  $0.019$          &  $0.019$           &  $0.019$                \\
E(J-H) (mag)                             &  $0.30\pm0.03$   &  $0.12\pm0.04$   &  $0.40\pm0.05$    &  $0.34\pm0.04$        \\
E(J-K) (mag)                             &  $0.50\pm0.05$   &  $0.20\pm0.07$   &  $0.61\pm0.07$    &  $0.55\pm0.07$        \\
E(B-V) (mag)                             &  $0.96\pm0.05$   &  $0.38\pm0.05$   &  $1.29\pm0.05$    &  $1.09\pm0.05$         \\
$R_{V}$                                  &  3.1             &  3.1             &  3.1              &  3.1                   \\
Distance modulus (mag)                   &  $15.10\pm0.20$  &  $13.80\pm0.10$  &  $15.10\pm0.20$   &  $15.10\pm0.20$        \\
Distance (Kpc)                           &  $2.90\pm0.20$   &  $4.80\pm0.20$   &  $3.6\pm0.10$    &  $3.8\pm0.10$         \\
$X$(Kpc)                         &  2.32            & -3.05            &  -2.45             &  3.76                  \\
$Y$(Kpc)                         &  10.24           &  12.20           &  11.13            &  9.10                 \\
$Z$(Kpc)                         &  0.04            &  -0.154          &  -0.14             &  -0.02                  \\
Total Luminosity(mag)                    &   $\sim 3.4$     &   $\sim 3.4$     &   $\sim 2.5$      &   $\sim 2.0$           \\
Cluster members                          &   225            &   353            &   350             &   395                  \\
IMF slope                                &   $1.38\pm0.17$  &   $1.27\pm0.10$  &   $1.37\pm0.08$   &   $1.29\pm0.13$        \\
Total mass ($M_{\odot}$                  &   $\sim 348 $    &   $\sim 595$     &   $\sim 763 $     &   $\sim 1088 $          \\
Average mass($M_{\odot}$)                &   $1.55$         & $1.68$           &   $2.18$          &   $2.75$               \\
Relaxation time(Myr)                     &   9.8            &   20.4           &   30.3             &   27.2                  \\
Dynamical evolution parameter ($\tau$)   &   $\sim58$       &   $\sim16$       &   $\sim3$        &   $\sim1.6$             \\
\hline
\end{tabular}
\label{para}
\end{center}
\end{table*}
\begin{table*}
   \centering
   \caption{Position and velocity components in the Galactocentric coordinate system. Here $R$ is the galactocentric
            distance, $Z$ is the vertical distance from the Galactic disc, $U$ $V$ $W$ are the radial tangential and the vertical 
            components of velocity respectively and $\phi$ is the position angle relative to the sun's direction.
}
   \begin{tabular}{ccccccccc}
   \hline\hline
   Cluster   & $R$ &  $Z$ &  $U$  & $V$  & $W$ & $\phi$   \\
   & (kpc) & (kpc) & (km/s) &  (km/s) & (km/s) & (radian)    \\
  \hline
   Czernik 14 & 10.71 & 0.07 & $16.16 \pm 0.81$  & $-229.41 \pm 0.67$ &  $01.05 \pm 0.29$ & 0.17    \\
   Haffner 14 & 11.26 & -0.15 & $19.16 \pm 0.50$ & $-226.74 \pm 1.87$ & $11.49 \pm 0.09$ & 0.39      \\
   Haffner 17 & 10.22 & -0.14 & $32.57 \pm 0.38$ & $-226.07 \pm 1.96$ & $-04.27 \pm 0.09$ & 0.33      \\
   King 10    & 10.16 & -0.01 & $-05.44 \pm 1.04$ & $-247.40 \pm 1.79$ & $04.04 \pm 0.10$ & 0.36      \\
\hline
  \end{tabular}
  \label{inp}
  \end{table*}

Fig. \ref{orbit} show orbits of the clusters Czernik 14, Haffner 14, Haffner 17 and King 10. In left panels, the motion of clusters
is described in terms of distance from Galactic center and Galactic plane, which shows two dimensional side view of the orbits. In 
right panels, cluster motion is described in terms of $x$ and $y$ components of Galactocentric distance, which shows a top view of orbits.

We also calculated the orbital parameters for the clusters and are listed in Table \ref{orpara}. Here $e$ is eccentricity, $R_{a}$ is
apogalactic distance, $R_{p}$ is the perigalactic distance, $Z_{max}$ is the maximum distance travelled by cluster from Galactic disc,
$E$ is the average energy of orbits, $J_{z}$ is $z$ component of angular momentum and $T$ is the time period of the clusters in the orbits.

\begin{table*}
  \centering
   \caption{Orbital parameters for the clusters obtained using the Galactic potential model.
   }
   \begin{tabular}{ccccccccc}
   \hline\hline
   Cluster  & $e$  & $R_{a}$  & $R_{p}$ & $Z_{max}$ &  $E$ & $J_{z}$ & $T$   \\
           &    & (kpc) & (kpc) & (kpc) & $(100 km/s)^{2}$ & (100 kpc km/s) & (Myr) \\ 
   \hline\hline  
   Czernik 14 &  0.00  & 10.83  & 10.87  & 0.07 & -09.84 & -24.57  & 291 \\
   Haffner 14  & 0.00 & 11.50  & 11.50  & 0.29 & -09.58 & -25.54  & 310  \\
   Haffner 17  & 0.00 & 10.63  & 10.70  & 0.15 & -10.15 & -23.11 & 282  \\ 
   King 10     & 0.01 & 10.29  & 10.09  & 0.07 & -09.73 & -25.15 & 256  \\
 \hline 
  \end{tabular}
  \label{orpara}
  \end{table*}

The orbits of the clusters under study follow a boxy pattern and eccentricities for all the clusters are zero. Hence they
trace a circular path around the Galactic center. From these orbits, we have determined the birth and present day position of
clusters in the Galaxy which are represented by the filled circle and filled triangle respectively in Fig. \ref{orbit}.

Czernik 14 and Haffner 14 are intermediate age open star clusters while Haffner 17 and King 10 are younger objects with eccentricity
$\sim$ 0. Orbits of these clusters are confined in a box of $ 9.0 < R_{gc} \leq 10.8 $ kpc, $ 9.6 < R_{gc} < 11.6 $ kpc, 
$ 8.2 < R_{gc} < 10.6 $ kpc
and $ 9.8 < R_{gc} < 10.5 $ kpc for the clusters Czernik 14, Haffner 14, Haffner 17 and King 10, respectively. This indicates that
all the clusters are outside the solar circle and not interacting within the inner region of the Galaxy. 
But all the clusters are orbiting near the Galactic disk, so they may be affected by the
tidal forces of the disk which leads to a shorter life of the clusters. Carraro \& Chiosi (1994) found that clusters which orbit
in the outer region of the Galaxy can survive more as compared to the clusters which are in inner Galaxy. The similar result was also
found by Rangwal et. al (2019) for the cluster NGC 2506. Webb
et al. (2014) found that clusters having circular orbits evolve slower as compared to the eccentric ones. The clusters in our 
sample have circular orbits and hence they evolve slowly. Orbital parameters determined in the present analysis are very 
much similar to the parameters found by Wu et al. (2009), except their orbits, are more eccentric than what we found in the present
analysis.

\section{Conclusions}

In the present paper, we have investigated four poorly studied open clusters namely Czernik 14, Haffner 14, Haffner 17 and King 10
using the multi-colour photometric database along with Gaia~DR2 astrometry. The reliable fundamental parameters have been estimated for these
clusters and are listed in Table \ref{para}. Our main findings of the present analysis are following:

\begin{enumerate}

\item The new center coordinates are estimated as ($\alpha^{\circ} = 3^{h} 17^{m} 00^{s}$, $\delta^{\circ}= 58^{\circ} 35^{\prime}
24^{\prime\prime})$ for Czernik 14, ($\alpha^{\circ} = 7^{h} 44^{m} 48^{s}$, $\delta^{\circ} =-28^{\circ} 22^{\prime} 12^{\prime\prime})$
for Haffner 14, ($\alpha^{\circ} = 7^{h} 51^{m} 31^{s}$, $\delta^{\circ} = -31^{\circ} 49^{\prime} 48^{\prime\prime})$ for Haffner 17 and 
($\alpha^{\circ} = 22^{h} 54^{m} 53^{s}$, $\delta^{\circ} = 59^{\circ} 10^{\prime} 12^{\prime\prime})$ for King 10.

\item  Cluster extent is determined as 3.5 arcmin (2.9 parsec), 3.7 arcmin (4.5 parsec), 6.2 arcmin (6.5 parsec) and 5.7 arcmin (6.3 parsec)
for Czernik 14, Haffner 14, Haffner 17 and King 10 respectively.

\item  We have estimated the mean proper motion in both RA and DEC directions as ($-0.42\pm0.02$, $-0.38\pm0.01) ~mas~yr^{-1}$
for Czernik 14, ($-1.82\pm0.009$, $1.73\pm0.008) ~mas~ yr^{-1}$ for Haffner 14, ($-1.17\pm0.007$, $1.88\pm0.006) ~mas~yr^{-1}$ for Haffner 17
and ($-2.75\pm0.008$, $-2.04\pm0.006) ~mas~yr^{-1}$ for King 10.

\item  Colour-colour diagrams have been constructed after combining Gaia~ DR2, 2MASS, APASS, Pan-STARRS1 and 
WISE database. The diagrams indicate that interstellar extinction law is normal towards the cluster's region.
Interstellar reddening $(E(B-V))$ have been determined as 0.96, 0.38, 1.29 and 1.09 mag for the clusters Czernik 14, Haffner 14,
Haffner 17 and King 10 using $2MASS$ colours.

\item  Distances to the clusters Czernik 14, Haffner 14, Haffner 17 and King 10 are determined as $2.9\pm0.2$, $4.8\pm0.2$ kpc, $3.6\pm0.1$ kpc,
and $3.8\pm0.1$ kpc respectively using CMDs. These distances are supported by the values estimated using mean parallax of the clusters. Ages of 
$570\pm60$, $320\pm35$, $90\pm10$ and $45\pm5$ Myr are determined for Czernik 14, Haffner 14, Haffner 17 and King 10 respectively
by comparing with the theoretical isochrones of Z=0.019 taken from Marigo et al. (2017).

\item The LFs and MFs are determined by considering the members selected from Gaia proper motion database. The overall
mass function slopes $x=1.38\pm0.17$, $1.27\pm0.10$, $1.37\pm0.08$ and $1.29\pm0.13$ are derived for Czernik 14, Haffner 14, Haffner 17 and
King 10 respectively. The MF slopes are in good agreement with the Salpeter (1955) value for the clusters under study. Total mass was estimated
as $\sim$348 $M_{\odot}$, $\sim$595 $M_{\odot}$,  $\sim$763 $M_{\odot}$ and $\sim$1088 $M_{\odot}$ for clusters Czernik 14, Haffner 14, Haffner 17
and King 10 respectively.

\item Evidence of mass-segregation was observed for these clusters using probable cluster members. The K-S test shows the confidence
level of mass-segregation as 91 $\%$, 88 $\%$, 75 $\%$ and 77 $\%$ for Czernik 14, Haffner 14, Haffner 17 and King 10 respectively. The 
cluster's age is higher than the relaxation time which indicates that all clusters are dynamically relaxed.

\item The mean radial velocities ($-55.12\pm1.04$ km/s for Czernik 14, $71.21\pm1.03$ km/s for Haffner 14, $49.97\pm2.09$ km/s for 
Haffner 17 and $-44.35\pm1.83$ km/s for King 10) are estimated from the Gaia~DR2 database. Galactic orbits and orbital parameters are determined 
using Galactic potential models. We found that these objects are orbiting in a boxy pattern. The different orbital parameters are 
listed in Table \ref{inp} and \ref{orpara} for the clusters under study.

\end{enumerate}

{\bf ACKNOWLEDGEMENTS}\\

This work has been financially supported by the Natural Science Foundation of China (NSFC-11590782, NSFC-11421303). 
This work has made use of data from the European Space Agency (ESA) mission Gaia (https://www.cosmos.esa.int/gaia), 
processed by the Gaia Data Processing and Analysis Consortium (DPAC, https://www.cosmos.esa.int/web/gaia/dpac/consortium). 
Funding for the DPAC has been provided by national institutions, in particular the institutions participating in the 
Gaia Multilateral Agreement.
%This work has made use of data from the European Space Agency (ESA) mission Gaia processed by Gaia Data processing 
%and Analysis Consortium (DPAC), (https://www.cosmos.esa.int/web/gaia/dpac/consortium). 
In addition to this, It is worthy to mention that, this work has been done by using WEBDA and the data products from 
the Two Micron All Sky Survey $(2MASS)$, which is a joint project of the University of Massachusetts
and the Infrared Processing and Analysis Center/California Institute of Technology, funded by the National Aeronautics and Space
Administration and the National Science Foundation (NASA).

	\end{document}